\documentclass[preprint2,tighten]{aastex6}
\pdfoutput=1

\shorttitle{Flux and polarization signals across the O$_2$ A--band}
\shortauthors{Fauchez et al.}

\begin{document}

\title{The O$_2$ A--band in fluxes and polarization of starlight \\
       reflected by Earth--like exoplanets}
       
\thanks{\bf Accepted for publication by ApJ, April 18th, 2017}

\author{Thomas Fauchez}
\affil{Laboratoire d'Optique Atmosph\`erique (LOA) \\
       UMR 8518, Universit\'e Lille 1, Villeneuve d’Ascq, France }

\author{Loic Rossi and Daphne M. Stam}
\affil{Faculty of Aerospace Engineering, Delft University of Technology \\
Kluyverweg 1, 2629 HS Delft, The Netherlands}


\begin{abstract}
Earth--like, potentially habitable exoplanets are prime targets in the search 
for extraterrestrial life. Information about their atmosphere and surface 
can be derived by analyzing light of the parent star reflected by the planet. 
We investigate the influence of the surface albedo $A_{\rm s}$, the optical 
thickness $b_{\rm cloud}$ and altitude of water clouds, and the mixing ratio 
of biosignature O$_2$ on the strength of the O$_2$ A--band 
(around 760 nm) in 
flux and polarization spectra of starlight reflected by Earth--like exoplanets. 
Our computations for horizontally homogeneous planets show that small mixing 
ratios ($\eta < 0.4$) will yield moderately deep bands in flux and moderate 
to small band strengths in polarization, and that clouds will usually decrease 
the  band depth in flux and the band strength in polarization. However, cloud 
influence will be strongly dependent on their properties such as optical 
thickness, top altitude, particle phase, coverage fraction, horizontal 
distribution. Depending on the surface albedo, and cloud properties, different 
O$_2$ mixing ratios $\eta$ can give similar absorption band depths in flux 
and band strengths in polarization, in particular  if the clouds have moderate 
to high optical thicknesses. Measuring both the flux and the polarization 
is essential to reduce the degeneracies, although it will not solve them, 
in particular not for horizontally inhomogeneous planets. Observations at 
a wide range of phase angles and with a high temporal resolution could help 
to derive cloud properties and, once those are known, the mixing ratio 
of O$_2$ or any other absorbing gas.
\end{abstract}




\section{Introduction}
\label{sect1}

After more than two decades of exoplanet detections, statistics shows 
that, on average, every star in the Milky Way has a planet, and that at least 
20$\%$ of the solar--type stars have a rocky planet in their habitable zone 
\citep[][]{Petigura26112013}, 
the region around a star where planets receive the right amount of energy 
to allow water to be liquid on their surface \citep[see, e.g.][]{kasting1993}
(assuming they have a solid surface)\footnote{Moons that orbit planets in a
habitable zone could  also be habitable}.
Recently, Proxima Centauri, the star closest to our Sun, was shown 
to host a potentially rocky planet in its habitable zone 
\citep[]{2016Natur.536..437A}.
Planets in habitable zones are prime targets in the search for extraterrestrial 
life because liquid water is essential for life as we know it. Whether or not
a rocky planet has liquid surface water  also depends on the thickness,
composition and structure of its atmosphere.
Narrowing down planets in our search for extraterrestrial life thus 
requires the characterisation of planetary atmospheres in terms of 
composition and structure, as well as surface pressure and albedo.
Of particular interest is the search for biosignatures, 
i.e. traces of present or past life, such as the atmospheric gases oxygen 
and methane, and for habitability markers, such as liquid surface water.

Gases like oxygen and methane are too chemically reactive to remain in 
significant amounts in any planetary atmosphere without continuous 
replenishment. The current globally averaged mixing ratio of 
biosignature (and greenhouse gas) methane is much smaller than that 
of dioxygen, i.e.\ only about 1.7~$\cdot~10^{-6}$. Also due to its distinct
sources, its distribution varies both horizontally and vertically across the
Earth and in time. The dioxygen mixing ratio in the current Earth's atmosphere
is about 0.21 and virtually altitude--independent. Although oxygenic
photosynthetic organisms appeared about 3.5~$\cdot~10^9$ years ago, the 
oxygen they produced was efficiently chemically removed from the atmosphere
by combining with dissolved iron in the oceans to form banded iron formations
\citep[][]{crowe2013}.
It is thought that when this oxygen sink became saturated, the atmospheric 
free oxygen started to increase in the so--called Great Oxygenation Event (GOE) 
around 2.3~$\cdot~10^9$~years ago. While after the GOE, the oxygen mixing ratio
remained fairly low and constant at about 0.03 for about 10$^9$~years, it
started to rise rapidly about 10$^9$~years ago to maximum levels of 0.35 about
2.80~$\cdot~10^8$~years ago. Since then, the ratio has leveled off to its
current value \citep[][]{crowe2013}.
The triatomic form of oxygen, ozone, is formed by photodissociation of 
dioxygen molecules. Ozone protects the Earth's biosphere from harmful 
UV--radiation by absorbing it. The ozone mixing ratio is variable 
and shows a prominent peak between about 20~and 30~km of altitude, 
the so--called ozone--layer.

In this paper, we investigate the planetary properties that determine the 
appearance of gaseous absorption bands in spectra of starlight reflected 
by exoplanets with Earth--like atmospheres. We concentrate on the 
so--called O$_2$ A--band, centered around 760~nm, 
the strongest absorption band of O$_2$ across the visible.
The advantage of concentrating on this band is not only that it appears to be 
a strong biosignature, but also that the range of absorption optical 
thicknesses across the band is large and thus probes virtually all altitudes
within an atmosphere (assuming it is well--mixed throughout the atmosphere). 
The identification of biosignatures like oxygen and methane in an
exoplanet signal will depend on the presence of the spectral features they 
leave in a planetary spectrum. The retrieval of the mixing ratio of an 
atmospheric gas will rely on the strength on the spectral features. This 
strength with respect to the continuum surrounding a feature will 
depend on the intrinsic strength of the feature, i.e.\ the absorption 
cross--section of the molecules and their atmospheric column number 
density (in molecules m$^{-2}$). It will also be affected by clouds in the
atmosphere, as they will cover (part) of the absorbing molecules, and as 
they will change the optical path lengths of the incoming photons and 
hence change the amount of absorption \citep[][]{fujii2013}.
The precise influence of clouds will depend on the (horizontal and vertical) 
distribution of the absorbing gases and on the cloud properties: their 
horizontal and vertical extent, cloud particle column number densities, 
the cloud particle microphysical properties, such as particle size distribution, 
composition, and even shape. Through the cloud particle microphysical 
properties, the influence of clouds on spectral features of atmospheric gases
will thus also depend on the wavelength region under consideration.

On Earth, the O$_2$ mixing ratio is known and constant up to high
altitudes. Therefore, as postulated by \cite{yamamoto1961} and demonstrated by 
\citet{1991JApMe..30.1245F} and \citet{1991JApMe..30.1260F}, the depth of the
O$_2$ A--band in spectra of sunlight reflected by a region of the Earth that 
is covered by an optically thick cloud layer allows to estimate cloud top 
altitudes. Because of the strength of the O$_2$ A--band, this method is
sensitive to both high and low clouds and appears to be insensitive to temperature inversions. The method is widely applied both 
to measurements taken from airplanes \citep[e.g.][]{lindstrot2006} and to
satellite data \citep[see][]{saiedy1965,1998GeoRL..25.3159V,koelemeijer2001,preusker2007,lelli2012,desmons2013}.
However, because this method only accounts approximately for the penetration 
and multiple scattering of photons inside the cloud, it tends to systematically
overestimate cloud top pressures (hence it underestimates cloud top altitudes) 
\citep[][]{vanbauce1998}. The retrieved pressure appears to be more 
representative to the pressure halfway the cloud 
\citep[see][]{vanbauce2003,wang2008,sneep2008,ferlay2010,desmons2013}.

In Earth remote--sensing, the retrieval of cloud top altitudes is important 
for climate research and especially for the retrieval of atmospheric column 
densities of trace gases, such as ozone and methane, that will be partly 
hidden from the view of Earth--orbiting satellites when clouds are present.
Not surprisingly, there is little interest in deriving O$_2$ mixing ratios. 

In exoplanet research, however, the O$_2$ mixing ratio will be unknown,
and absorption band depths cannot be used to derive cloud top altitudes.
Indeed, the direct detection of exoplanetary radiation to investigate the 
depth of gaseous absorption bands is extremely challenging, because of the huge 
flux contrast between a parent star and an exoplanet and the small angular 
separation between the two. \citet{2013Sci...339.1398K} were the first to 
succeed in capturing a thermal spectrum of one of the exoplanets around the 
star HR~8799 through spatially separating it from its star. The spectrum of 
this young and hot, and thus thermally bright, planet shows molecular lines 
from water and carbon monoxide. Because of their moderate temperatures, 
potentially habitable exoplanets will not be very luminous at infrared 
wavelengths and the relatively small size of rocky exoplanets will require 
highly optimized telescopes and instruments for their characterization.
Examples of current instruments that aim at spatially resolving large, 
gaseous, old and cold exoplanets from their parent star and characterizing 
them from their directly detected signals are SPHERE (Spectro-Polarimetric 
High-contrast Exoplanet Research)
\citep[see][and references therein]{2006Msngr.125...29B} on ESO's Very Large 
Telescope (VLT), GPI (Gemini Planet Finder) \citep[see][]{2014AAS...22322902M} 
on the Gemini North telescope, CHARIS (Coronographic High Angular Resolution 
Imaging Spectrograph) \citep[see][]{groff2014} on the Subaru telescope,
and HROS (High--Resolution Optical Spectrograph) for the future TMT (Thirty
Meter Telescope) \citep[][]{2006SPIE.6269E..1VF,2006SPIE.6269E..30O}.
The future European Extremely Large Telescope (E--ELT) also has the
characterization of Earth--like exoplanets as one of its main science cases.

Both SPHERE and GPI can measure not only thermal fluxes that their target 
planets emit and fluxes of starlight that the planets reflect, but they can 
also measure the state of polarization of the planetary radiation. 
In particular, SPHERE has a polarimetric optical arm that is based
on the ZIMPOL (Z\"{u}rich Imaging Polarimeter) technique \citep[][]{2005ASPC..343...89S,2004SPIE.5492..463G} 
(IRDIS, an infrared arm of SPHERE, has polarimetric capabilities designed for 
observations of circumstelllar matter, but potentially of use for exoplanet 
detection, too). Polarimetry is also a technique that will be used in EPICS, 
the Earth--like Planet Imaging Camera System \citep[][]{2011IAUS..276..343G,2010SPIE.7735E.212K}, that is being planned for the E--ELT. 
First detections of polarimetric signals of exoplanets have been claimed 
\citep[see][and references therein]{2015ApJ...813...48W,2016MNRAS.459L.109B}.

There are several advantages of using polarimetry in exoplanet research.
Firstly, light of a solar--type star can be assumed unpolarized 
\citep[see][]{1987Natur.326..270K} when integrated across the stellar disk,
while starlight that has been reflected by a planet will usually be (linearly)
polarized 
\citep[see e.g.][]{2000ApJ...540..504S,2004A&A...428..663S,2008A&A...482..989S}.
Polarimetry can thus increase the much--needed contrast between a planet and its 
parent star \citep[][]{2006SPIE.6269E..26K} and facilitate the direct detection
of an exoplanet. Secondly, detecting a polarized object in the vicinity of a 
star would immediately confirm the planetary nature of the object, as stars or
other background objects will have a negligible to low degree of polarization.
Thirdly, the state of polarization of the starlight (in particular as functions
of the planetary phase angle and/or wavelength) that has been reflected by the
planet is sensitive to the structure and composition of the planetary 
atmosphere and surface, and could thus be used for characterizing the planet, 
e.g.\ by detecting clouds and hazes and their composition.
A famous example of this application of polarimetry is the derivation of 
the size and composition of the cloud droplets that form the ubiquitous
Venus clouds from disk--integrated polarimetry of reflected sunlight at three 
wavelengths and across a broad phase angle range by \citet{1974JAtS...31.1137H} 
(thus similar observations as would be available for direct exoplanet 
observations, with the exoplanet's phase angle range depending on the orbital 
inclination angle). These cloud particle properties, that were later 
confirmed by in--situ measurements, could not be derived from the spectral 
and phase angle dependence of the sunlight's reflected flux, because flux 
phase functions are generally less sensitive to the microphysical properties 
of the scattering particles. For exoplanets, \citet{2012A&A...548A..90K} and
\citet{2007AsBio...7..320B} have numerically shown that the primary rainbow 
of starlight that has been scattered by liquid water cloud particles 
on a planet should be observable for relatively small water cloud coverage 
($10\%-20\%$), even when the liquid water clouds are partly covered by ice 
water clouds (which themselves do not show the rainbow feature).
In Earth--observation, the PARASOL/POLDER instrument--series 
\citep[][]{2007ApOpt..46.5435F,1994Deschamps} uses polarimetry to determine 
the phase of the (water) clouds it observes 
\citep[see e.g.][]{2000JGR...10514747G}.

In this paper, we not only investigate influences on the O$_2$ A--band in 
flux spectra of starlight that is reflected by exoplanets, but also in 
polarization spectra. Indeed, gaseous absorption bands not only show up 
in flux spectra of light reflected by (exo)planets, they usually also appear 
in polarization spectra \citep[see][for examples in the Solar System]{2008ApOpt..47.3467B,2007A&A...463.1201J,2004A&A...428..663S,2001GeoRL..28..519A,1999JGR...10416843S}. 

There are two main reasons why absorption bands appear in polarization 
spectra, despite polarization being a relative measure, i.e.\ the polarized
flux divided by the total flux. Firstly, with increasing absorption, the 
reflected light contains less multiple scattered light, which usually has a
lower polarization than the singly scattered light. The relative increase
of the contribution of singly scattered light to the reflected signal thus
increases its degree of polarization.
Secondly, with increasing absorption, the altitude at which most of the
reflected light has been scattered increases. If different altitude regions
of the atmosphere contain different types of particles, with different single
scattering polarization signatures, the polarization will vary across an 
absorption line, with the degree of polarization in the deepest part
of the line representative for the particles in the higher atmospheric
layers, and the polarization in the continuum representative for the 
particles in the lower, usually denser atmospheric layers.
For an in--depth explanation of these effects, see \citet{1999JGR...10416843S}. 
Note that while attenuation through the Earth's atmosphere will change
the flux of an exoplanets, it does not change the degree
of polarization across gaseous absorption bands in a spectrum of a planet or 
exoplanet. This is an additional advantage of using polarimetry for 
the detection of gaseous absorption bands with ground--based telescopes,
in particular when (exo)planet observations are pursued in wavelength
regions where the Earth's atmosphere itself absorbs light.

The results presented in this paper can not only be used to investigate
the retrieval of trace gases and cloud properties of exoplanets. 
They will also be useful for the design and optimization of spectrometers
for exoplanetary detection and characterization: the optical
response of mirrors, lenses, and e.g. gratings usually depends on the 
degree and direction of the light that is incident on them, and when
observing a polarized signal, such as starlight that has been reflected
by an exoplanet, the detected flux signal will depend on the degree and
direction of polarization of the incoming light. In particular, the 
detected depth of a gaseous absorption band, and hence the gaseous 
mixing ratio that will be derived from it, thus depends on the polarization
across the band. Even if a telescope's and/or instrument's 
polarization sensitivities are fully known, detected fluxes can only be 
accurately corrected for polarization sensitivities if the polarization 
of the observed light is measured as well 
\citep[see][for examples of such corrections]{2000JGR...10522379S}.
In the absence of such polarization measurements, numerical simulations 
such as those presented in this paper can help to assess the uncertainties.

The structure of this paper is as follows.
In Sect.~\ref{sect2}, we describe our method for calculating the flux and 
polarization of starlight that is reflected by an exoplanet, including our 
disk--integration technique and how we handle the spectral computations.
In Sect.~\ref{sect3}, we present our numerical results for cloud--free, 
completely cloudy and partly cloudy exoplanets.
Finally, in Sects.~\ref{sect:discussion} and~\ref{sect4}, we discuss and 
summarize our results.


\section{Calculating reflected starlight}
\label{sect2}


\subsection{Flux vectors and polarization}
\label{sect2.1}

\begin{figure}
\figurenum{1}
\epsscale{1.15}
\plotone{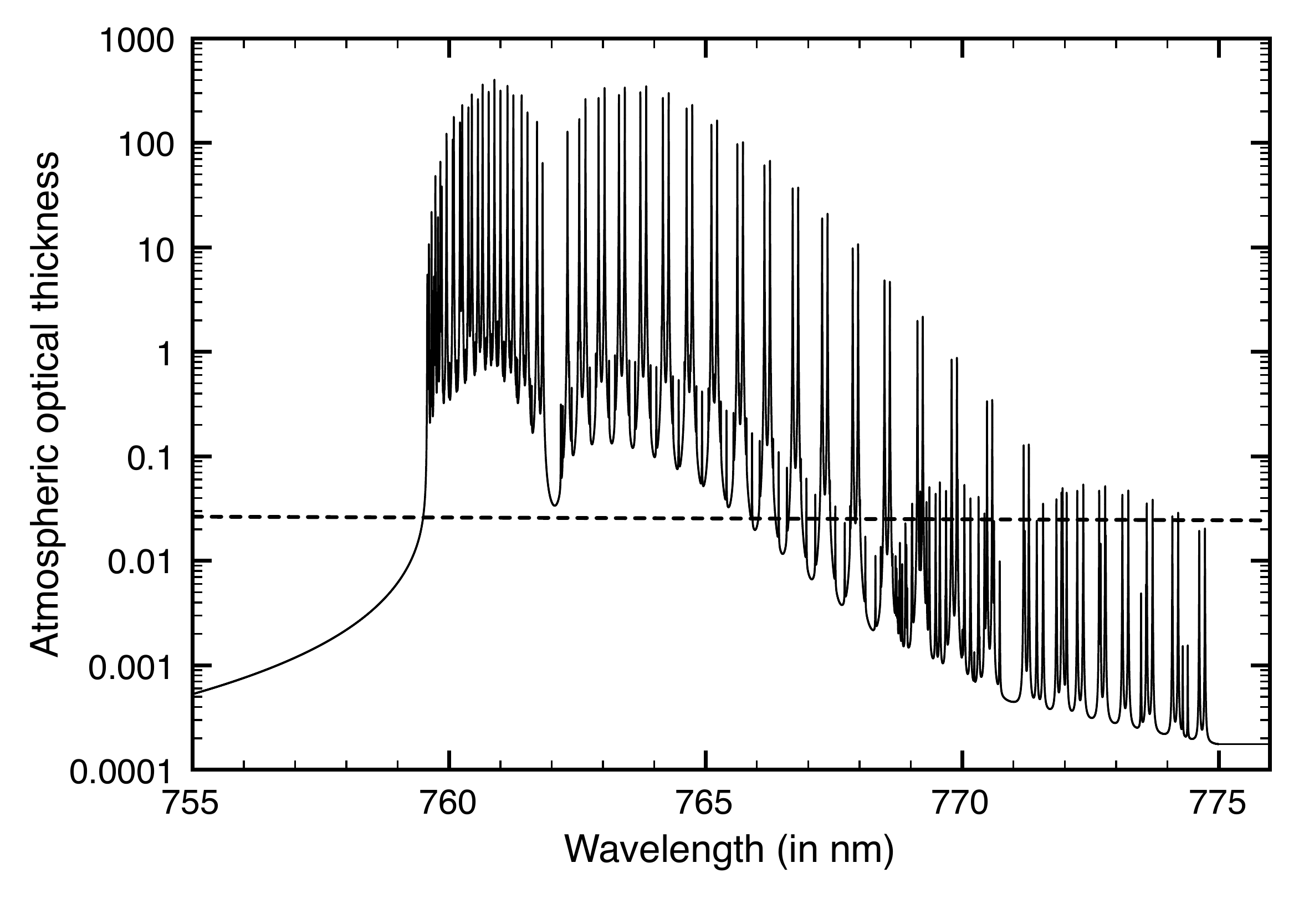}
\caption{Solid line: the gaseous absorption optical thickness of the
         Earth's atmosphere across the O$_2$ A--band, computed for a model
         atmosphere, using a mid--latitude summer pressure--temperature 
         profile \citep[][]{1972McClatchey} and absorption line parameters 
         from \citet{2005JQSRT..96..139R}. 
         Dashed line: the computed gaseous scattering optical thickness of 
         the same model atmosphere. Its value decreases from 0.027 at 755~nm 
         to 0.024 at~775 nm. Details on the used absorption line profiles and  
         computation can be found in \citet{2000JQSRTStam}.}
\label{fig1}
\end{figure}

The flux and state of polarization of starlight that is reflected by a 
spatially unresolved exoplanet and that is received by a distant observer, 
is fully described by a flux (column) vector, as follows 
\begin{equation}
   \pi {\bf F} =
   \pi \left[ F, Q, U, V \right],
\label{eq_stokesvector}
\end{equation}
with $\pi F$ the total flux, 
$\pi Q$ and $\pi U$ the linearly polarized fluxes, 
defined with respect to a reference plane,
and $\pi V$ the circularly polarized flux
\citep[for details on these parameters, see e.g.][]{2004Hovenier,
1974SSRv...16..527H}.
We use the planetary scattering plane, i.e. the plane through the centers of the 
planet, the star and the observer, as the reference plane for 
parameters $Q$ and $U$. 

Integrated over the stellar disk, light of a solar--type star
can be assumed to be virtually unpolarized \citep[][]{1987Natur.326..270K}.
We thus describe its flux vector as  
${\bf \pi F_0}= \pi F_0 \left[ 1, 0, 0, 0 \right] = \pi F_0 {\bf 1}$, 
with $\pi F_0$ the stellar flux measured perpendicular to the 
direction of propagation of the light, and 
${\bf 1}$ the unit (column) vector.

Integrated over the illuminated and visible part of a planetary
disk, the starlight that is reflected by a planet will usually be
linearly polarized with the degree of polarization depending on the 
properties of the planetary atmosphere and surface (if present)
\citep[see, e.g.][]{2008A&A...482..989S,2006A&A...452..669S,2004A&A...428..663S}.
The reflected starlight can also be partly circularly polarized, because
our model atmospheres contain not only Rayleigh scattering gases but also
cloud particles (see Sect.~\ref{sect2.2}). While Rayleigh scattering alone 
doesn't circularly polarize light, light that has been scattered once and 
is linearly polarized, can get circularly polarized when it is 
scattered by cloud particles. The circularly polarized flux $V$ of
a planet is usually very small 
\citep[see][]{1971Natur.232..165K,1974SSRv...16..527H,1978Icar...33..217K}, 
in particular when integrated over the planetary disk
\citep[][in preparation]{RossiStam2017}.
In the following, we therefore neglect $V$. This doesn't introduce 
significant errors in $F$, $Q$, and $U$ \citep[see][]{2005HovenierStam}.

We define the degree of linear polarization of the reflected light as
\begin{equation}
   P = \frac{\sqrt{Q^2 + U^2}}{F},
\label{eq_polarization}
\end{equation}
which is independent of the choice of reference plane. 
In case $U=0$, which is true for planets that are mirror--symmetric with
respect to the reference plane, the direction of polarization can be
included into the definition of the degree of polarization:
\begin{equation}
  P_{\rm s}= -Q/F
\label{eq_polarization2}
\end{equation}
If $U=0$ and $Q < 0$, the light is polarized parallel to
the reference plane and $P_{\rm s} \geq 0$, while if $U=0$ and $Q \geq 0$,
the light is polarized perpendicular to the reference plane and 
$P_{\rm s} < 0$.\\


\subsection{The planetary model atmospheres and surfaces}
\label{sect2.2}

\begin{figure*}
\figurenum{2}
\plottwo{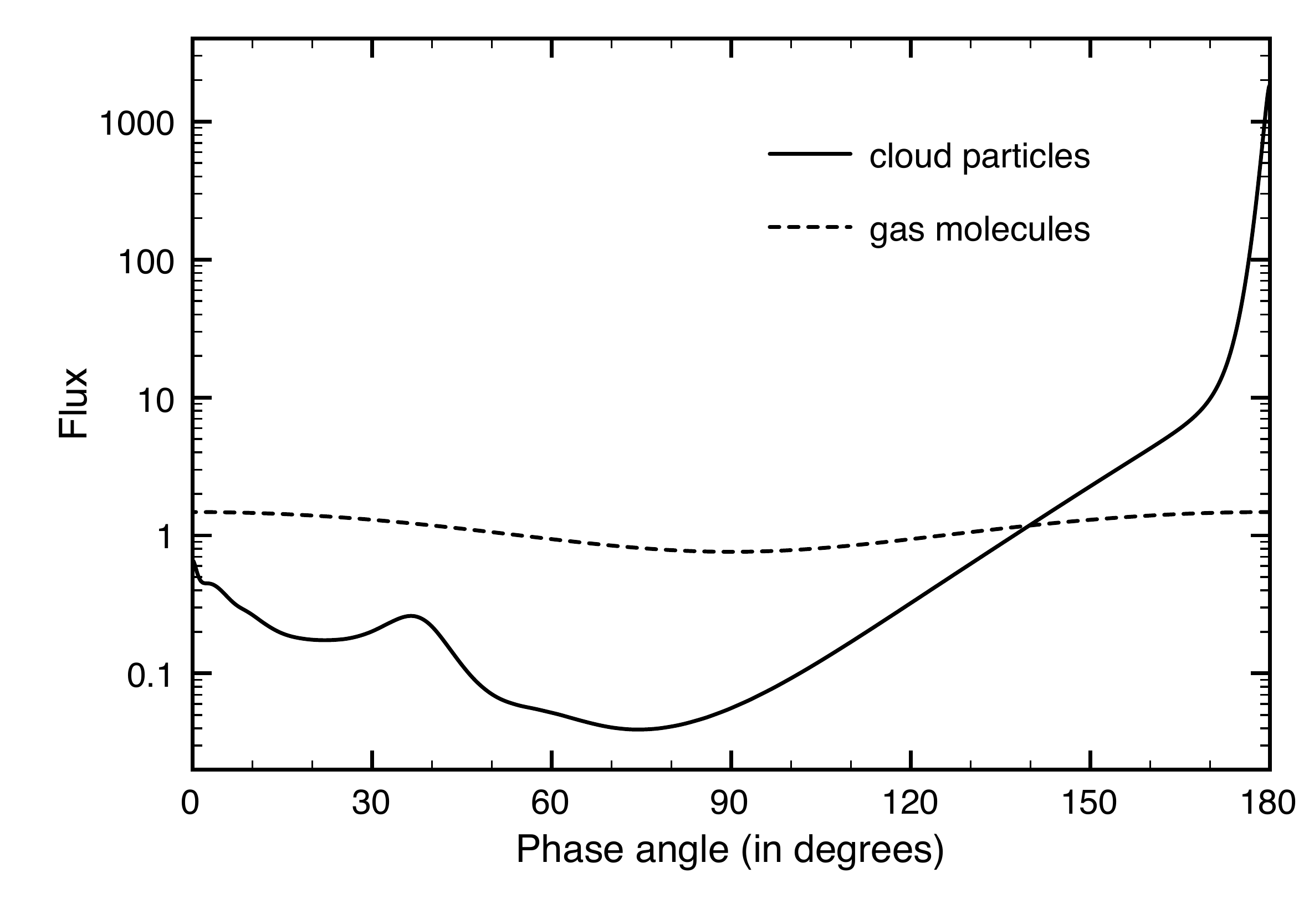}{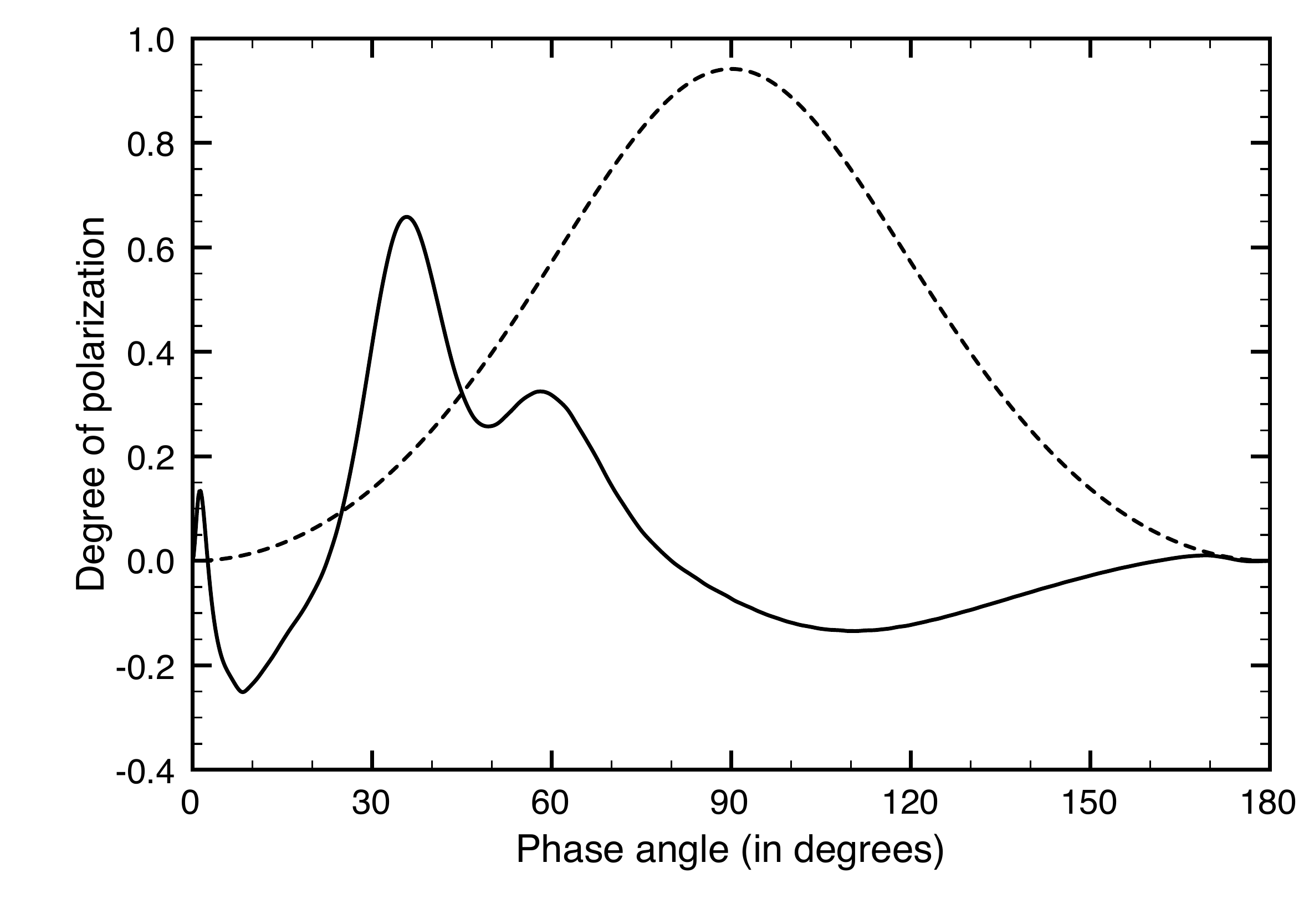}
\caption{The total flux (left) and degree of linear polarization (right) of
         unpolarized incident light at $\lambda= 765$~nm, that is singly
         scattered by samples of the cloud particles (solid lines) and 
         samples of the Rayleigh scattering gas molecules
         (dashed lines) as functions of the phase angle (i.e. 180$^\circ$ - 
         the single scattering angle). The flux and polarization peak around
         35$^\circ$ in the cloud particle curves is the primary rainbow.}
\label{fig2}
\end{figure*}

The atmospheres of our model planets are composed of stacks of locally
horizontally homogeneous layers, containing gas molecules and, optionally, 
cloud particles. We assume the gas is terrestrial air and use 
pressure--temperature profiles representative for the Earth
\citep[][]{1972McClatchey}. 

For our model planets, we calculate $b_{\rm abs}$, the gaseous absorption 
optical thickness of the atmosphere, as the integral of the mixing ratio $\eta$ 
of the absorbing molecules times the gaseous number density (in m$^{-2}$) 
times the absorption cross--section $\sigma_{\rm abs}$ (in m$^2$) 
along the vertical direction. Both $\eta$ and $\sigma_{\rm abs}$ usually 
depend on the ambient pressure and temperature, and thus on the altitude.
Figure~\ref{fig1} shows the computed $b_{\rm abs}$ of the Earth's atmosphere
across the wavelength region with the $O_2$ A--band, with a 
spectral resolution high enough to resolve individual absorption lines. 
We have calculated this $b_{\rm abs}$ following \citet{2000JQSRTStam},
assuming that $O_2$ is well--mixed, with $\eta=0.21$.
Note that $b_{\rm sca}$, the gaseous scattering optical thickness of the 
Earth's atmosphere is about 0.0255 in the middle of the absorption band.

Figure~\ref{fig2} shows the flux and degree of linear polarization 
of unpolarized incident light that is singly scattered by 
a sample of gas molecules. Here, we used the Rayleigh
scattering matrix described by \citet{1974SSRv...16..527H} with a 
depolarization factor of 0.03.
The depolarization factor modifies the isotropic Rayleigh scattering 
matrix (that applies to molecules that are perfect dipoles) to that of 
most molecules found in planetary atmospheres, whose scattering exhibits 
some anisotropy \citep[for details, see][]{1981JOSA...71.1142Y}.
Although Fig.~\ref{fig2} pertains to singly scattered light, we use the
phase angle (i.e.\ $180 - \Theta$, with $\Theta$ the single scattering
angle) to facilitate the comparison with planetary light curves later on.

The cloud particles are spherical and consist of liquid water with a 
refractive index of 1.335. The cloud particles are distributed in size 
according to a log--normal size distribution 
\citep[see Eq.~2.56 in][]{1974SSRv...16..527H}, with an effective radius 
of 6.0~$\mu$m and an effective variance of~0.5. We calculate the single
scattering properties of the cloud particles using Mie--theory and the 
algorithm described by \citet{1984A&A...131..237D}.
Figure~\ref{fig2} shows the flux and degree of linear polarization
of unpolarized incident light that is singly scattered by 
a sample of the cloud particles at $\lambda=765$~nm. 
Because we only consider the ~20~nm wide wavelength region of the 
O$_2$ A--band, we ignore any wavelength dependence of the single 
scattering properties of cloud particles.

The surface below the atmospheres is locally horizontally homogeneous and
reflects Lambertian, i.e. isotropic and unpolarized, with a surface albedo
$A_{\rm s}$. While our model atmospheres and surfaces are {\em locally}
horizontally homogeneous, our model exoplanets can be {\em globally} 
horizontally inhomogeneous, for example, they can be covered by patchy 
clouds (see Sect.~\ref{sect3.3}).


\subsection{Integration across the planetary disk}

We perform the calculations of the starlight that is reflected by a spherical 
model planet with the same adding--doubling algorithm used by 
\citet{2008A&A...482..989S}, except here we use a (more computing time--
consuming) disk--integration algorithm that also applies to horizontally
inhomogeneous exoplanets, for example, those with patchy clouds.
We integrate across the illuminated and visible part of the planetary
disk as follows: \\

\noindent {\bf 1.} 
We divide the disk into equally sized, square 'detector' pixels. The more pixels, 
the higher the accuracy of the integration, in particular for large phase angles, 
but the longer the computing time. We use 100~pixels along the planet's equator
for every phase angle $\alpha$. Numerical tests show that with this number of
pixels, convergence is reached at all phase angles. \\

\noindent {\bf 2.} For each pixel and a given $\alpha$, we compute the
illumination and viewing geometries for the location on the planet in 
the center of the pixel. The local illumination geometries are $\theta_0$, 
the angle between the local zenith direction and the direction to the star, 
and $\phi_0$, the azimuthal angle of the incident starlight (measured in the
local horizontal plane). The local viewing geometries are $\theta$, the 
angle between the local zenith direction and the direction to the observer, 
and $\phi$, the azimuthal angle of the reflected starlight (measured in 
the local horizontal plane). For each pixel, we also compute $\beta$, 
the angle between the local meridian plane (which contains both the local 
zenith direction and the direction towards the observer) and the planetary
scattering plane. \\

\noindent {\bf 3.} For each pixel, atmosphere and surface properties, and 
the geometries for the location on the planet in the center of the pixel, 
we compute the locally reflected starlight with our adding--doubling 
algorithm, and rotate this flux vector from the local meridian plane to 
the planetary scattering plane \citep[][]{1983A&A...128....1H}. 
All rotated flux vectors are summed to obtain the disk--integrated 
flux vector. From that vector, the degree of polarization is obtained. \\

To avoid having to perform separate radiative transfer calculations
for pixels with different illumination and viewing geometries but with 
the same planetary atmosphere and surface, we calculate, for every
atmosphere--surface combination on a model planet,
the (azimuthal angle independent) coefficients of the Fourier--series 
in which the locally reflected flux vector can be expanded
\citep[see][]{1987A&A...183..371D} for a range of values of $\theta_0$ 
and $\theta$ (because polarization is included, each coefficient is 
in fact a column vector). With these pre-computed Fourier--coefficients,
we can efficiently evaluate the flux vector of the locally reflected 
starlight for each pixel and for every $\alpha$.

We normalize each disk--integrated flux vector such that the reflected 
total flux at $\alpha=0^\circ$ equals the planet's geometric albedo. 
With this normalization, and given the stellar luminosity, planetary 
orbital distance and radius, and the distance between the planet and the
observer, absolute values of the flux vector arriving at the observer can 
straightforwardly be calculated \citep[see Eqs.~5 and 8 of][]{2004A&A...428..663S}.
Because the degree of polarization is a relative measure, it is independent 
of absolute fluxes.


\subsection{Spectral computations}

\begin{figure*}
\figurenum{3}
\plottwo{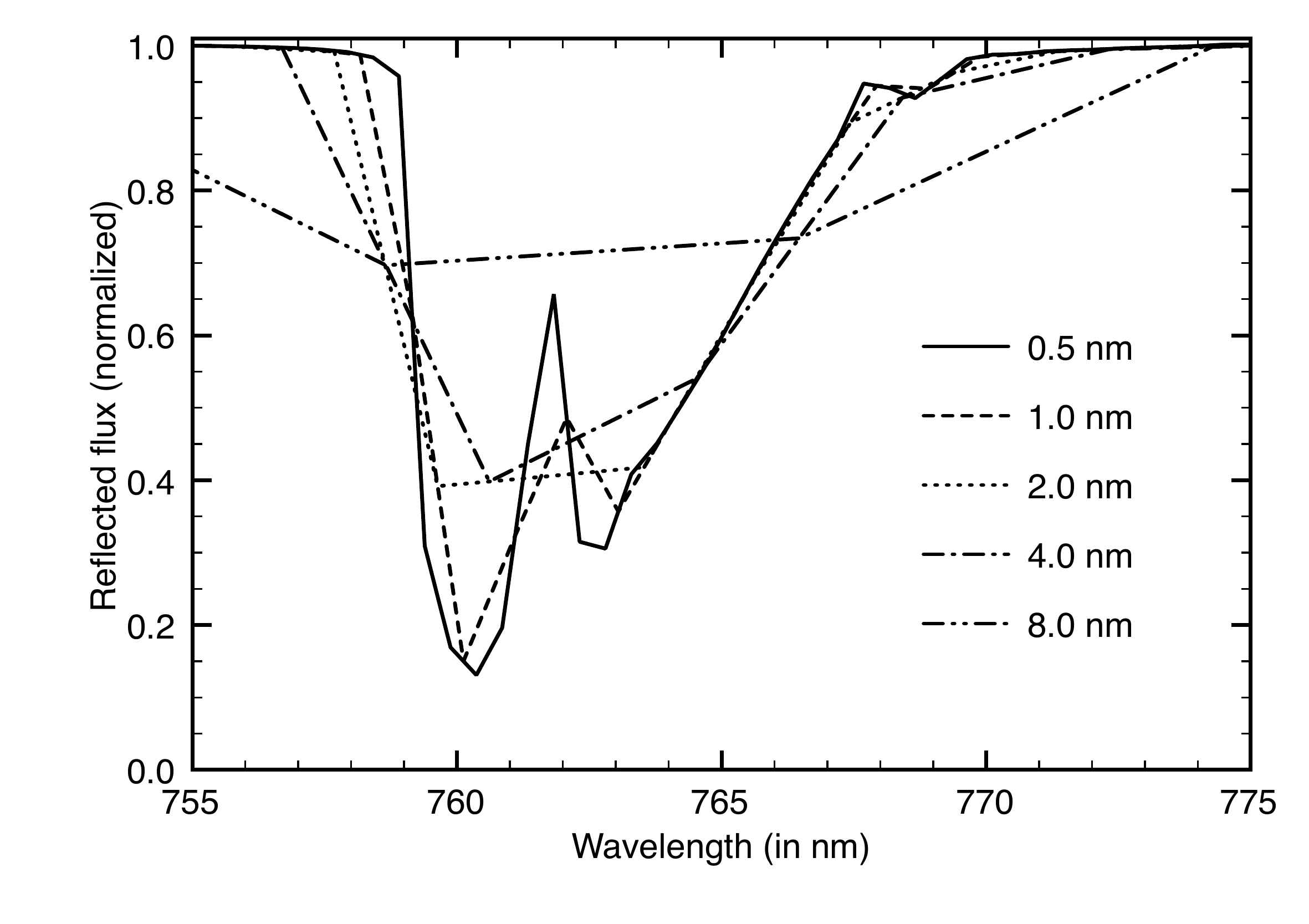}{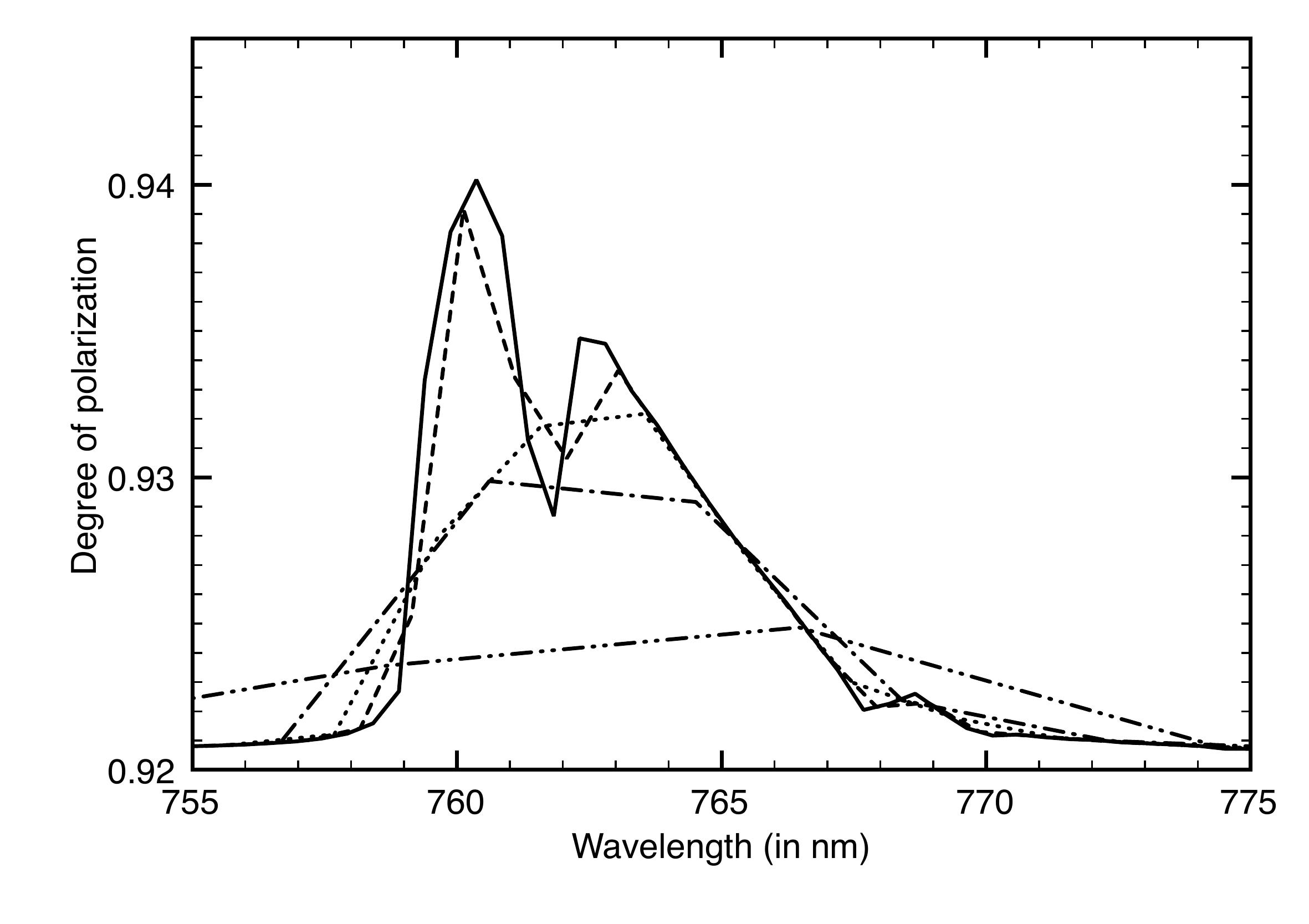}
\caption{Flux $F$ (left) and degree of polarization $P_{\rm s}$ (right) 
         of starlight
         reflected by a cloud--free planet with a black surface at 
         $\alpha=90^\circ$ for spectral bin widths $\Delta\lambda$ ranging 
         from 0.5~to 0.8~nm.}
\label{fig3}
\end{figure*}

When measuring starlight that has been reflected by an Earth--like exoplanet,
the spectral resolution across a gaseous absorption band will likely be 
much lower than shown in Fig.~\ref{fig1}. The most accurate simulations 
of low spectral resolution observations across a gaseous absorption band 
would require radiative transfer calculations at a spectral line resolving 
resolution (so--called line--by--line calculations), followed by a convolution 
with the actual instrumental spectral response function. Performing
line--by--line calculations while fully including polarization and multiple 
scattering for planets with cloudy atmospheres, and integrating across the 
planetary disk for various phase angles, 
requires several hours of computing time. 
Our approach to compute low spectral resolution spectra across the 
O$_2$ A--band is therefore based on the correlated $k$--distribution method
(from hereon the c$k$--method). 

A description of the c$k$--method without polarization was given by 
\citet{1991JGR....96.9027L}. The application of the c$k$--method for 
polarized radiative transfer has been described by \citet{2000JQSRTStam}.
That paper also includes a detailed comparison between simulations using the 
line--by--line method and the c$k$--distribution method, to assess the 
accuracy of the latter. According to \citet{2000JQSRTStam}, the absolute
errors in the degree of polarization across the O$_2$ A--band for sunlight
that is locally reflected by cloudy, Earth--like atmospheres due to using 
the c$k$--distribution method are largest in the deepest part of the band, 
but still less than 0.0025 for a spectral bin width $\Delta \lambda$ of 0.2~nm 
(the line--by--line method yields a slightly stronger polarization in the band). 
The errors decrease with increasing $\Delta \lambda$, and are much smaller
for cloud--free atmospheres because they arise due to scattering.
And in cloud--free Earth--like atmospheres, at these wavelengths, there is very 
little scattering.

For a given atmosphere--surface combination, we compute and store the 
Fourier coefficients of the locally reflected flux vectors for the range of 
atmospheric absorption optical thicknesses $b_{\rm abs}$ shown in 
Fig.~\ref{fig1}, i.e. 
from 0 (the continuum) to 400 (the strongest absorption line). Then, given 
an instrumental spectral resolution, we use the stored coefficients to 
efficiently calculate the reflected vector of the spectral bins $\Delta \lambda$ 
across the absorption band, given the distribution of $b_{\rm abs}$ in each
spectral bin. The integration across each $\Delta \lambda$ is performed with
Gauss--Legendre integration.
We use a block--function for the instrumental response function per spectral bin. 
As described in \citet{2000JQSRTStam}, the c$k$--distribution method can be
combined with other response functions. A block--function, however, 
captures the variation across the band without introducing more free parameters.

Figure~\ref{fig3} shows the disk--integrated reflected flux $F$ and degree of 
polarization $P_{\rm s}$ across the band of a cloud--free planet with a black 
surface 
for $\Delta \lambda$ ranging from 0.5~to 0.8~nm. The difference between
$F$ and $P_{\rm s}$ in the deepest part of the band and in the continuum 
decreases with increasing $\Delta \lambda$ as more wavelengths where 
$b_{\rm abs}$ is small fall within the spectral bin. This difference also 
depends on the O$_2$ mixing ratio, on the presence,
thickness and altitude of clouds, and on the surface albedo 
(see Sect.~\ref{sect3} for a detailed explanation of the shapes of the curves).
In the following, we use $\Delta \lambda= 0.5$~nm and 
50~Gaussian abscissae per spectral bin 
\citep[similar to what was used by][]{2000JQSRTStam}.


\section{Numerical results}
\label{sect3}


\subsection{Cloud--free planets}
\label{sect3.1}

\begin{figure*}[ht!]
\figurenum{4}
\plottwo{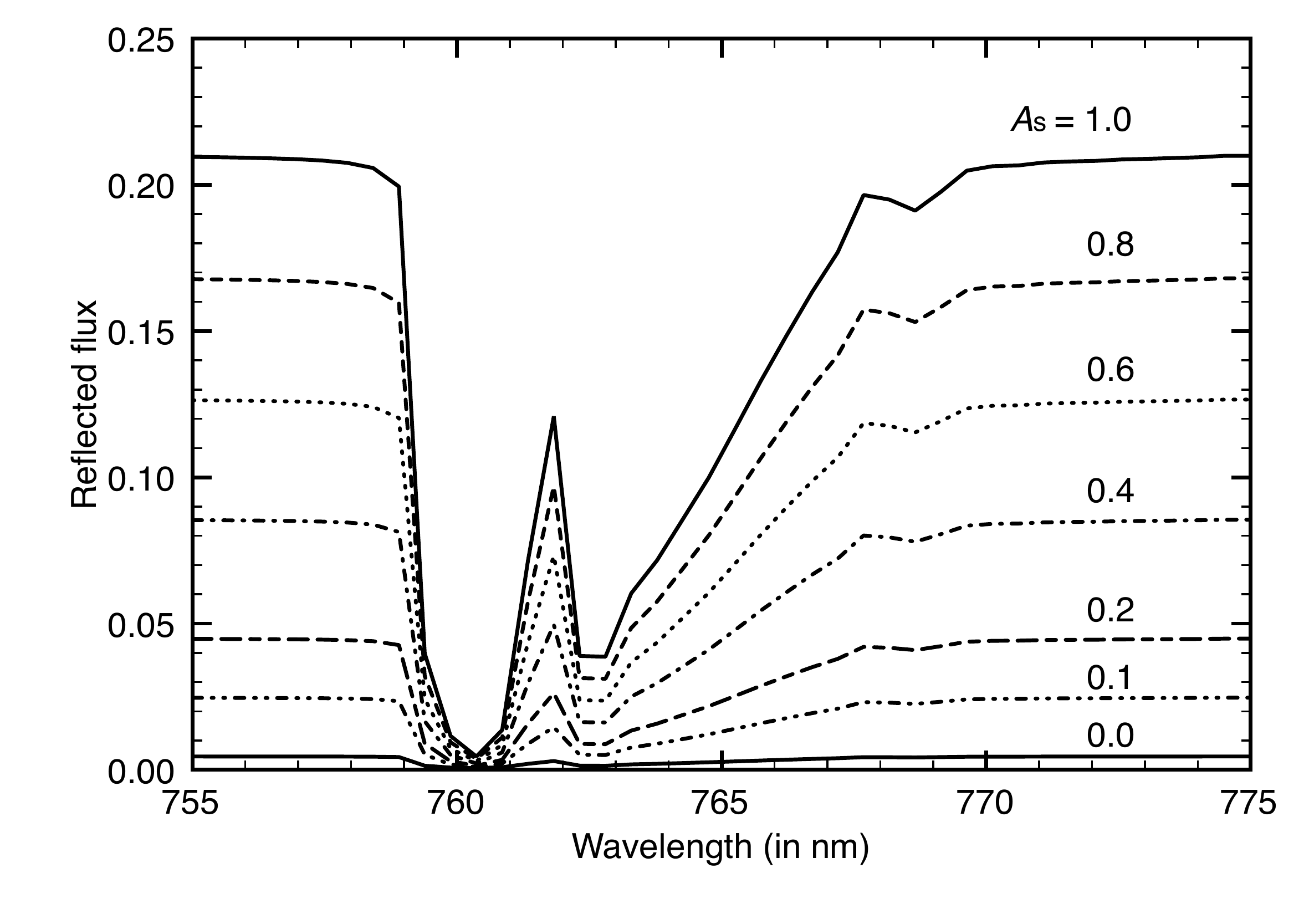}{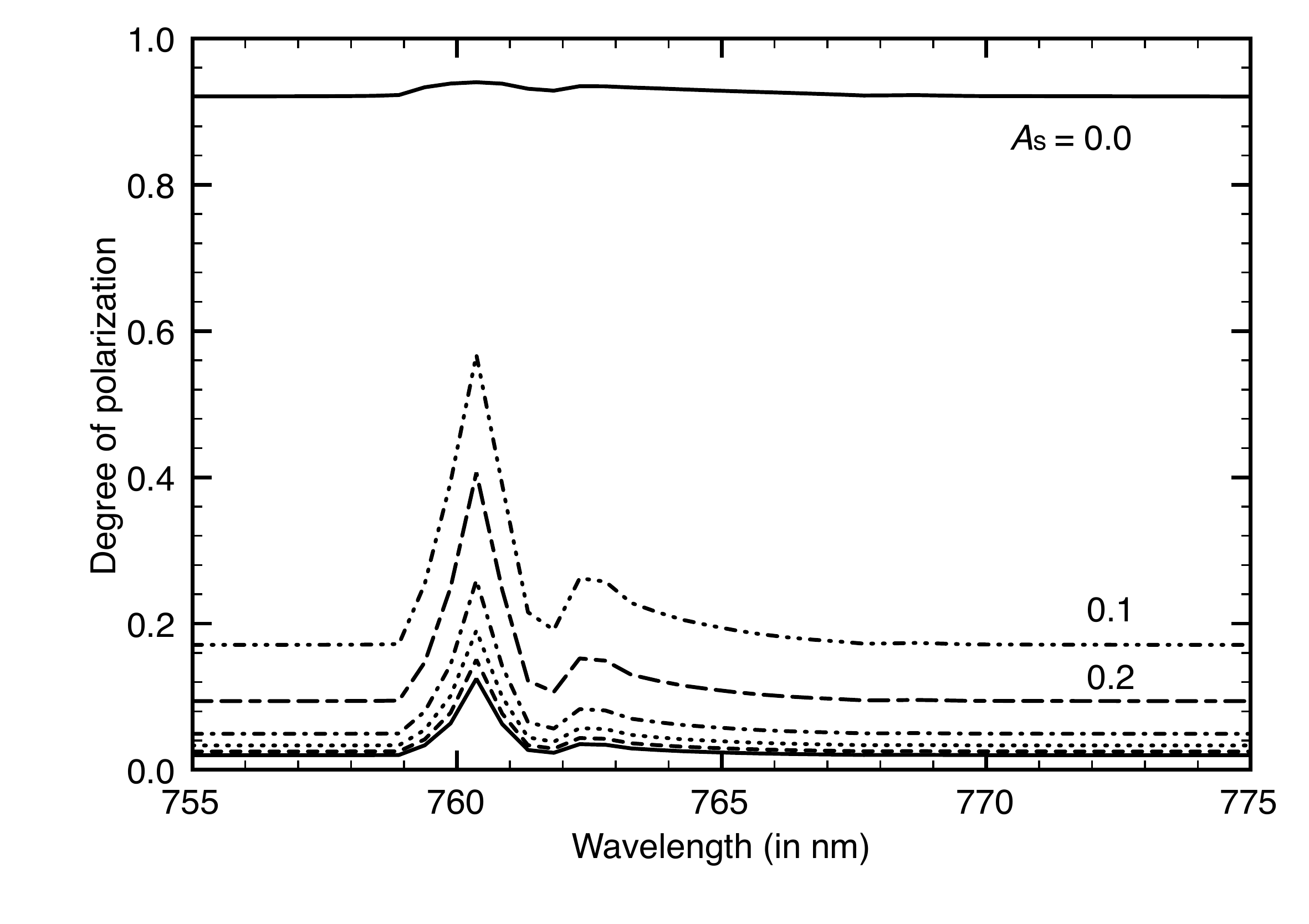}
\caption{Flux $F$ (left) and degree of polarization $P_{\rm s}$ (right) 
         of starlight
         reflected by cloud--free exoplanets at $\alpha=90^\circ$
         for different surface albedo's $A_{\rm s}$:
         0.0 (thin solid line, cf. Fig.~\ref{fig3} and note the different 
         polarization scale), 0.1 (dot--dot--dashed line),
         0.2 (short - long dashed line), 0.4 (dot--dashed line), 
         0.6 (dotted line), 0.8 (dashed line), 1.0 (thick solid line).}
\label{fig4}
\end{figure*}

\subsubsection{The influence of the surface albedo $A_{\rm s}$}

The first results to discuss are those for starlight reflected by model 
planets with gaseous, cloud--free atmospheres above surfaces with different 
albedo's. Figure~\ref{fig4} shows $F$ and $P_{\rm s}$ for surface albedo's 
$A_{\rm s}$
ranging from~0.0 to~1.0, at a phase angle $\alpha$ of $90^\circ$.
This is a very advantageous phase angle for the direct detection of an exoplanet,
because there its angular distance to its host star is largest. This phase 
angle also occurs at least twice every planetary orbit, independently of 
the orbital inclination angle.

For these cloud--free planets, $F$ and $P_{\rm s}$ in the continuum depend
strongly 
on $A_{\rm s}$, because the atmospheric gaseous scattering optical thickness 
$b_{\rm sca}$ in this spectral region is very small (see Fig.~\ref{fig1}).
The continuum $P_{\rm s}$ decreases with increasing $A_{\rm s}$, because of 
the increasing contribution of unpolarized light that has been reflected by 
the surface to the planetary signal: even if $A_{\rm s}$ is only slightly 
larger than zero, the surface contribution already strongly influences this 
signal.

For these cloud--free planets with gaseous atmospheres, polarization 
$P_{\rm s}$ will increase with increasing $b_{\rm abs}$, because the 
contribution of 
multiple scattered light, with usually a low degree of polarization, to the 
planetary signal will decrease.
Also, at wavelengths where the atmospheric absorption optical thickness
$b_{\rm abs}$ is large (see Fig.~\ref{fig1}), virtually no light will reach the
surface to subsequently emerge from the top of the atmosphere. At those 
wavelengths, $F$ and $P_{\rm s}$ are insensitive to $A_{\rm s}$. In Fig.~\ref{fig4}, 
$F$ and $P_{\rm s}$ in the deepest part of the absorption band 
(around 760.5~nm) do 
depend on $A_{\rm s}$, because they pertain to spectral bins and every 
bin includes wavelengths where $b_{\rm abs}$ is small, and thus light that 
has been reflected by the surface.

\begin{figure}
\figurenum{5}
\epsscale{1.1}
\plotone{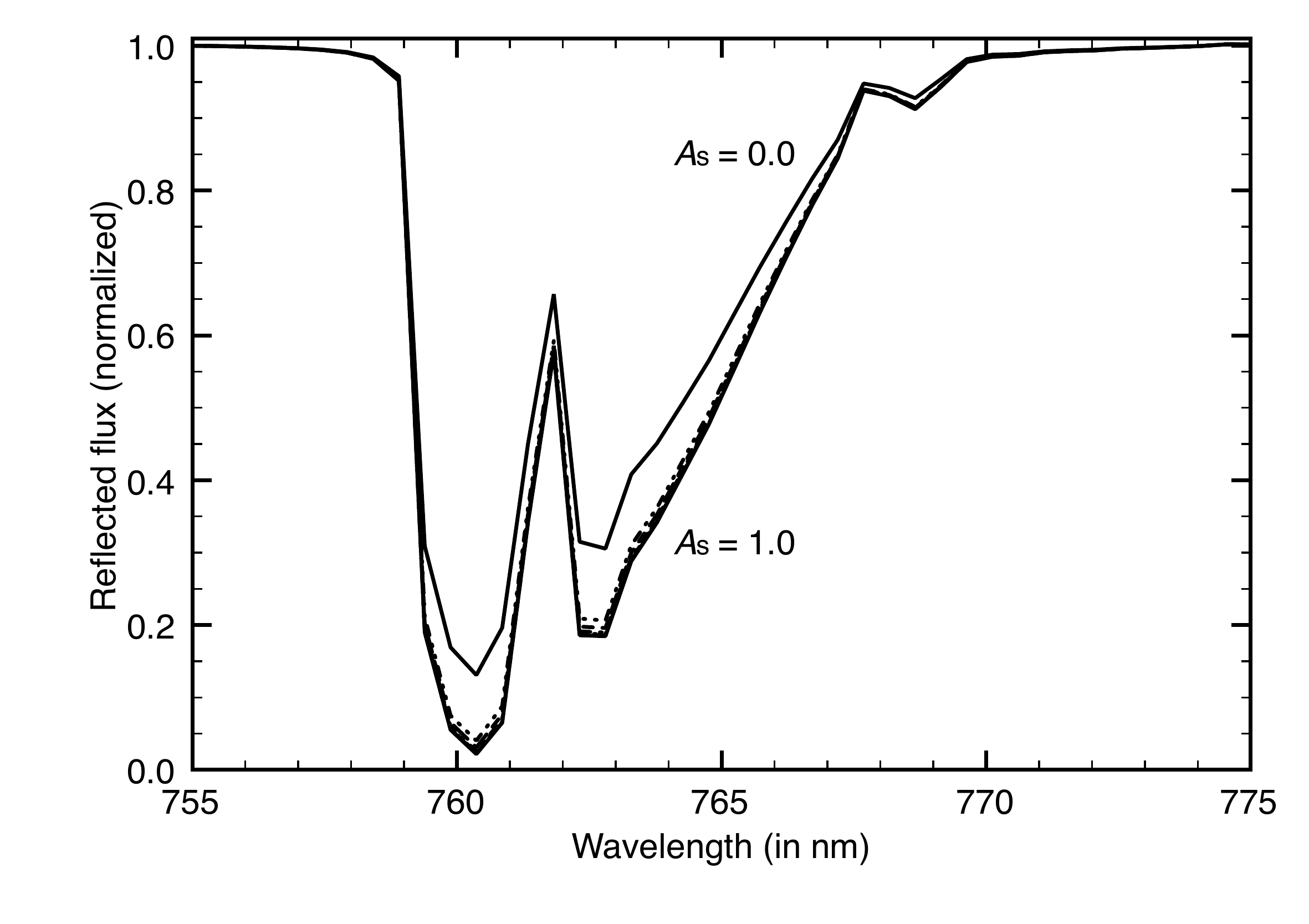}
\caption{The reflected flux curves of Fig.~\ref{fig4} normalized at           
         $\lambda=755$~nm. The shallowest band pertains to $A_{\rm s}=0.0$,
         and the deepest to $A_{\rm s}=1.0$.}
\label{fig5}
\end{figure}

In exoplanet observations, absolute fluxes, such as shown in Fig.~\ref{fig4},
might not be available because of missing knowledge about e.g.\ the planet 
radius, and/or the distances involved. Such observations could, however, 
provide relative fluxes, as shown in Fig.~\ref{fig5}.
For a cloud--free atmosphere, the relative depth of the absorption band 
appears to be very insensitive to $A_{\rm s}$, provided $A_{\rm s} > 0.0$
(the curve for $A_{\rm s}=0.05$, which is not shown in the figure, would 
fall only slightly above that for $A_{\rm s}=0.1$). The curves in 
Fig.~\ref{fig5} pertain to horizontally homogeneous planets, but this 
insensitivity of the relative band depth to $A_{\rm s}$ also holds 
for cloud--free planets with a range of albedo's across their surfaces.   

This insensitivity of the relative band depth in flux can be explained by 
how the reflected flux varies with $b_{\rm abs}$, shown in Fig.~\ref{fig6}.
Because of the small values of $b_{\rm sca}$ in this spectral region,
and thus the small amount of multiple scattering, the flux that reaches 
the surface and then the top of the atmosphere, depends almost linearly 
on $A_{\rm s}$ for every value of $b_{\rm abs}$. The insensitivity is
thus independent of the spectral resolution. The slight increase of
the relative band depth with increasing $A_{\rm s}$ visible in Fig.~\ref{fig5},
is due to the slight increase in multiple scattered light and hence a
slight increase of the absorption. With a higher surface pressure, and 
thus a larger $b_{\rm sca}$ and more multiple scattering, the relative 
band depth will show a larger sensitivity to~$A_{\rm s}$.

Figure~\ref{fig6} also shows that in polarization spectra, the widths of
individual absorption lines decrease with increasing surface albedo $A_{\rm s}$.
The explanation for this narrowing of absorption lines in polarization spectra
is that if for a given value of $b_{\rm abs}$,
$A_{\rm s}$ is increased, more unpolarized light is added to the reflected 
starlight, decreasing $P_{\rm s}$ \citep[see Eq.~13 in][]{1999JGR...10416843S}.

\subsubsection{The influence of the planetary phase angle $\alpha$}

Figure~\ref{fig7} shows $F$ and $P_{\rm s}$ in the continuum (755~nm) 
as functions of the planetary phase angle $\alpha$, for different values of
$A_{\rm s}$. 
For comparison, the curves for $b_{\rm abs}=400$ (the deepest absorption 
lines in the O$_2$ A--band) are also shown ($F$ is virtually zero). 
These curves (that are independent of $A_{\rm s}$) can be used to estimate 
the expected strength of individual absorption lines as compared to the 
continuum. 

\begin{figure*}
\figurenum{6}
\plottwo{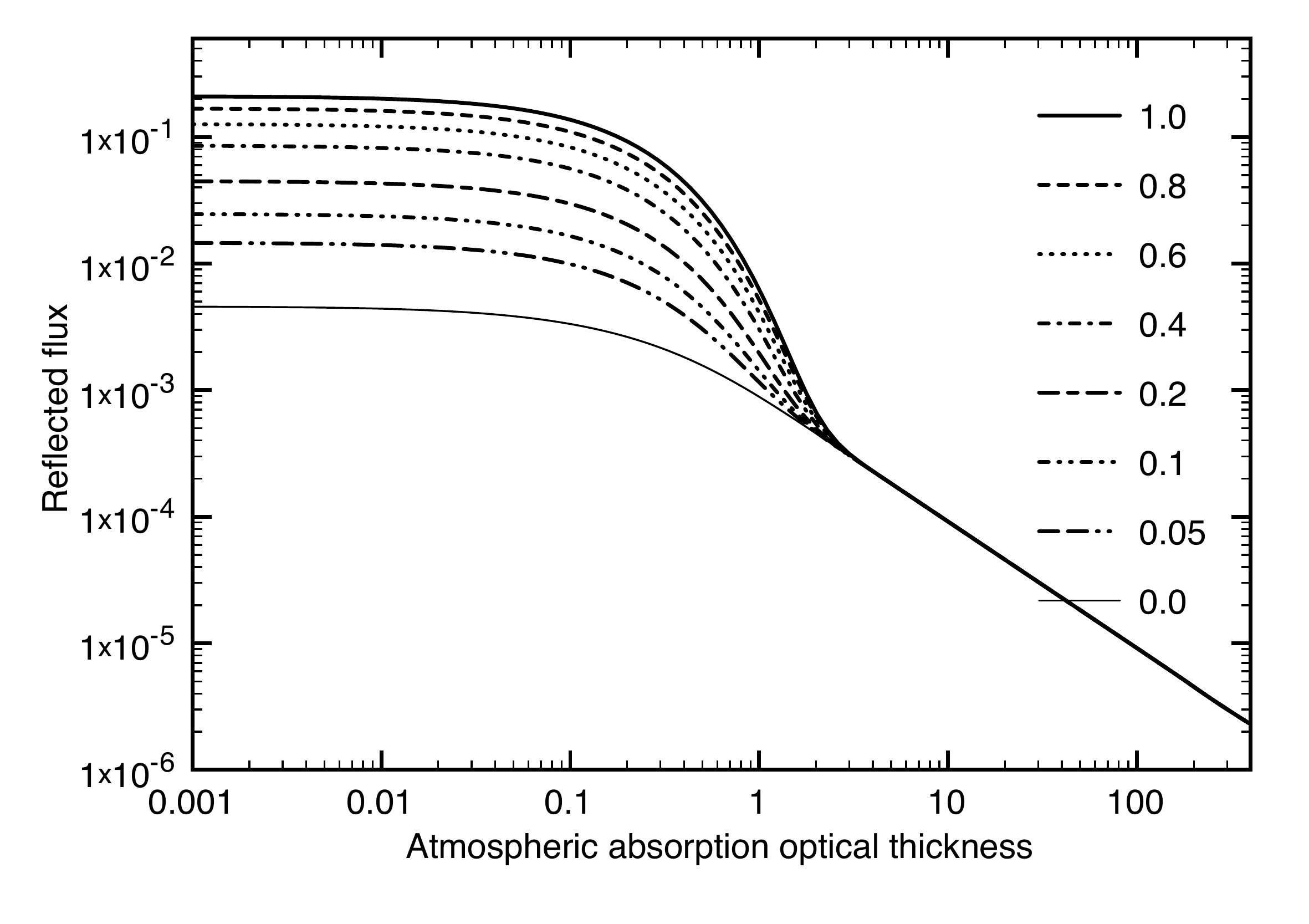}{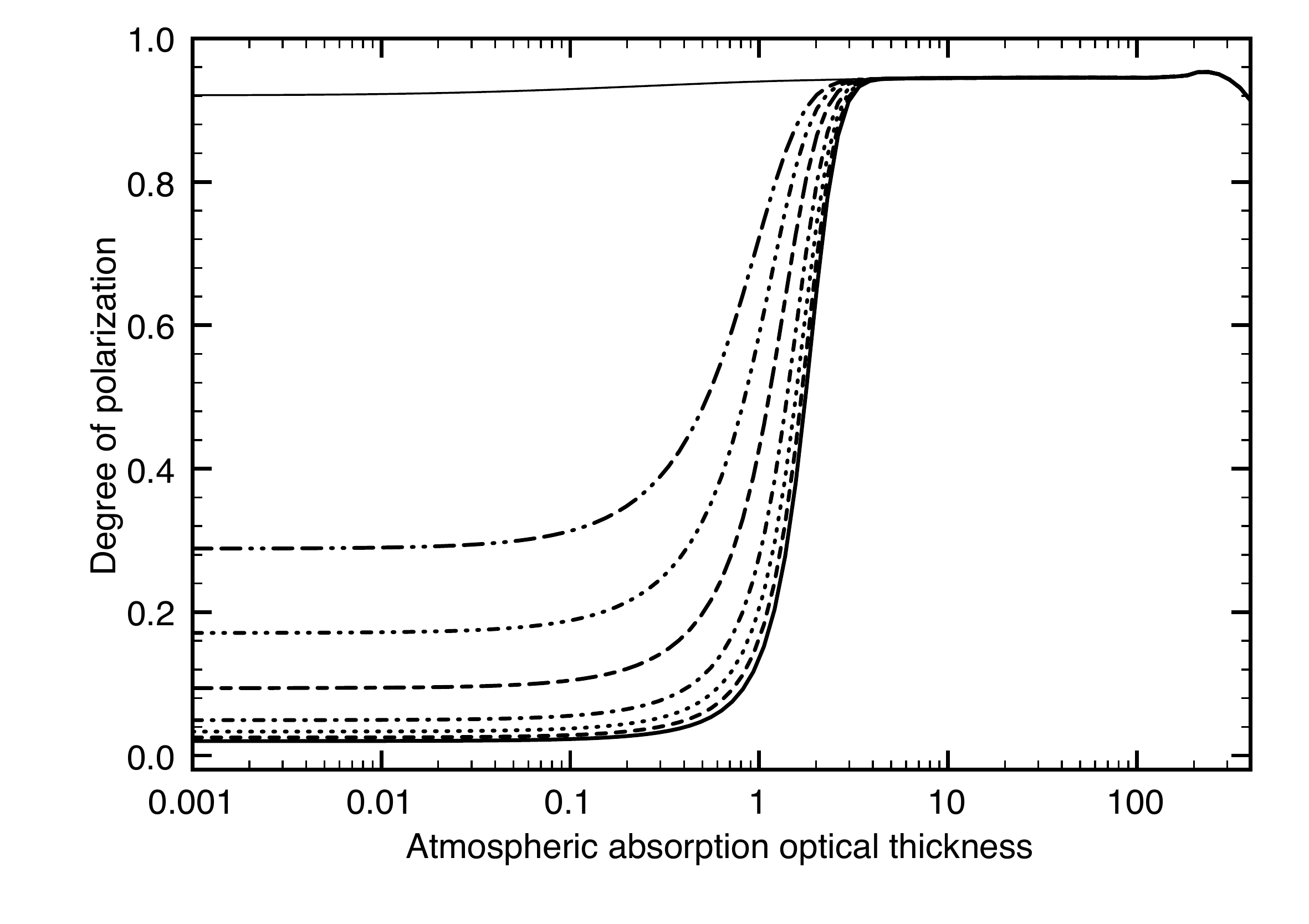}
\caption{$F$ (left) and $P_{\rm s}$ (right) of starlight
         reflected by cloud--free exoplanets at $\alpha=90^\circ$
         as functions of $b_{\rm abs}$ for different values of $A_{\rm s}$.}
\label{fig6}
\end{figure*}
\begin{figure*}
\figurenum{7}
\plottwo{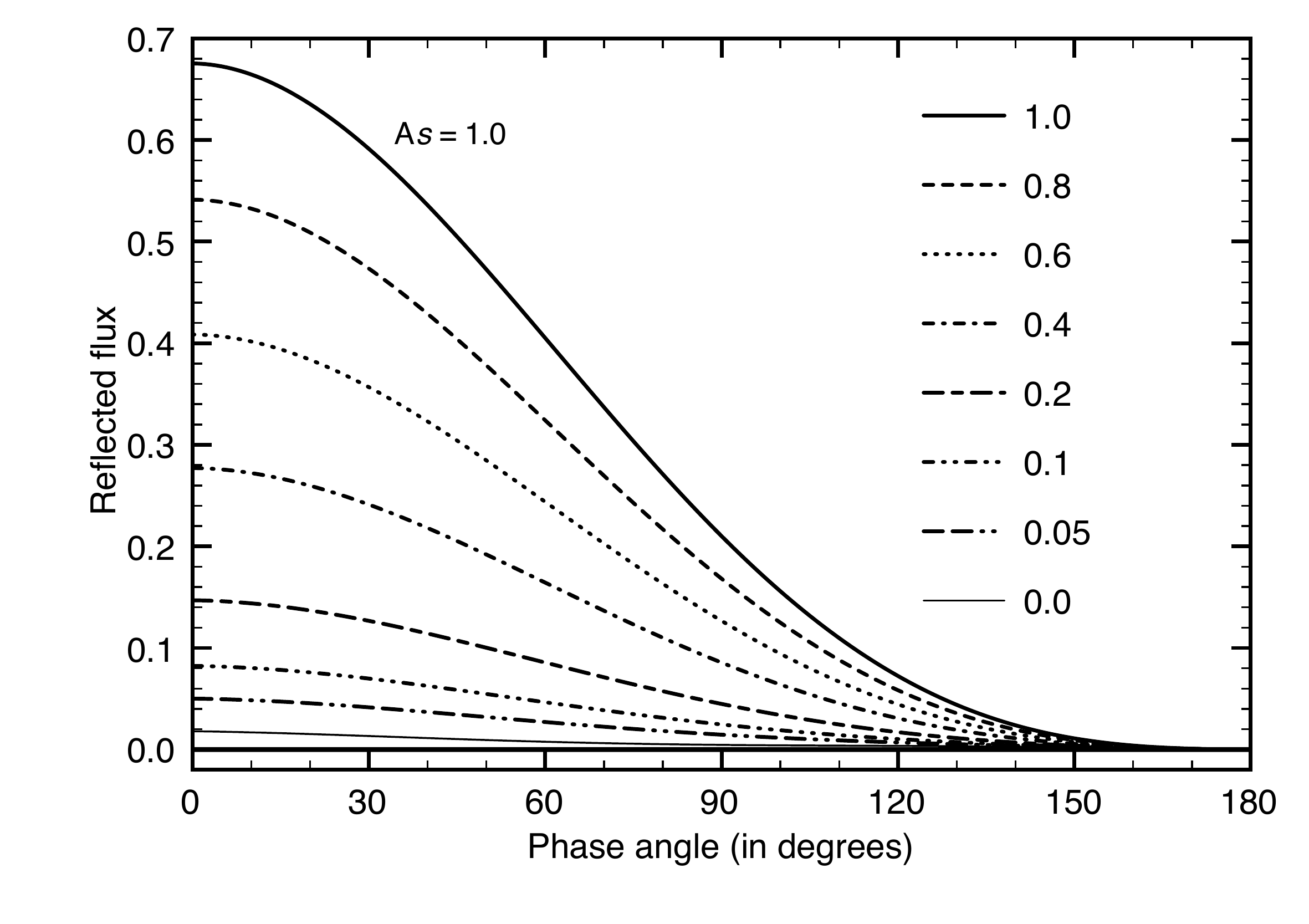}{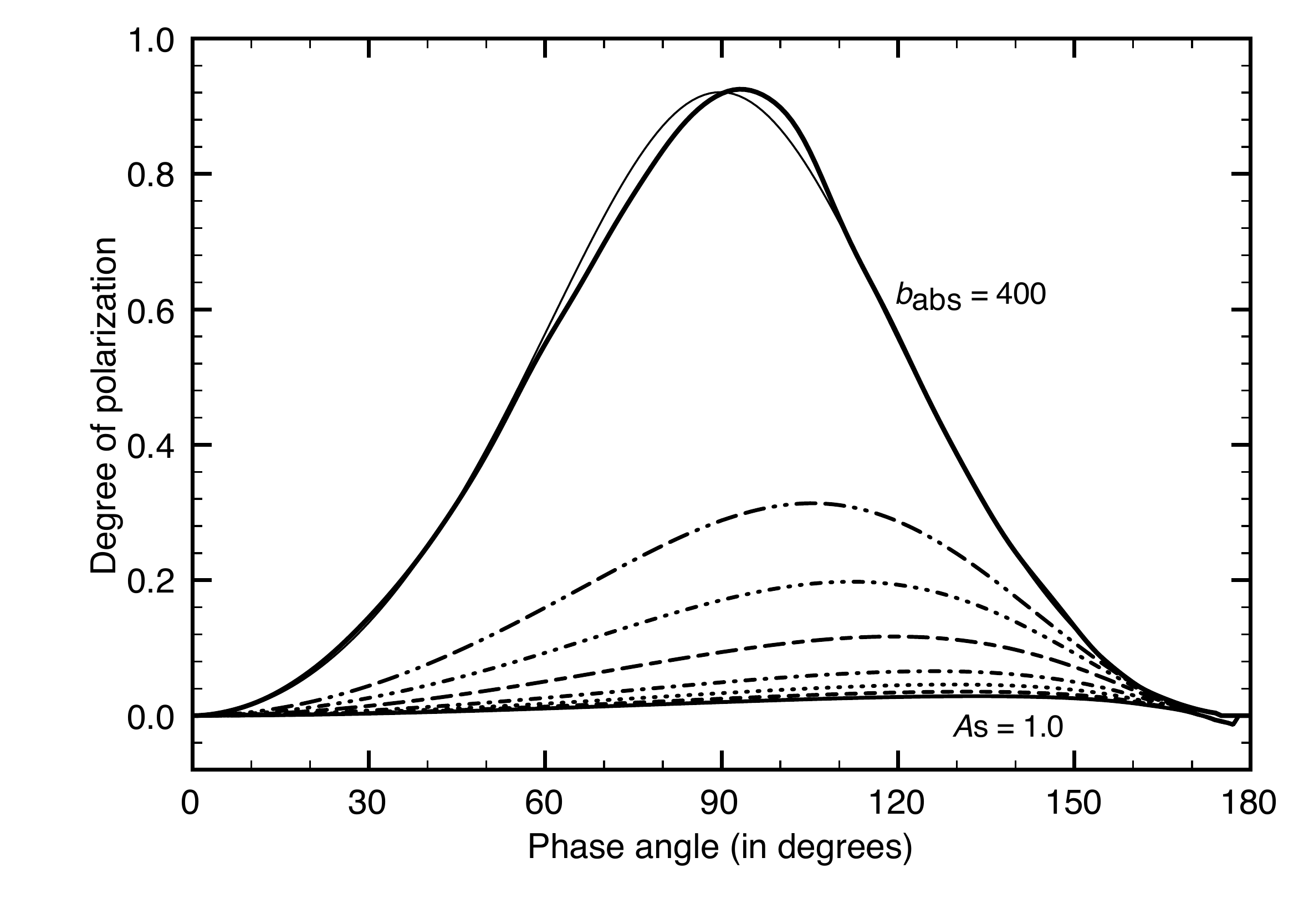}
\caption{$F$ (left) and $P_{\rm s}$ (right) of starlight reflected at 
         $\lambda=755$~nm (the continuum) by cloud--free exoplanets as 
         functions of $\alpha$ for the same $A_{\rm s}$ as in Fig.~\ref{fig4},
         except with curves for $A_{\rm s}=0.05$ added. 
         The thick solid curve peaking at $\alpha=93^\circ$ in $P_{\rm s}$
         ($F$ is virtually zero) pertains to $b_{\rm abs}=400$.}
\label{fig7}
\end{figure*}
The continuum $F$ decreases smoothly with increasing $\alpha$, while the
continuum polarization curves have a maximum that shifts from 90$^\circ$ when 
the surface is black, to larger values of $\alpha$ with increasing $A_{\rm s}$.
Both for the planet with the black surface and for the 
planet with $b_{\rm abs}=400$, $P_{\rm s}$ peaks at or near $\alpha=90^\circ$,
because for those planets the signal is mostly determined by light 
singly scattered by gas molecules (there is almost no multiple scattering 
in the atmosphere and there is no contribution from the surface), and
at that phase angle, the scattering angle of the singly scattered 
light is 90$^\circ$, precisely where Rayleigh scattered light has the 
highest polarization (see Fig.~\ref{fig2}).
At 755~nm, the model atmosphere has a very small gaseous scattering
optical thickness, the reflected flux is thus mainly determined by $A_{\rm s}$,
and the maximum of $P_{\rm s}$ decreases rapidly with increasing $A_{\rm s}$
and shifts to larger values of $\alpha$.

From the polarization curves in Fig.~\ref{fig7}, it can be seen that for 
cloud--free planets with reflecting surfaces, the polarization inside the 
absorption band will be higher than in the continuum at all phase angles.
For black planets and $60^\circ < \alpha < 90^\circ$, $P_{\rm s}$ should be 
slightly lower inside the deepest absorption lines than in the continuum.
However, as can be seen in Fig.~\ref{fig6} showing the flux
and $P_{\rm s}$ as functions of $b_{\rm abs}$ at $\alpha= 90^\circ$
for various values of $A_{\rm s}$, lower values of $P_{\rm s}$ will only
occur in the deepest absorption lines (where the flux is extremely small). 
Thus, for cloud--free planets and without an absorption line--resolving 
spectral resolution, $P_{\rm s}$ in the band is expected to be higher than
$P_{\rm s}$ in the continuum at all phase angles and for all $A_{\rm s}$.


\subsubsection{The influence of the mixing ratio $\eta$}

\begin{figure*}
\figurenum{8}
\plottwo{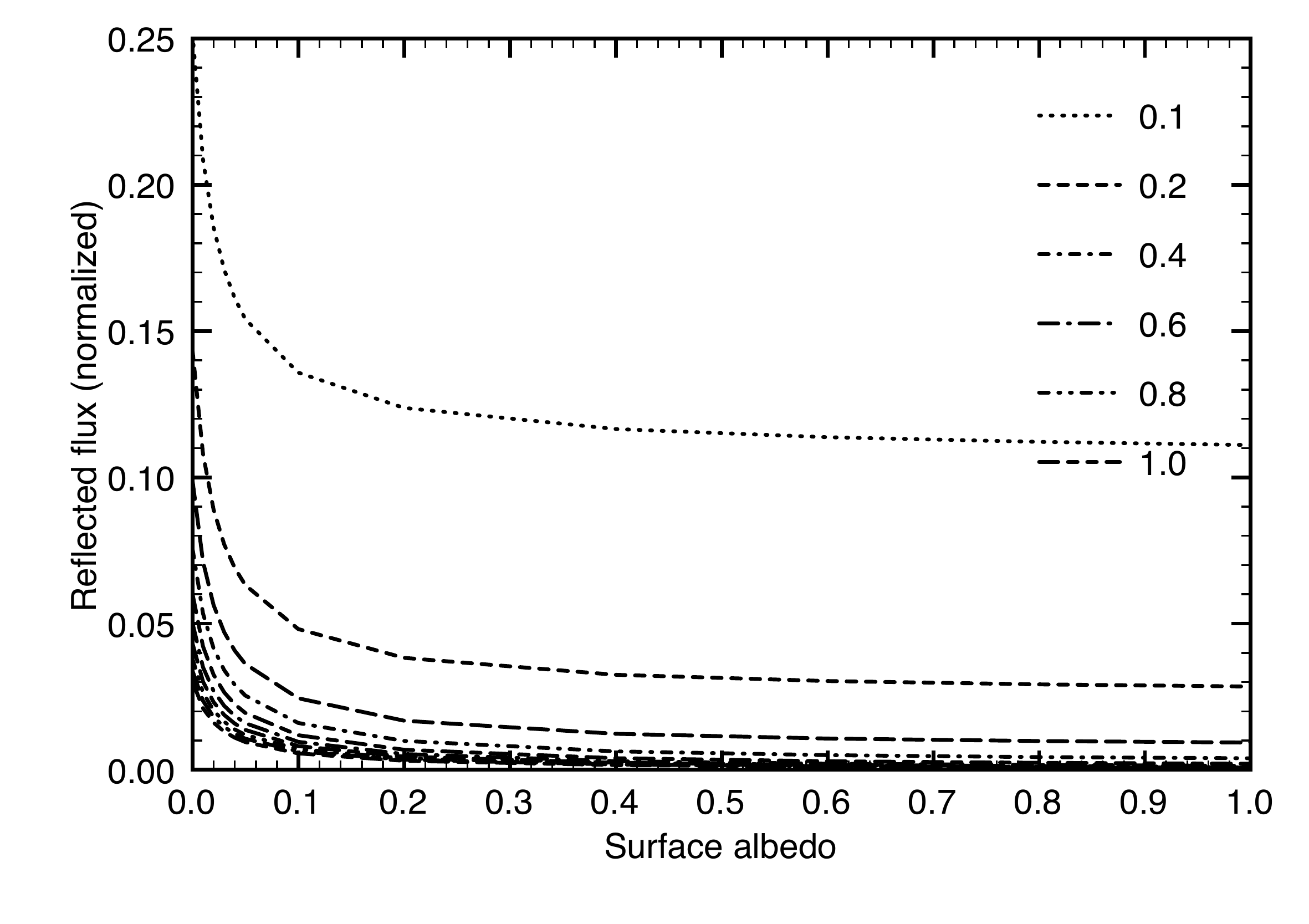}{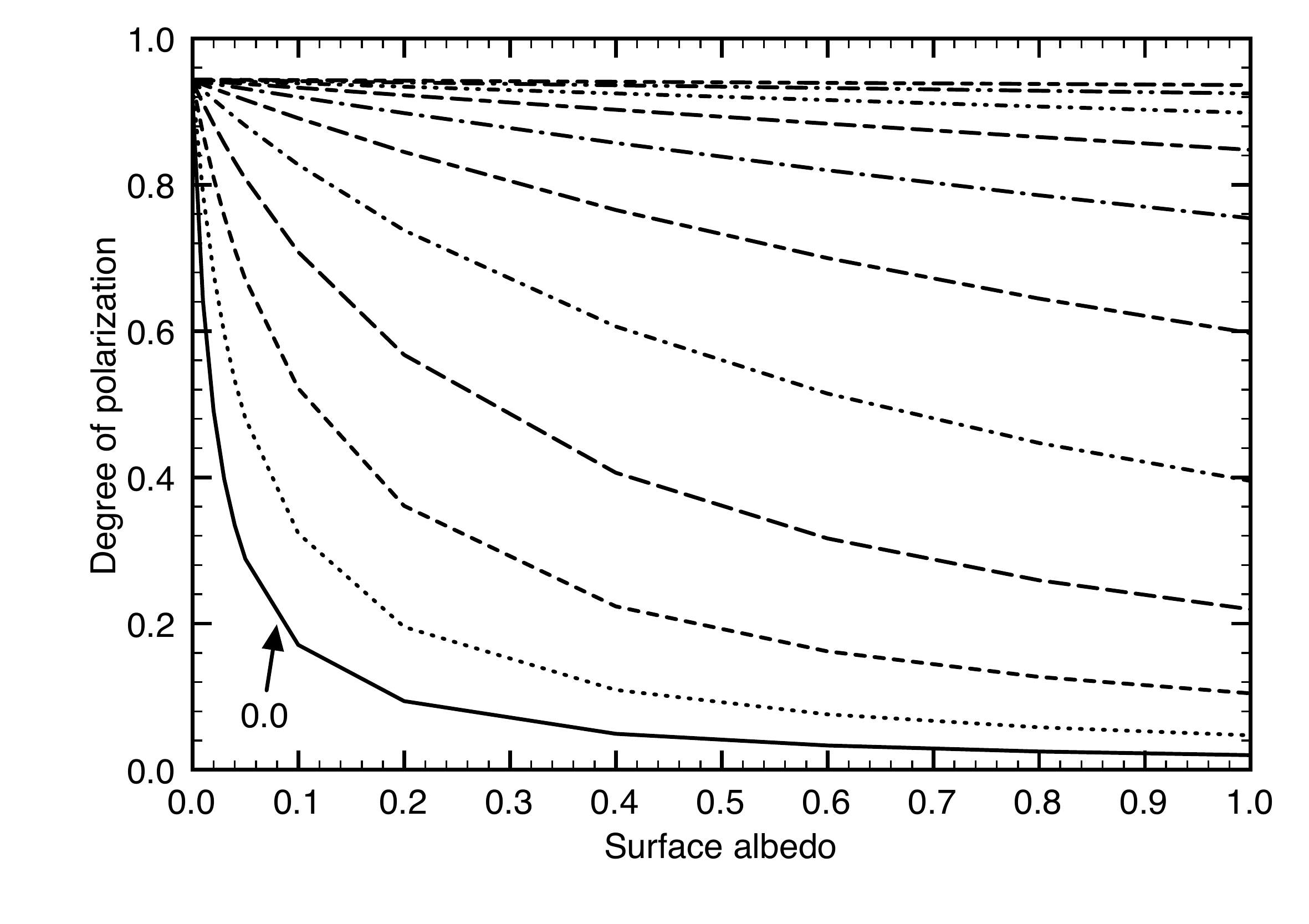}
\caption{$F$ (left), normalized to its continuum value, and $P_{\rm s}$ 
         (right) of starlight reflected by cloud--free exoplanets in the 
         deepest part of the O$_2$ A--band (the spectral bin around 
         $\lambda=760.4$~nm), at $\alpha=90^\circ$,
         as functions of $A_{\rm s}$ for $\eta$
         ranging from 0.0 (thick solid line, not shown in the flux
         as it equals 1.0) to 1.0 (long -- short -- short -- long dashes)
         in steps of 0.1. The short dashed line represents $\eta=0.2$,
         closest to the Earth's value.}
\label{fig8}
\end{figure*}

The strength of the O$_2$ A--band for a cloud--free atmosphere also depends 
on the O$_2$ mixing ratio $\eta$. Figure~\ref{fig6} gives insight in the 
influence of $\eta$, because $b_{\rm abs}$ depends linearly on $\eta$
(for a given surface pressure). To better illustrate the observable signals, 
Fig.~\ref{fig8} shows $F$ and $P_{\rm s}$ in the spectral bin covering 
the deepest part of the absorption band (around 760.4~nm), for $\eta$ up to 
1.0 (a pure O$_2$ atmosphere) as functions of $A_{\rm s}$. 
The curve for $\eta=0.0$ equals the continuum. The phase angle is 90$^\circ$.

The reflected flux curves in Fig.~\ref{fig8} show that
{\em 1.} the band depth (i.e. 1.0 - curve) increases with increasing
$\eta$, with saturation starting for $\eta > 0.4$ independent of $A_{\rm s}$, 
and {\em 2.} the insensitivity of the band depth to $A_{\rm s}$
(see Fig.~\ref{fig5} for $\eta=0.21$) holds for all values of $\eta$,
except for (near)--black surfaces.
The polarization curves show that, with the 0.5~nm wide spectral bin, 
the band strength (i.e. $\mid$~curve~-~continuum~$\mid$, with the continuum 
$P_{\rm s}$ given by the solid line in Fig.~\ref{fig8}) is smallest when
the surface is (near)--black for all $\eta$.

\begin{figure*}
\figurenum{9}
\plottwo{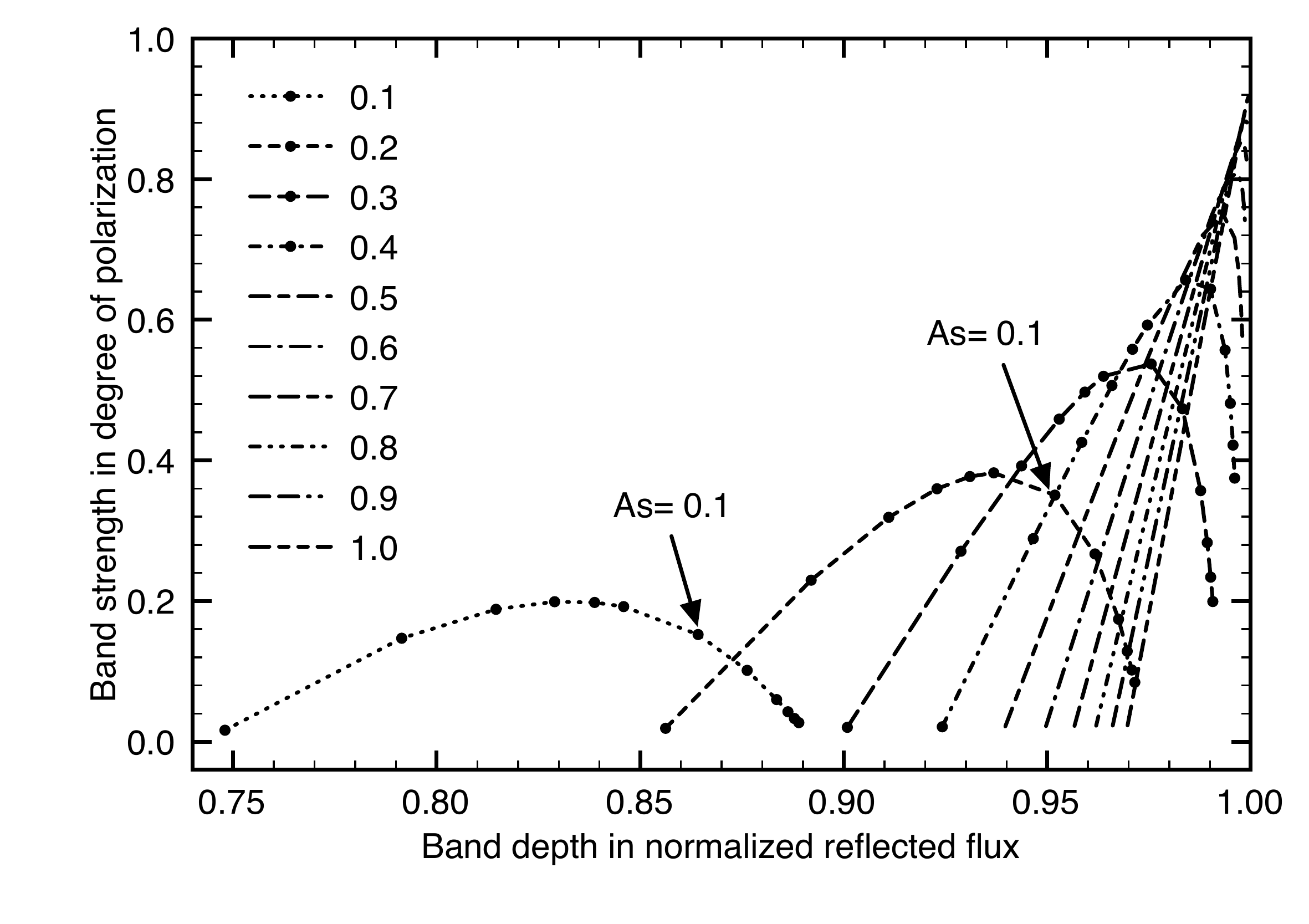}{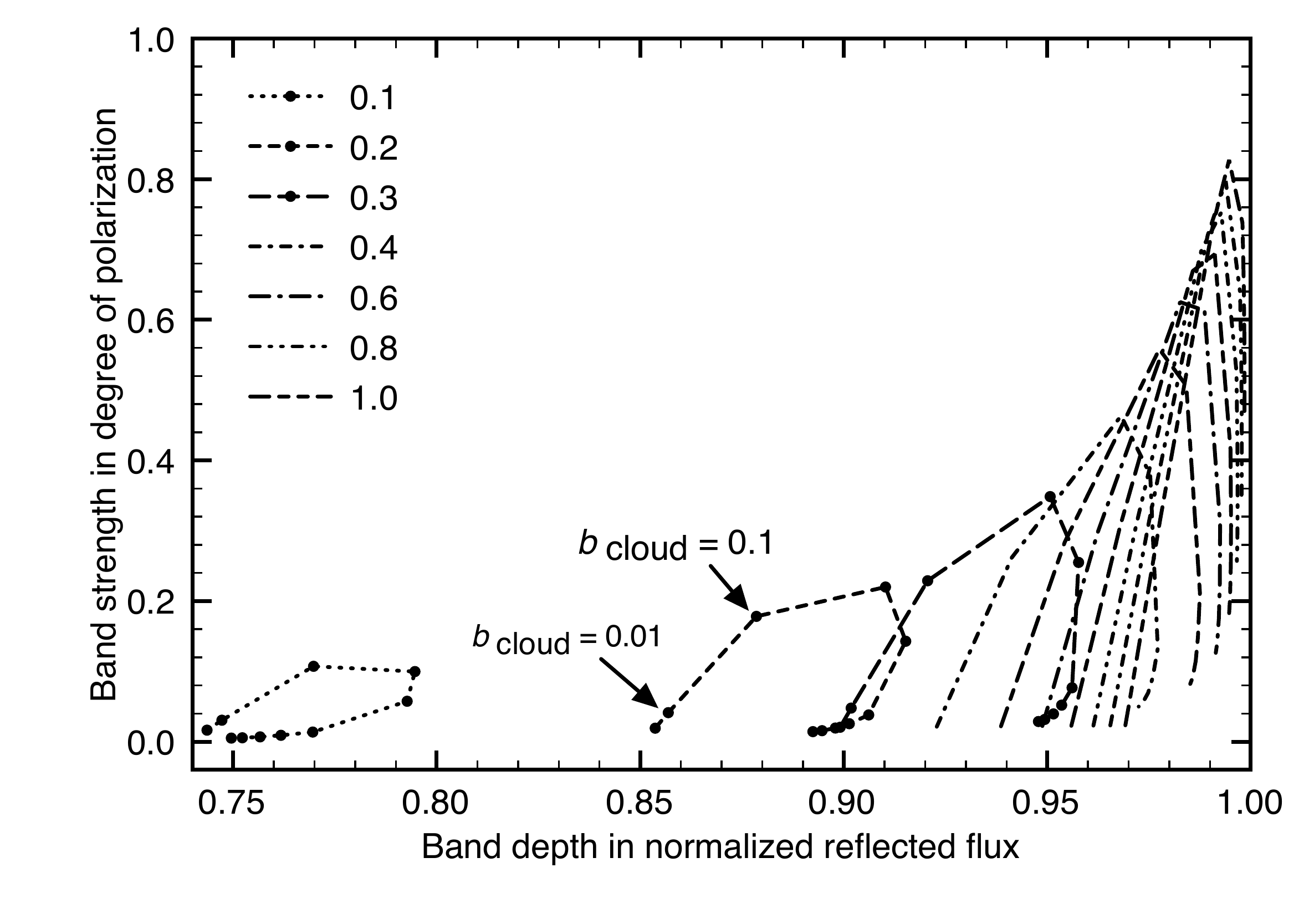}
\plottwo{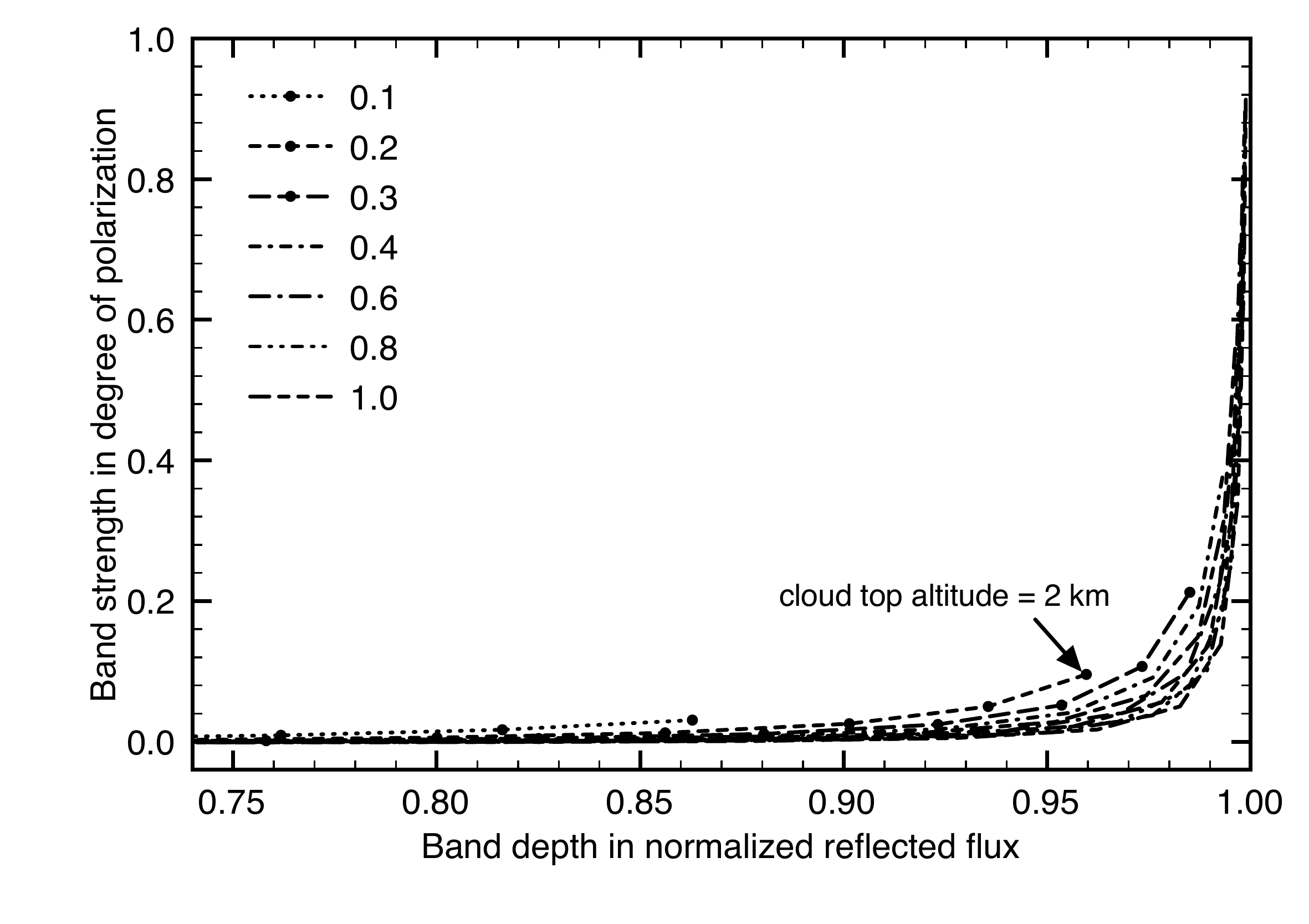}{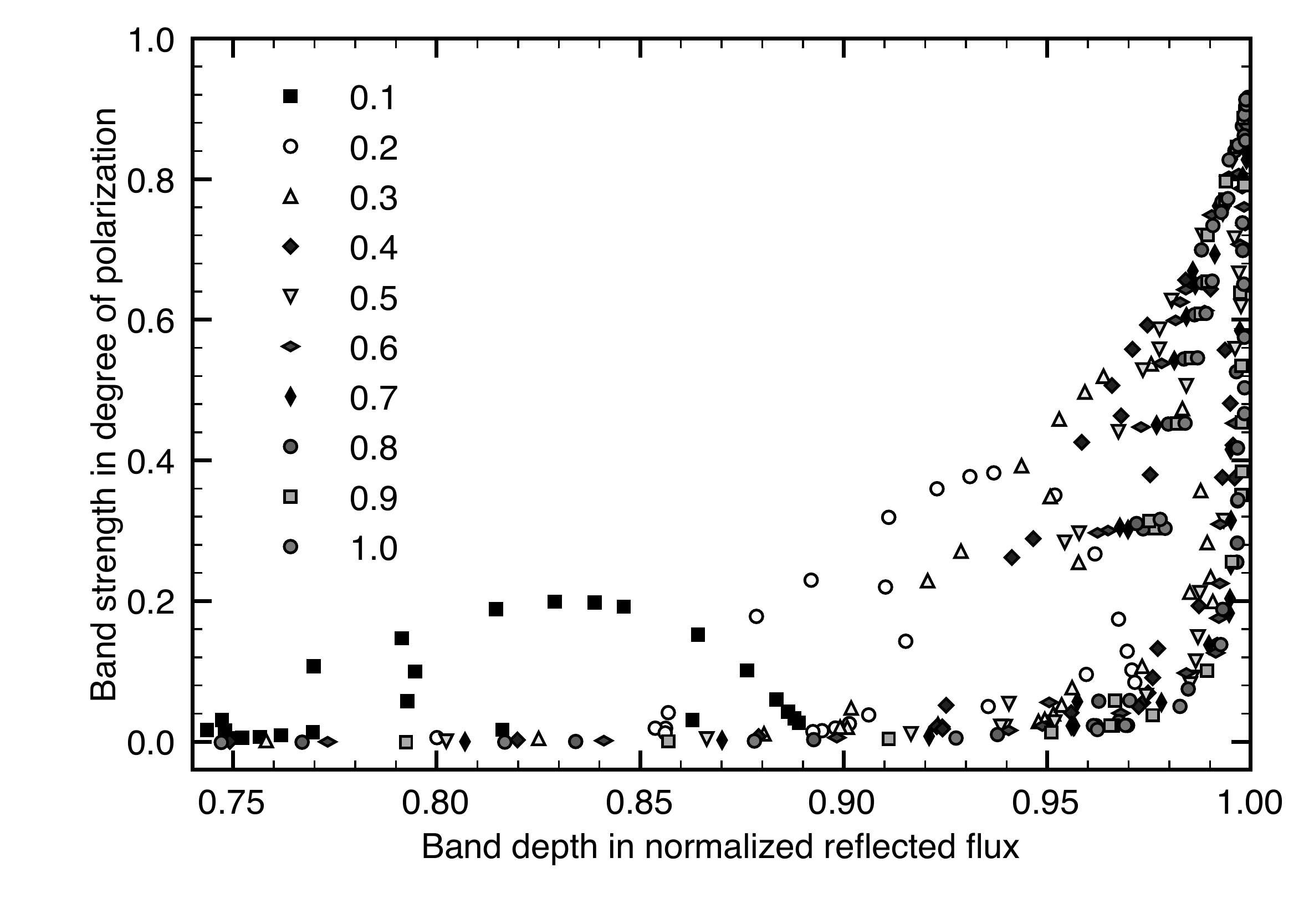}
\caption{Scatter plot of $\Delta P$ versus $\Delta F$, with
         $\Delta P = P_{\rm 760.4~nm} - P_{\rm 755~nm}$ and
         $\Delta F = 1.0 - F_{\rm 760.4~nm}/F_{\rm 755~nm}$
         ($F_{\rm x~nm}$ indicating the flux in the spectral bin around x~nm). 
         Data is shown for $\eta$ ranging from 0.1 to 1.0 in steps of 0.1
         (not all values appear in all legends).
         Top--left: Data points derived from Fig.~\ref{fig8}, thus for  
         $A_{\rm s}$ equal to 0.0 (on the left of each line), 0.01,
         0.02, 0.03, 0.04, 0.05, 0.1, 0.2, 0.4, 0.6, 0.8, and 1.0 
         (on the right of each line).
         Top-right; Data points derived from Fig.~\ref{fig12}, thus for 
         $b_{\rm cloud}$ equal to 0.0 (on the left of each line), 
         0.01, 0.1, 0.5, 1.0, 5.0, 10.0, 50.0, and 100.0.
         Bottom--left: Data points derived from Fig.~\ref{fig16}, thus
         for cloud top altitudes equal to 2.0 (on the right of each line), 4.0,
         6.0, 8.0, 10.0, 12.0~km (on the left of each line,
         note that we did not include flux band depths $< 0.74$).
         Bottom--right: All data points from the other three plots combined.}
\label{fig9}
\end{figure*}

For a cloud--free planet with a depolarizing surface, high polarization 
is usually associated with low fluxes. The relation between the band 
strength in $P_{\rm s}$ and the band depth in the normalized $F$, derived 
from Fig.~\ref{fig8}, is shown in the top--left of Fig.~\ref{fig9} 
(data for $\eta= 0.0$ have been omitted). The plot clearly shows 
that the band depth in $F$ is most sensitive to $A_{\rm s}$ when the 
surface is dark ($A_{\rm s} < 0.1$) and $\eta$ is small. Also, large 
polarization band strengths ($> 0.5$) correlate with large band depths 
in $F$. Indeed, if one would measure a large band strength in $P_{\rm s}$
combined with a small band depth in $F$, one would have an indication that 
the planetary atmosphere contains clouds and/or haze particles in addition 
to the gaseous molecules (see Sect.~\ref{sect3.2}).

The data points in Fig.~\ref{fig9} have been calculated assuming an 
Earth--like surface pressure and thus an Earth--like value for the 
atmospheric gaseous scattering optical thickness $b_{\rm sca}$. 
Calculations show that data points for other (still small) values of 
$b_{\rm sca}$ fall between those shown in Fig.~\ref{fig9}: for a 
surface pressure (and hence $b_{\rm sca}$) twice as high as the Earth's, 
the data points for, for example, $\eta=0.2$ are similar to those for 
$\eta=0.4$, except for slightly different values of $A_{\rm s}$. 
Thus, measuring a certain combination of band strength in $P_{\rm s}$
and band depth in $F$ without knowing the surface pressure
and $A_{\rm s}$ would not directly allow the retrieval of $\eta$. 

\subsection{Completely cloudy planets}
\label{sect3.2}

\subsubsection{The influence of the cloud optical thickness $b_{\rm cloud}$}
\label{sect3.2.1}

\begin{figure*}[t]
\figurenum{10}
\plottwo{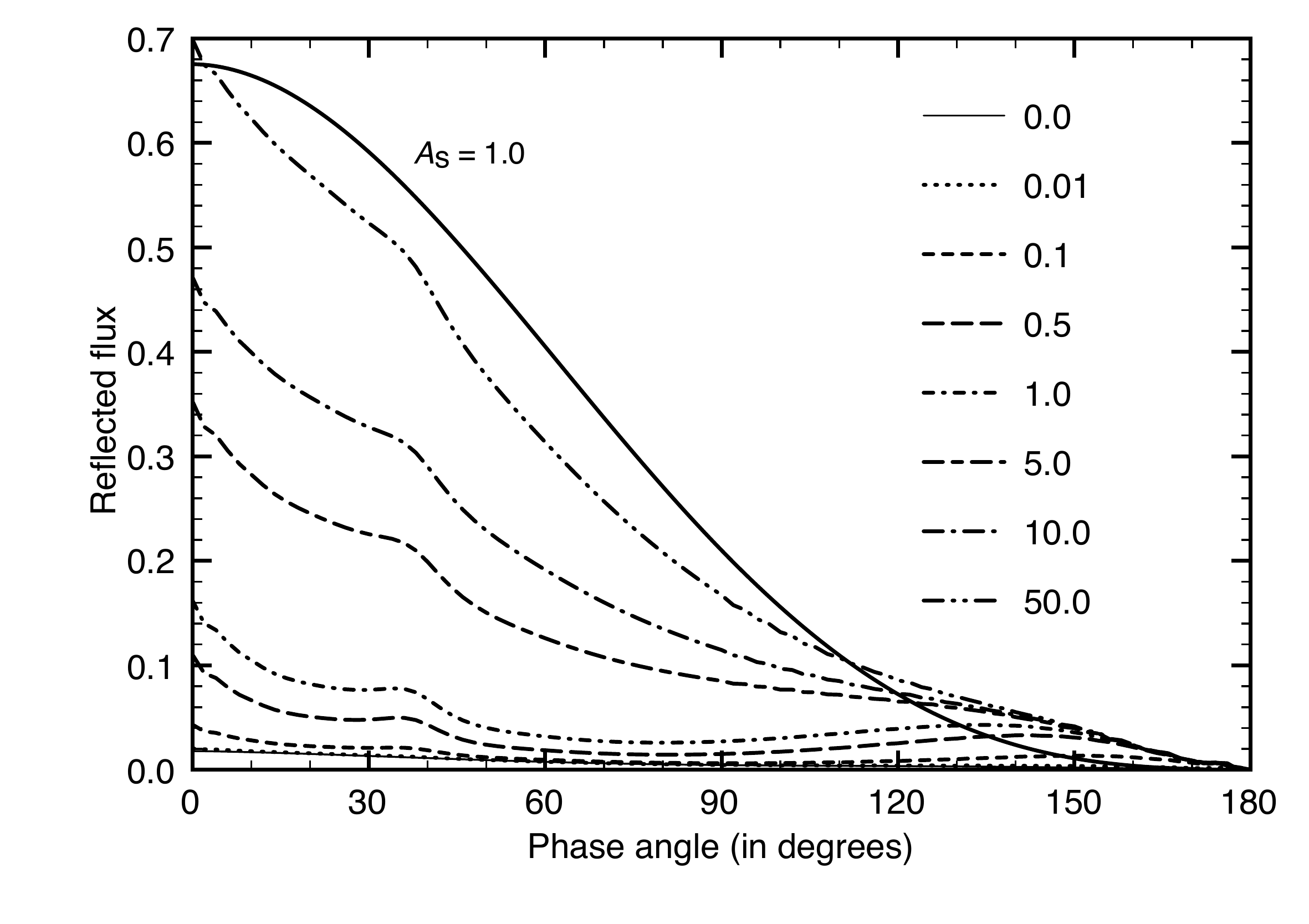}{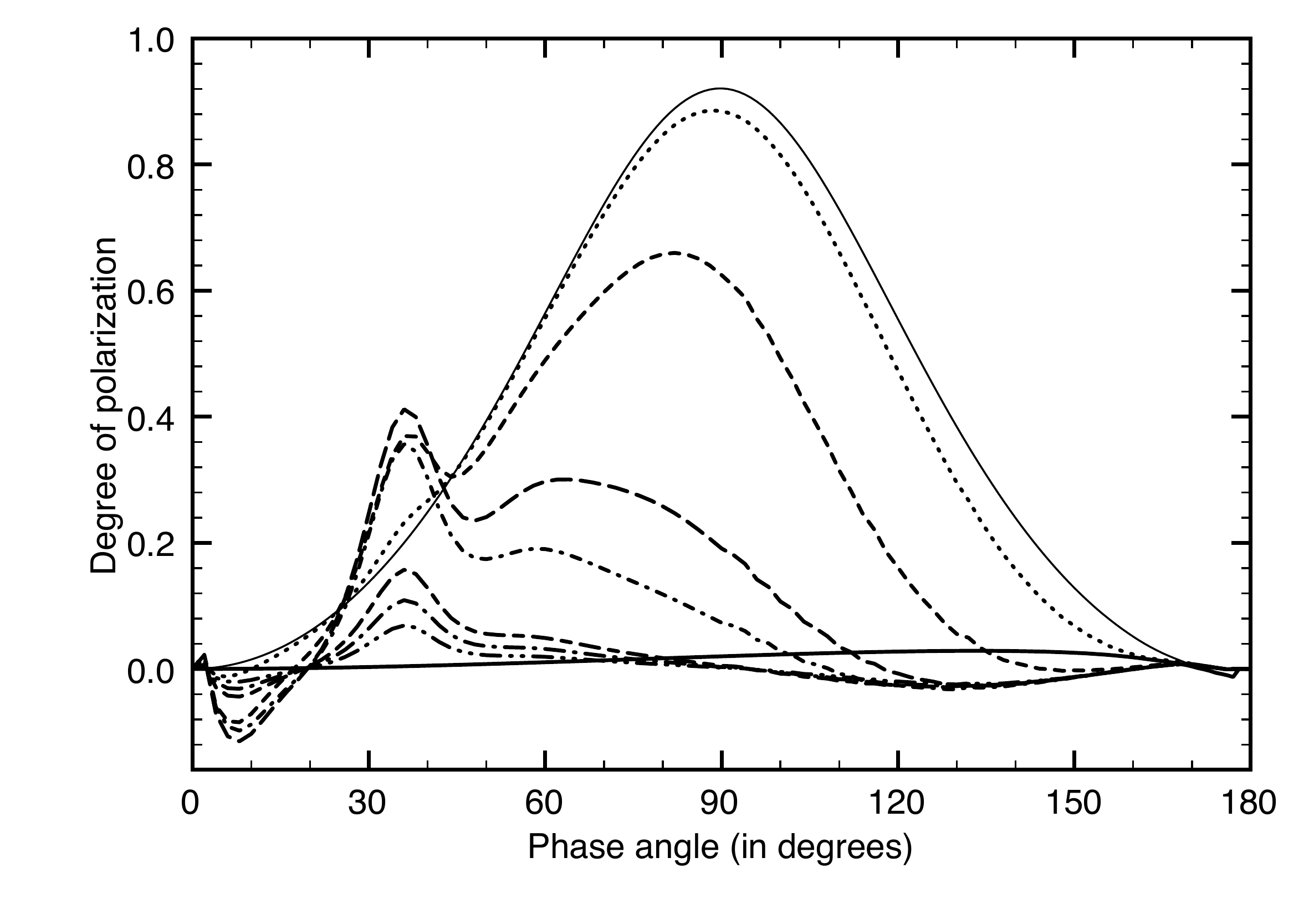}
\caption{$F$ (left) and $P_{\rm s}$ (right) of starlight
         reflected at $\lambda=755$~nm (the continuum) by completely cloudy
         exoplanets with $A_{\rm s}= 0.0$, as functions of $\alpha$. 
         The optical thickness $b_{\rm cloud}$ (defined at 765~nm) 
         varies from 0.01 to 50.0. 
         The curves for cloud--free planets ($b_{\rm cloud}=0.0$) with 
         a black and a white surface ($A_{\rm s}=1.0$) are also included.
         The cloud layer extends from 4~to 6~km.}
\label{fig10}
\end{figure*}

The Earth has an average cloud coverage of about 60$\%$, with cloud optical 
thicknesses $b_{\rm cloud}$ ranging from almost 0 to over 100 in the visible 
\citep[][]{marshak2005}. Figure~\ref{fig10} shows the influence of 
$b_{\rm cloud}$ on $F$ and $P_{\rm s}$ in the continuum around the 
O$_2$ A--band as functions of $\alpha$, for $A_{\rm s}=0.0$. These curves 
show the background on which the absorption band could be measured.
The cloud is a horizontally homogeneous layer of 2~km thick with its 
top at 6~km, embedded in the gaseous atmospheres of Sect.~\ref{sect3.1}.

The curves in Fig.~\ref{fig10} show how increasing $b_{\rm cloud}$ 
brightens a planet at all phase angles, except for optically thin 
clouds ($b_{\rm cloud} \leq 1.0$), where the planets are darkest at 
$60^\circ < \alpha < 90^\circ$. This is due to the single scattering phase 
function of the cloud particles (Fig.~\ref{fig2}). This single scattering 
phase function is also the explanation for the 'primary rainbow': the 
shoulder in $F$ and the local maximum in $P_{\rm s}$ around $\alpha=38^\circ$.
The rainbow feature could help to identify water clouds on exoplanets
\citep[see e.g.][]{2012A&A...548A..90K,2007AsBio...7..320B}.

The Rayleigh scattering polarization maximum around 90$^\circ$,
disappears with increasing $b_{\rm cloud}$, as the scattering by the cloud 
particles increasingly dominates the reflected signal. In particular, around 
$\alpha=90^\circ$, the single scattering polarization of cloud particles 
is almost zero (see Fig.~\ref{fig2}). At small phase angles, $P_{\rm s}$ of 
the cloudy planets is negative, and the polarization direction is thus parallel 
to the reference plane. The directional change from parallel to perpendicular 
to the reference plane occurs around $\alpha=20^\circ$.
Another directional change occurs at a larger phase angle: the larger
$b_{\rm cloud}$, the closer this phase angle is to 80$^\circ$
for the liquid water particles that form our clouds (see Fig.~\ref{fig2}).

\begin{figure}
\figurenum{11}
\epsscale{1.2}
\plotone{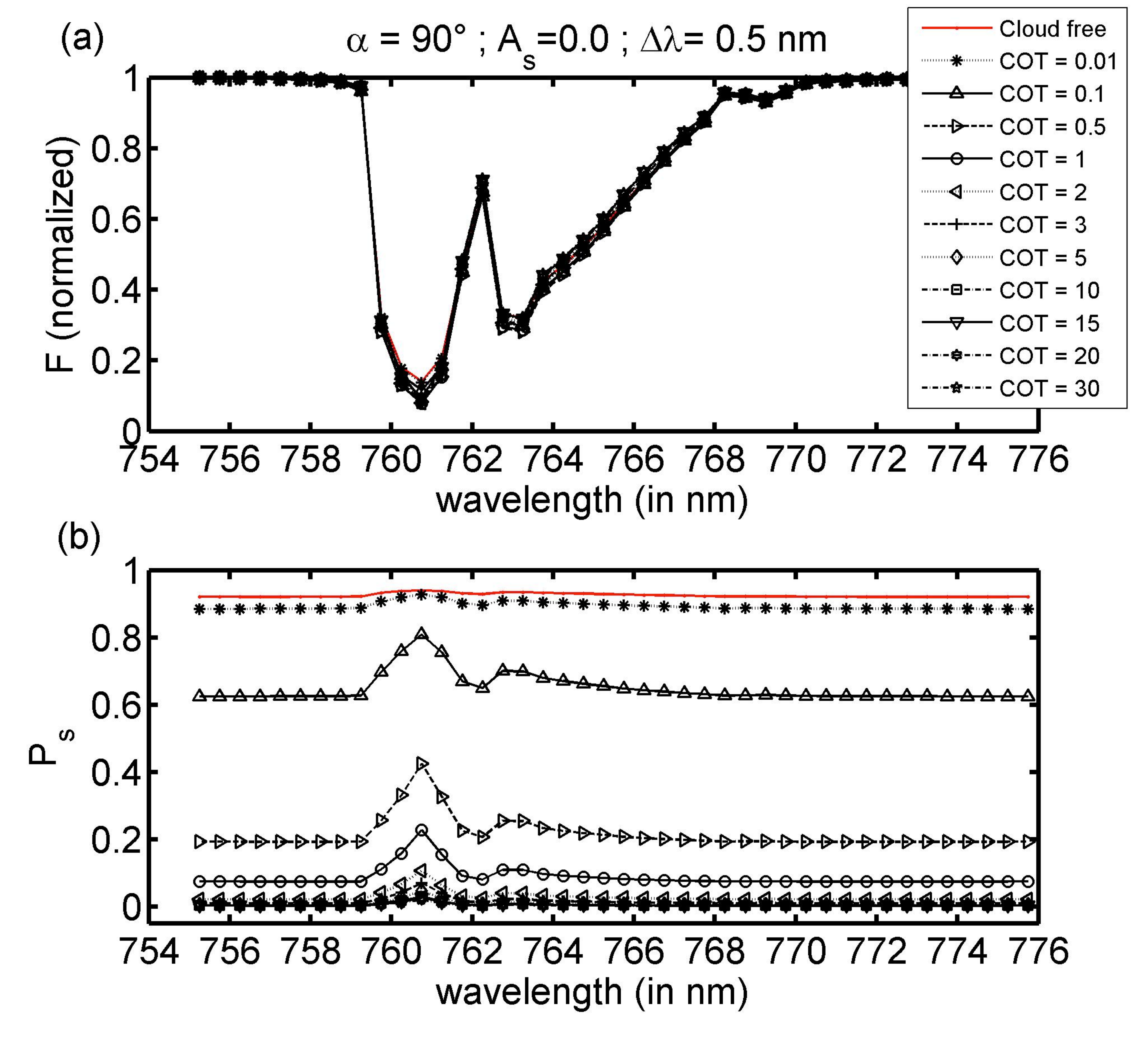}
\caption{$F$ (top) and $P_{\rm s}$ (bottom) of starlight
         reflected by exoplanets that are completely covered by clouds from
         4 to 6~km of altitude with cloud optical thicknesses (defined at
         765~nm) ranging from 0.0 (cloud--free) to 30. 
         The flux curves have been normalized at 755~nm.}
\label{fig11}
\end{figure}

Fig.~\ref{fig11} shows $F$ (normalized at 755~nm; for the absolute differences
in the continuum flux, see Fig.~\ref{fig10}) and $P_{\rm s}$
across the O$_2$ A--band for the same model planets, at $\alpha=90^\circ$. 
The flux band depth depends only weakly on $b_{\rm cloud}$,
because of two opposing effects:
1.\ increasing $b_{\rm cloud}$ decreases the average photon path
length (and thus the absorption), because more photons are scattered
back to space, and 2.\ increasing $b_{\rm cloud}$ increases the average
photon path length (and thus the absorption), through the increase of 
multiple scattering in the atmosphere. Which effect dominates, depends 
on the values of $b_{\rm abs}$ within a spectral bin, the illumination 
and viewing geometries, and the cloud micro- and macrophysical properties. 
For our model planets, the net effect is a flux band depth that is insensitive 
to $b_{\rm cloud}$.

\begin{figure*}
\figurenum{12}
\plottwo{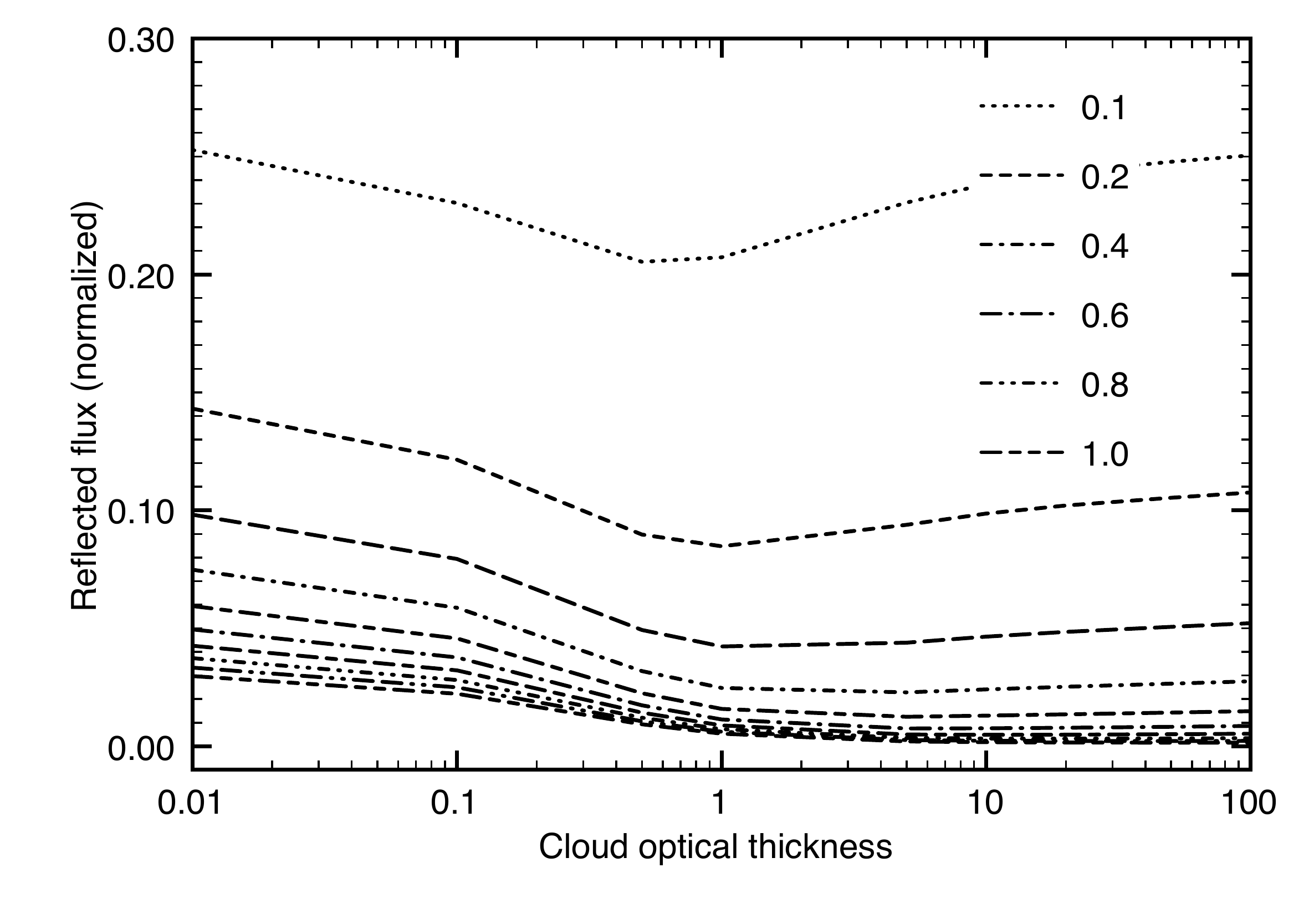}{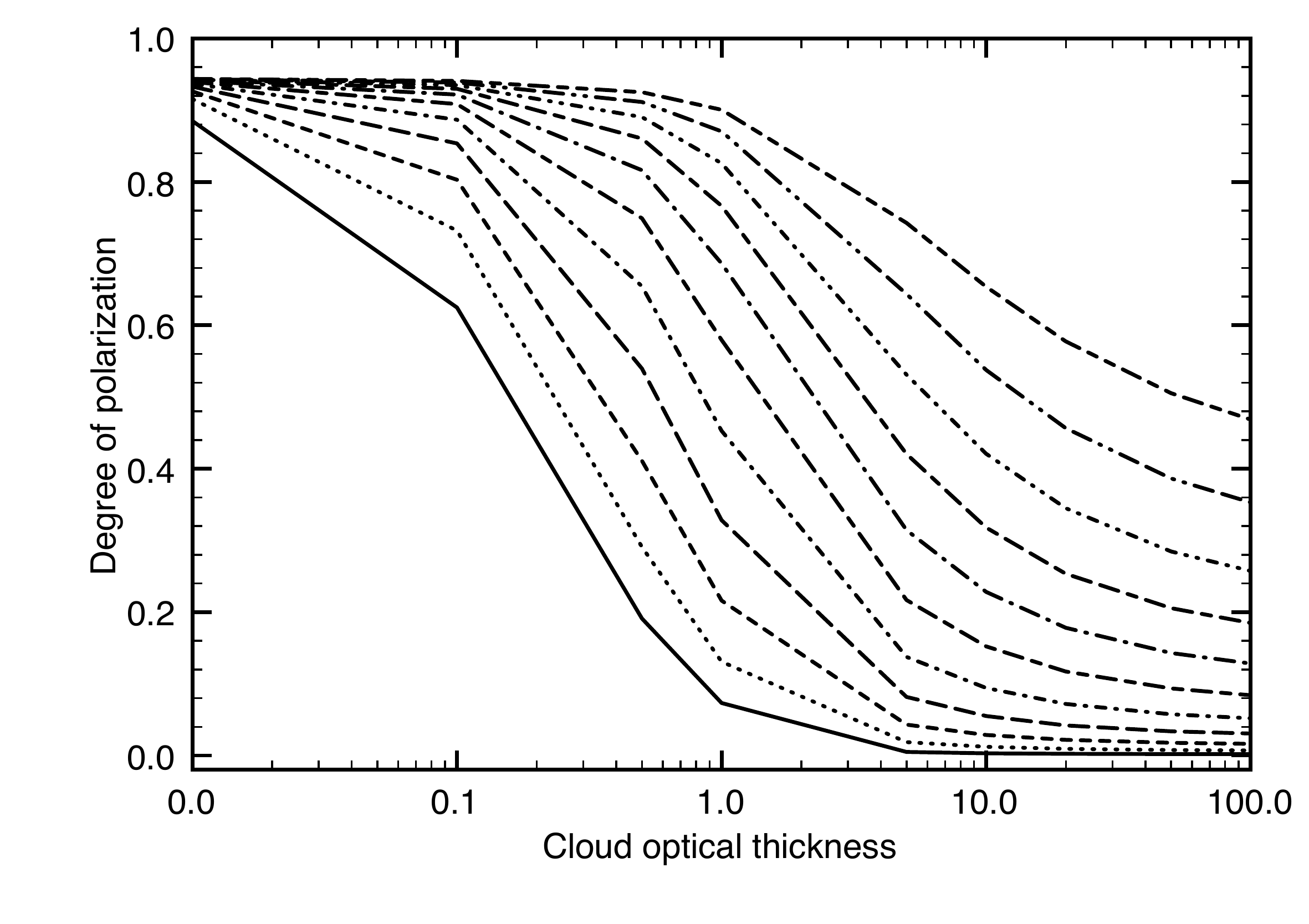}
\caption{Similar to Fig.~\ref{fig8}, except as functions of $b_{\rm cloud}$.
         The cloud is 2~km thick and has its top at 6.0~km. 
         The surface is black and $\alpha=90^\circ$. The thick, solid 
         line pertains to $\eta=0.0$.}
\label{fig12}
\end{figure*}

This insensitivity holds for any O$_2$ mixing ratio $\eta$, as can be seen 
in Fig.~\ref{fig12}, where we plotted $F$ and $P_{\rm s}$ in 
in the spectral bin covering the deepest part of the band as 
functions of $b_{\rm cloud}$ and $\eta$.
Indeed, for $b_{\rm cloud} > 1.0$ and $\eta > 0.2$, the flux band depth
is fairly constant with $b_{\rm cloud}$ for every $\eta$.
Because of the insensitivity to $b_{\rm cloud}$,
$\eta$ could be derived from the band depth of the normalized flux,
although we would have to know the cloud top altitude 
(see Sect.~\ref{sect3.2.2}).

The strength of the absorption band in $P_{\rm s}$ is sensitive for 
$b_{\rm cloud}$ up till about 5 (for larger $b_{\rm cloud}$ and $\eta=0.21$, 
the band has virtually disappeared). The continuum $P_{\rm s}$ decreases with 
increasing $b_{\rm cloud}$ (see also Fig.~\ref{fig10}). $P_{\rm s}$ in the band
also decreases, but at a different rate: 
at wavelengths where $b_{\rm abs}$ is large, the clouds are invisible and
$P_{\rm s}$ will be high (see the $b_{\rm abs}=400$ line Fig.~\ref{fig7}),
while at wavelengths with little absorption, $P_{\rm s}$ will behave
similarly as in the continuum. With unresolved absorption lines, 
as in Fig.~\ref{fig11}, the band strength is a mixture of high and
low polarization. According to Fig.~\ref{fig12}, the $P_{\rm s}$ band 
strength (the absolute difference between each curve and the
$\eta=0.0$ curve), increases with increasing $\eta$,
yielding band strengths of tens of percents in $P_{\rm s}$.

Figure~\ref{fig9} includes a scatter plot based on Fig.~\ref{fig12},
that shows the relation between the $P_{\rm s}$ band strength and 
the (normalized) $F$ band depth for various $b_{\rm cloud}$.
Large $P_{\rm s}$ band strengths clearly correlate with large
$F$ band depths. Thus, measuring a strong $P_{\rm s}$ band in 
combination with a shallow $F$ band at $\alpha=90^\circ$ 
would not be explainable by a cloud layer of liquid water particles.
A reflecting surface instead of $A_{\rm s}=0.0$, would decrease 
$P_{\rm s}$ even more (except for very thick clouds, where $A_{\rm s}$ 
is irrelevant), and would thus not improve the explanation. 

\begin{figure*}
\figurenum{13}
\plottwo{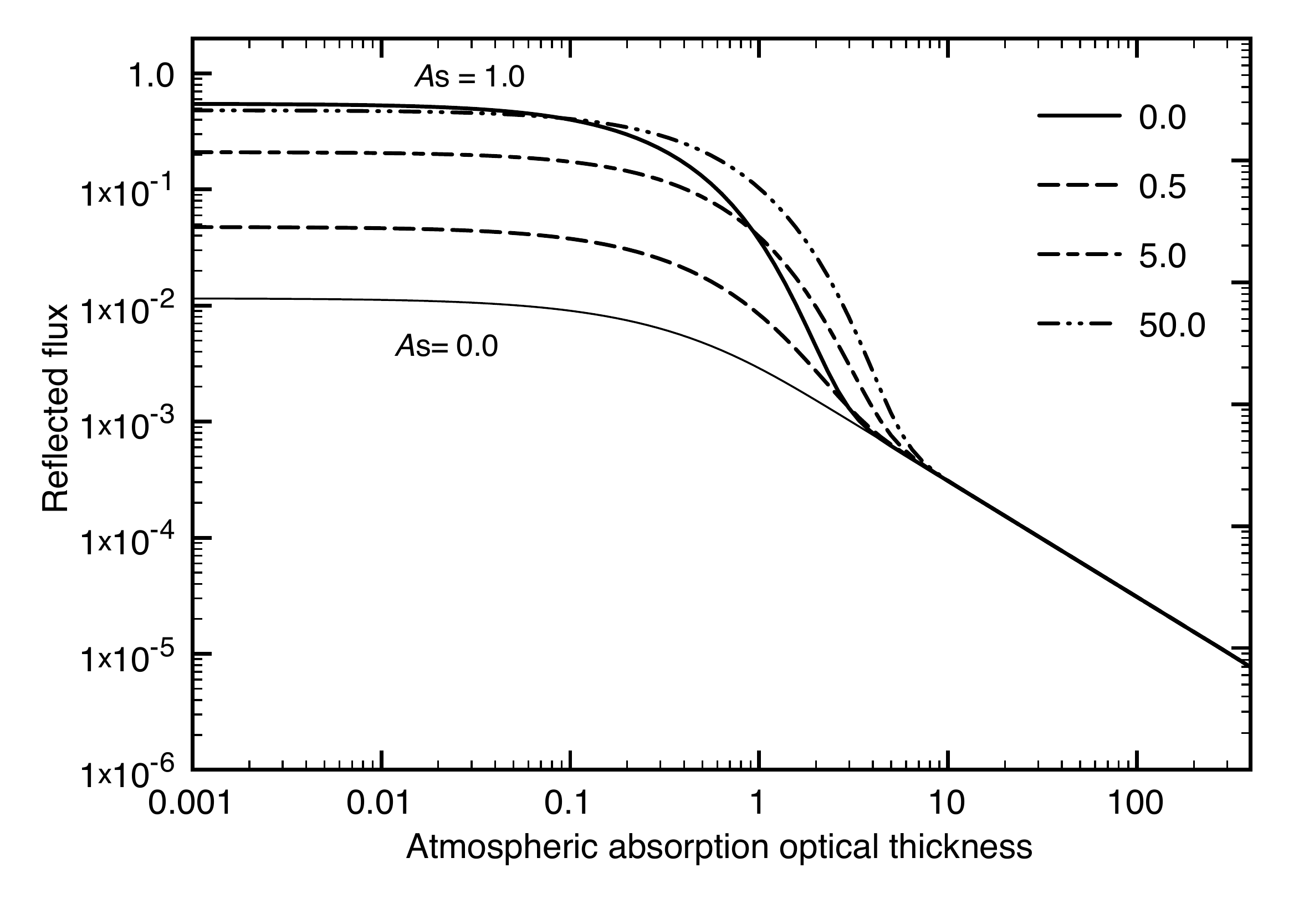}{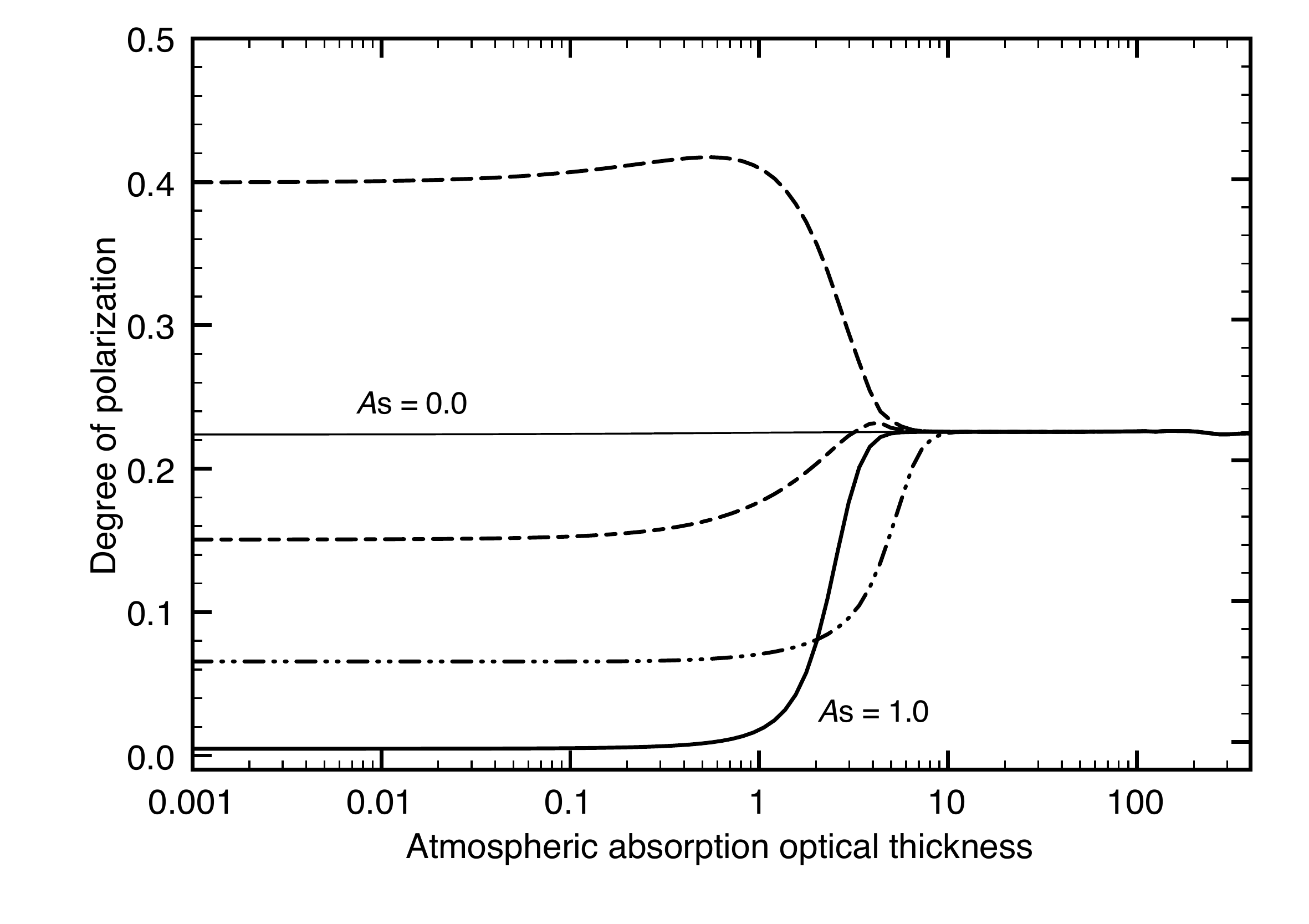}
\caption{Similar to Fig.~\ref{fig6}, except for $\alpha=38^\circ$,
         the 'rainbow' angle, and $b_{\rm cloud}$ equal to 0.5, 5.0, and 50.0.
         For comparison, the curves for cloud-free planets with black and
         white surfaces have also been included.}
\label{fig13}
\end{figure*}

The previous discussion focused on $\alpha=90^\circ$.
From Fig.~\ref{fig10}, it can be seen that around $\alpha=38^\circ$, 
the continuum $P_{\rm s}$ of planets with optically thin clouds is 
higher than that of a cloud--free, black planet.
Because with increasing $b_{\rm abs}$, the signals of
all planets tend towards those of cloud--free planets, 
$P_{\rm s}$ inside the deepest lines of the O$_2$ A--band could be 
{\em lower} than in the weaker lines and the continuum for dark, 
e.g. ocean covered, planets with thin clouds, around $\alpha=38^\circ$.
This can be seen in Fig.~\ref{fig13}.

\begin{figure}
\figurenum{14}
\epsscale{1.1}
\plotone{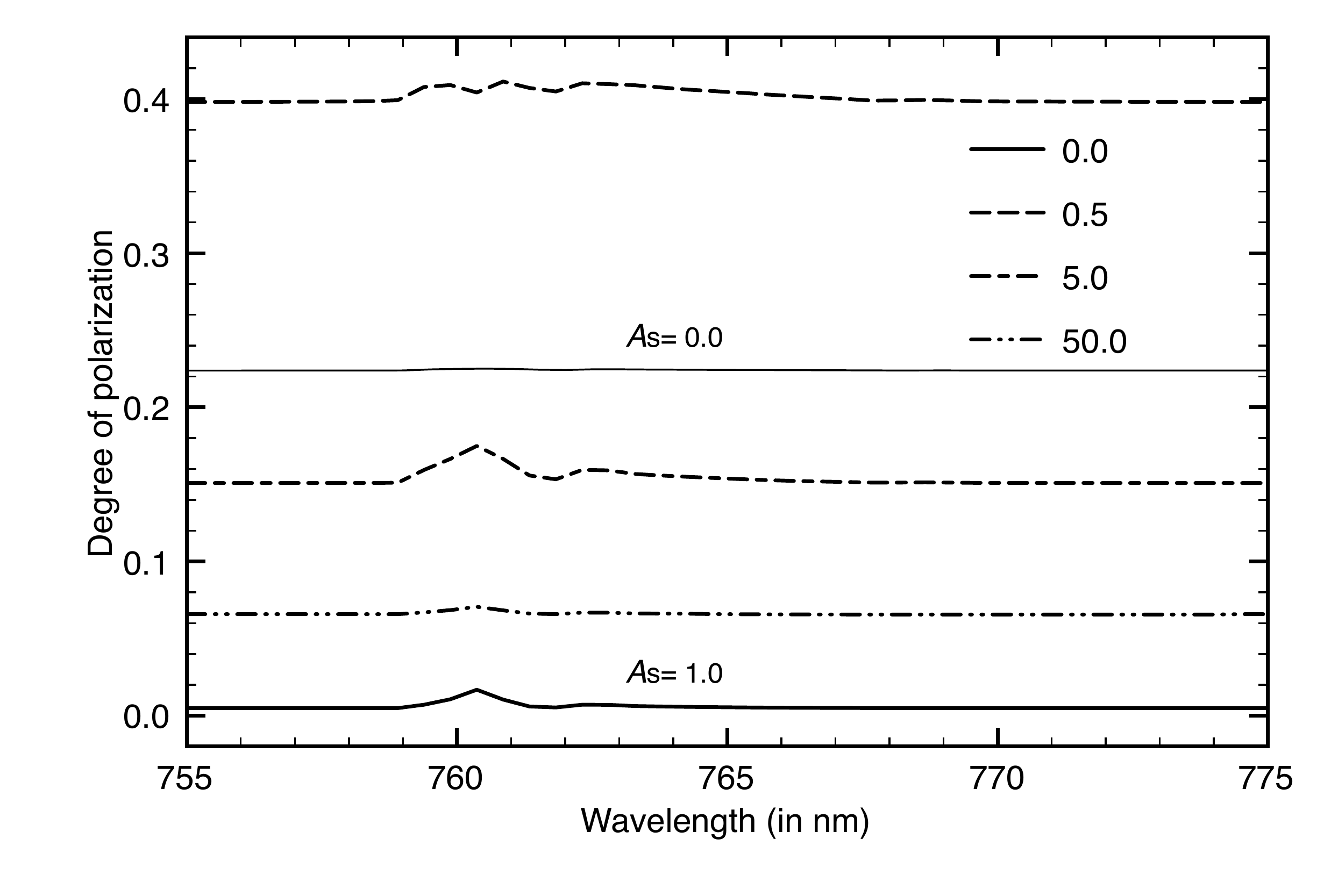}
\caption{Similar to Fig.~\ref{fig12}, except for $\alpha=38^\circ$,
         the 'rainbow' angle, and $b_{\rm cloud}$ equal to 0.5, 5.0, and 50.0.
         For comparison, the curves for black ($A_{\rm s}=0.0$) and
         white ($A_{\rm s}=1.0$), cloud--free exoplanets have also been included.}
\label{fig14}
\end{figure}

When absorption lines are not resolved, the effect will be smaller than
shown in Fig.~\ref{fig13}, because wavelengths with large $b_{\rm abs}$
will be mixed with wavelengths with small $b_{\rm abs}$.
Figure~\ref{fig14} shows the band in $P_{\rm s}$ With our spectral bin 
width of 0.5~nm, at $\alpha=38^\circ$.
For $b_{\rm cloud}=0.5$, $P_{\rm s}$ in the spectral bins with 
the strongest absorption lines is indeed lower than in adjacent bins, 
creating a band with a collapsed top.
A non--zero surface albedo will lower the continuum $P_{\rm s}$ 
for planets with optically thin clouds, and diminish this inversion
in the O$_2$ A--band.
Figure~\ref{fig14} also shows the lack of a band feature on a cloud--free
planet with a dark surface (because of the small $b_{\rm sca}$), 
and the difference in $P_{\rm s}$ of a planet with a thick cloud 
($b_{\rm cloud}=50.0$) and a cloud--free planet with a white surface, 
that will have similar fluxes.

Figure~\ref{fig13} also shows that the widths of resolved absorption lines
depend on $b_{\rm cloud}$. In $F$, optically
thicker clouds narrow lines down for $0.1 \leq b_{\rm abs} \leq 10$
as compared to those in the spectra of cloud--free planets.
In $P_{\rm s}$, lines can exhibit various shapes: narrowed down 
(for $b_{\rm cloud}= 50.0$), widened up with a tiny dip (0.01 or 1~\%) 
when $b_{\rm abs} > 4$ (for $b_{\rm cloud}= 5.0$),
or with higher $P_{\rm s}$ (0.02 or 2~\%) at the edges, 
with a narrow dip starting at $b_{\rm abs} \approx 0.6$.
The precise values of $b_{\rm abs}$ mentioned above depend on 
$\alpha$, as can be seen when comparing the curves for the 
cloud--free planet in Fig.~\ref{fig13} (with $\alpha=38^\circ$)
with those in Fig.~\ref{fig6} (where $\alpha=90^\circ$).
Examples of variations of line shapes in $P_{\rm s}$ in the presence
of aerosol can be found in \citet{1999JGR...10416843S}.

\subsubsection{The influence of the cloud top altitude $z_{\rm top}$}
\label{sect3.2.2}

Because on Earth, the mixing ratio of O$_2$ is well--known and more or less 
constant with altitude, the depth of the O$_2$ A--band in the flux that is
reflected to space can be used, to a first approximation, to derive the 
altitude of the top of a cloud layer.
Indeed, this is the basis of several cloud top altitude retrieval 
techniques \citep[][]{1991JApMe..30.1245F,1991JApMe..30.1260F}.
In this section, we investigate the relation between cloud top altitude,
$z_{\rm top}$, and band depths and strengths for our model planets.

\begin{figure}
\figurenum{15}
\epsscale{1.2}
\plotone{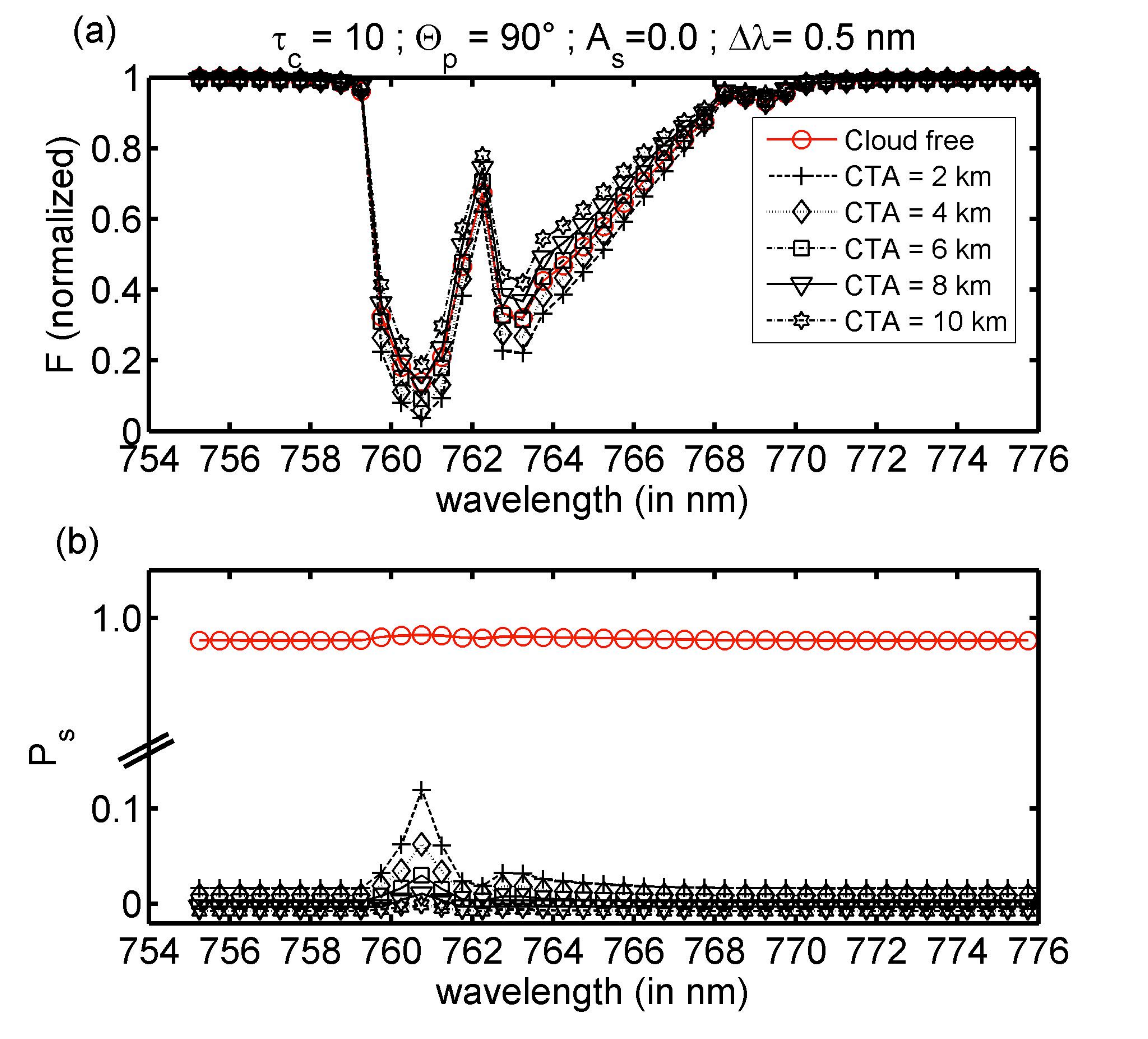}
\caption{$F$ (top) and $P_{\rm s}$ (bottom) of starlight reflected by 
         planets that are completely covered by a horizontally homogeneous 
         cloud layer with $b_{\rm cloud}= 10$ that is 2~km thick. The cloud
         top altitude ($z_{\rm top}$ or CTA in the plot) ranges from 
         2.0 to 10.0~km. The red curves are for cloud--free planets with
         $A_{\rm s}=0.0$ (cf. Fig.~\ref{fig4}). 
         All flux curves have been normalized at 755~nm.}
\label{fig15}
\end{figure}

Figure~\ref{fig15} shows the (normalized) $F$ and $P_{\rm s}$ across 
the O$_2$ A--band for exoplanets with black surfaces that are completely 
covered by horizontally homogeneous clouds with $b_{\rm cloud}= 10$ 
and $z_{\rm top}$ ranging from 2.0~to 10.0~km. The figure shows that 
the higher the cloud, the shallower the band in $F$. This is 
to be expected because the higher the cloud, the shorter the mean photon path
through the atmosphere, and thus the less absorption. 
For the lowest cloud, the lowest (normalized) $F$ is only 6~\% of the continuum. 
For the highest cloud, it is almost 20~\% of the continuum.
Interestingly, the $F$ band depth for the cloud--free planet is
similar to that for a cloudy planet with $z_{\rm top}=6$~km. Because
for the cloud--free planet with $A_{\rm s}=0.0$, 
only Rayleigh scattering contributes, its absolute flux will of 
course be much smaller than that of the cloudy planet (see Fig.~\ref{fig10}).
A cloud--free planet with $A_{\rm s}=1.0$, that would have a similar 
absolute flux as a cloudy planet, would have an $F$ band depth 
similar to that of the planet with $z_{\rm top}=2$~km (see Fig.~\ref{fig5}).

The continuum $P_{\rm s}$ decreases with $z_{\rm top}$ and changes from 
perpendicular to parallel to the reference plane when $z_{\rm top}$
exceeds about 7~km (this altitude will depend on $b_{\rm cloud}$).
The negative $P_{\rm s}$ is due to the negative single scattering 
polarization of the cloud particles at $\alpha=90^\circ$
(see Fig.~\ref{fig2}), and the higher the clouds, the stronger their
contribution to the planetary $P_{\rm s}$.
Measurements by the POLDER--instrument \citep[]{1994Deschamps}
of the continuum polarization of sunlight reflected by regions on Earth
are indeed being used to derive cloud top altitudes
\citep[][]{1994ITGRS..32...78G,2000JQSRT..64..173K}, 
and a similar approach was used for cloud top altitude retrieval on Venus 
using Pioneer Venus orbiter data and sulfuric acid model clouds 
\citep[][]{1998JGR...103.8557K}. Because of the abundance of photons 
in these Earth and Venus observations, high polarimetric accuracies can 
be reached. For example, the accuracy of POLDER is about 0.02 in degree of
polarization \citep[][]{toubbe1999}.

The cloud top altitude also affects $P_{\rm s}$ in the band: the higher 
the cloud, the shallower the band. The reason for this change in band 
strength is that with increasing $z_{\rm top}$, the absorption optical 
thickness above the cloud decreases, and while $P_{\rm s}$ in the deepest
absorption lines will remain high, it will decrease in the wings
of these lines, see Fig.~\ref{fig14} for $\alpha=38^\circ$ 
(for optically thin clouds, it will increase when $\alpha=38^\circ$).

\begin{figure*}
\figurenum{16}
\plottwo{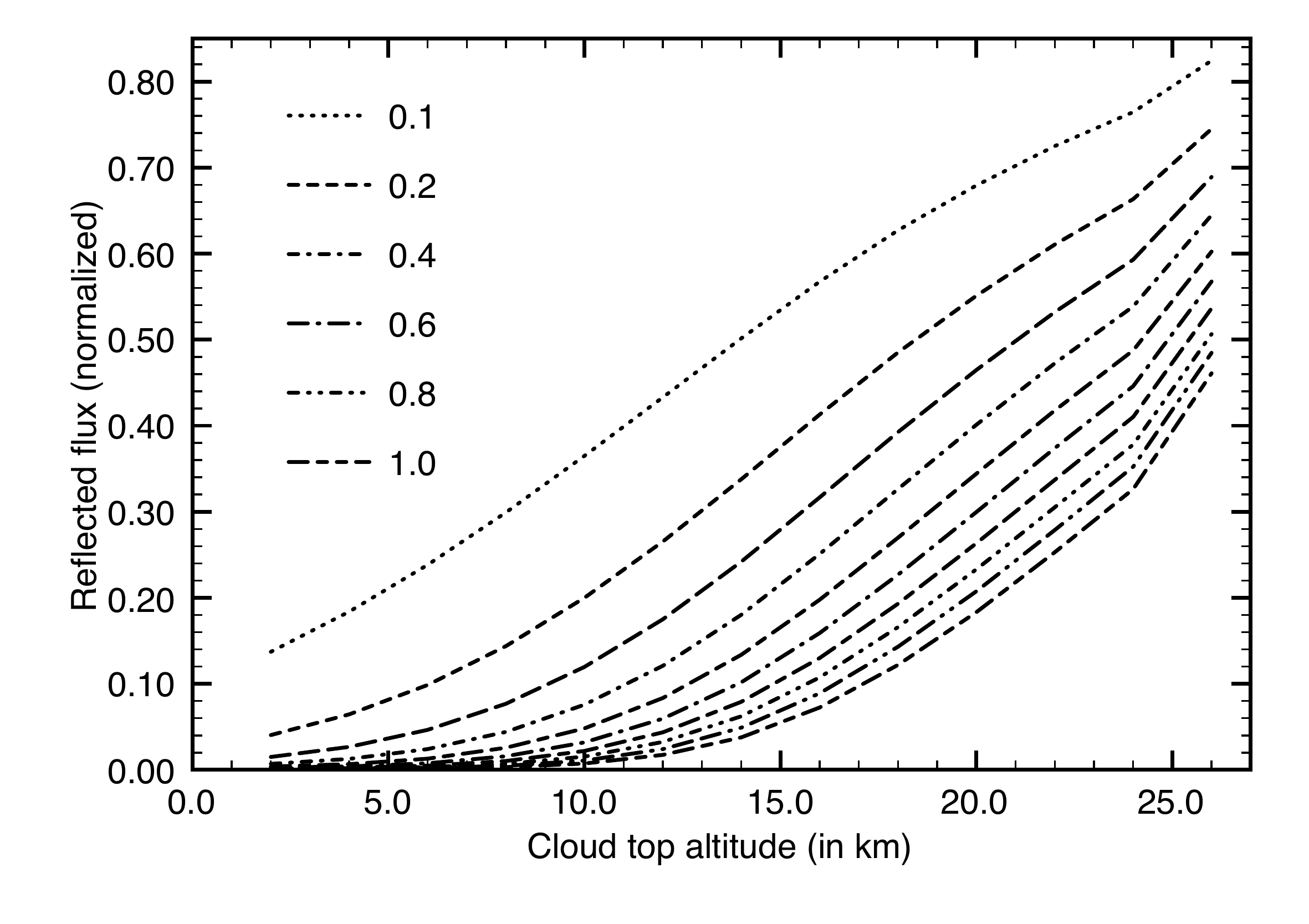}{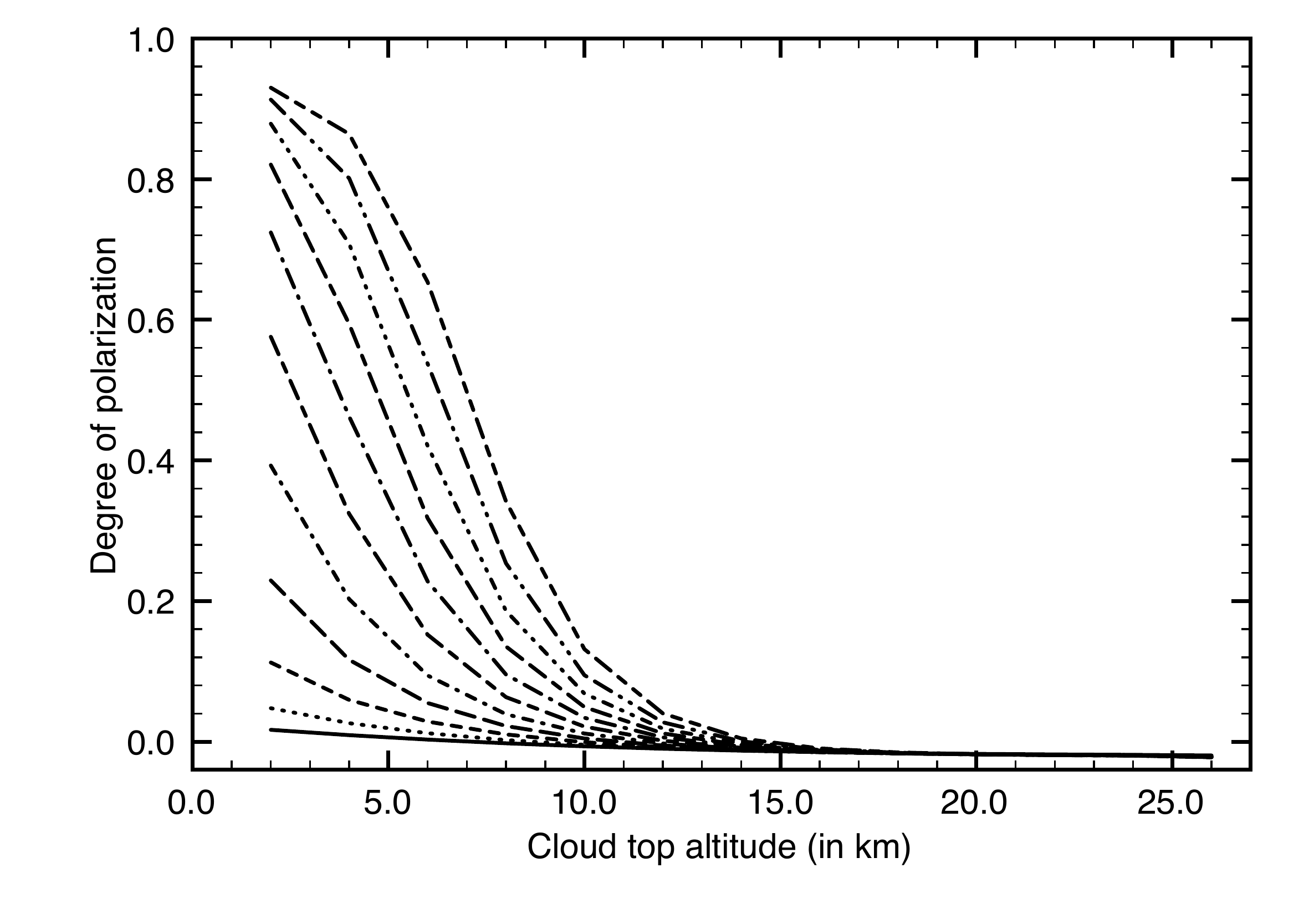}
\caption{Similar to Fig.~\ref{fig8}, except as functions of $z_{\rm top}$,
         for $b_{\rm cloud}=10$, a geometrical thickness of 2~km, and
         $A_{\rm s}=0.0$. The O$_2$ mixing ratio $\eta$
         ranges from 0.0 (thick solid line, not shown in the flux
         as it equals 1.0) to 1.0, in steps of 0.1.}
\label{fig16}
\end{figure*}

The effect of the O$_2$ mixing ratio $\eta$ on $F$ and $P_{\rm s}$ 
in the spectral bin covering the deepest part of the O$_2$ A--band,
for various values of $z_{\rm top}$
is shown in Fig.~\ref{fig16} (recall that the continuum $F$ and $P_{\rm s}$
equal the case for $\eta=0.0$, only shown in the polarization plot). 
On Earth, the tops of optically thick clouds will usually not have 
$z_{\rm top}$ higher than about 14~km (except at equatorial latitudes 
where the tropopause can reach altitudes of about 18~km),
but because cloud formation depends strongly on the atmospheric 
temperature profile, which will depend on the planet, we have 
included larger values for $z_{\rm top}$. It can be seen how the band
depth in $F$ decreases with increasing $z_{\rm top}$: for $\eta > 0.5$,
$F$ in the band is close to zero unless $z_{\rm top} > 10$~km, and thus 
insensitive to $z_{\rm top}$. There is an ambiguity between $z_{\rm top}$
and $\eta$: an observed band depth in $F$ could be fitted with different
combinations of $z_{\rm top}$ and $\eta$.
The band strength in $P_{\rm s}$ (i.e. the difference between the non--solid
curves and the solid, $\eta=0.0$, curve), decreases with increasing
$z_{\rm top}$, and is close to zero for small values of $\eta$ and/or 
$z_{\rm top} > 12$~km.
Here, there is a similar ambiguity, except that the smaller $\eta$,
the smaller the maximum band strength in $P_{\rm s}$: large band
strengths can only be explained by large $\eta$'s and/or low clouds.

The band strength in $P_{\rm s}$ as a function of the band depth in 
(normalized) $F$ is also shown in the lower left panel in Fig.~\ref{fig9}.  
This figure makes clear that measuring both the band depth in $F$
and the band strength in $P_{\rm s}$ would allow the
retrieval of both $z_{\rm top}$ and $\eta$, although that would
require very accurate measurements and (in the absence of absolute
flux and polarization measurements) assumptions about the
planet's surface albedo, especially with small values of $b_{\rm cloud}$, 
and the cloud coverage.


\subsection{Partly cloudy planets}
\label{sect3.3}

The model planets that we used previously, were either cloud--free or fully 
covered by a horizontally homogeneous cloud layer. These planets provide 
straightforward insight into the influence of clouds on the depth and 
strength of the O$_2$ A--band in, respectively, flux and polarization. 
The lower right panel of Fig.~\ref{fig9} contains all data points of the 
other panels, and thus shows the band strength in $P_{\rm s}$ as function 
of the band depth in $F$ for different values of $A_{\rm s}$, $\eta$, 
$b_{\rm cloud}$, and $z_{\rm top}$.
The cloud of data points in this panel shows that the larger $\eta$,
the larger the possible range of band depths in $F$ and band strengths in 
$P_{\rm s}$, and that for an Earth--like $\eta$ of 0.21, the band
will show up in $P_{\rm s}$ only for dark (but not too dark) surfaces
and optically thin clouds ($b_{\rm cloud} < 5.0$) (at $\alpha=90^\circ$,
and with our spectral bin width of 0.5~nm).
A horizontally homogeneous planet is however an extreme case. Here, we will 
use horizontally inhomogeneous model planets, with patchy surfaces and 
patchy clouds, such as found on Earth.

\begin{figure}
\figurenum{17}
\epsscale{1.2}
\plotone{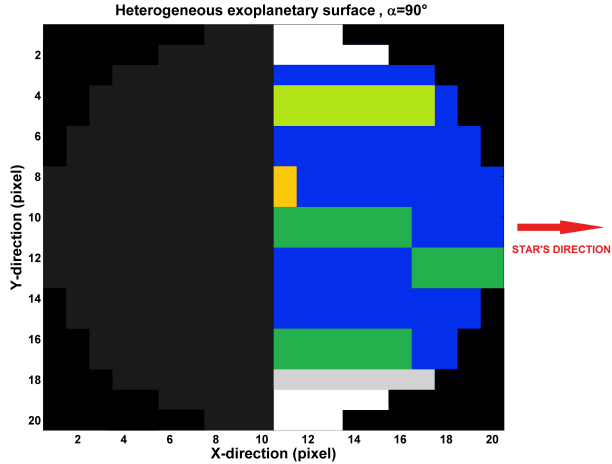}
\caption{Our model planet with its heterogeneous surface, and without
         clouds, at $\alpha=90^\circ$. The colors of the pixels 
         represent the following surface types: 
         White = fresh snow regions (32 pixels @ $A_{\rm s}$=0.9); 
         light--grey = old and/or melted snow (15 pixels @ $A_{\rm s}$= 0.6);
         light--green = grassy lands (39 pixels @ $A_{\rm s}$= 0.25);
         dark--green = forests (50 pixels @ $A_{\rm s}$= 0.15);
         yellow = sandy desert (30 pixels @ $A_{\rm s}$= 0.4);
         blue = ocean (150 pixels @ $A_{\rm s}$= 0.06).}
\label{fig17}
\end{figure}

For our horizontally inhomogeneous model planets, we use the following values
for the surface albedo $A_{\rm s}$: 0.90 (representative for fresh snow), 
0.60 (old and/or melted snow), 0.40 (sandy lands), 0.25 (grassy lands), 
0.15 (forests), and 0.06 (oceans). Figure~\ref{fig17} shows an example of a
(pixelated) cloud--free model planet, at $\alpha=90^\circ$, covered by five
types of surfaces. The geometric albedo of this planet is 0.24.

All our surfaces are depolarizing,
we thus do not include e.g. Fresnel reflection to describe the ocean
surface. In \citet{2008A&A...482..989S}, the disk--integrated signal
of a cloud--free Earth--like planet with a flat Fresnel reflecting 
interface on top of a black surface (including glint) was found to have 
a $P_{\rm s}$ that was about 0.04 {\em lower} than that of the same planet 
without the Fresnel reflecting interface. For fully cloudy planets 
($b_{\rm cloud}$=10), there was virtually no difference in the disk--integrated 
$P_{\rm s}$.
Waves on the ocean surface would very likely reduce the influence
of the Fresnel reflection, because of the randomizing effect of 
the variation in their directions, shapes and heights.


To model a patchy cloud pattern, we choose a cloud coverage (CC), i.e.\ 
the fraction of pixels that are cloudy. Note that these 
pixels are smaller than the pixels shown in Fig.~\ref{fig17}.
Next, we distribute an initial number of cloudy pixels (iCC, with iCC 
much smaller than the total number of cloudy pixels) randomly across the
planetary disk. The remaining cloudy pixels are distributed across the
remainder of the disk with the probability that a pixel will be cloudy 
increasing with the number of cloudy neighboring pixels.  
This method allows us to create patchy clouds with the size of the
patches depending on the values of CC and iCC. In particular, 
the larger the difference between iCC and CC, the larger the patches.
The user can assign different cloud properties 
($b_{\rm cloud}$, $z_{\rm top}$, geometrical thickness, and particle
micro--physics) to different cloudy pixels.

\begin{figure}
\figurenum{18}
\epsscale{1.2}
\plotone{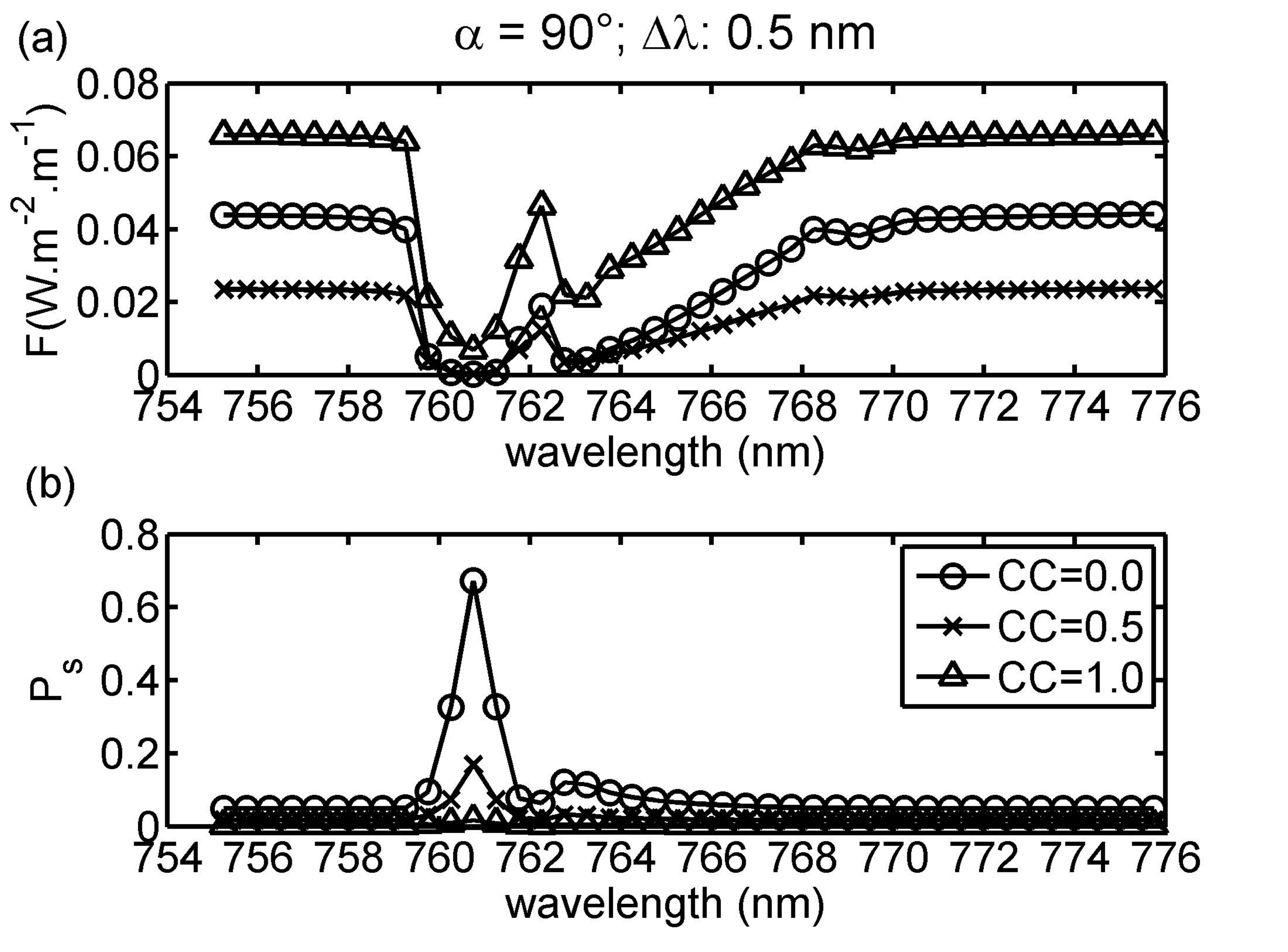}
\caption{Similar to Fig.~\ref{fig15}, except for the heterogeneous model 
         planet shown in Fig.~\ref{fig17} ($\alpha=90^\circ$) without 
         clouds (CC= 0.0), with half of the pixels covered by clouds 
         (CC=0.5) and with all pixels covered by clouds (CC=1.0). The 
         clouds have the following properties:  
         50\% with $b_{\rm cloud}= 50$, 16.7\% with $b_{\rm cloud}= 5$, 
         16.7\% with $b_{\rm cloud}= 10$, all between 4 and 6~km altitude,
         16.5\% with $b_{\rm cloud}= 10$, between 8 and 10~km.}
\label{fig18}
\end{figure}

Figure~\ref{fig18} shows $F$ and $P$\footnote{For horizontally 
heterogeneous planets, the disk--integrated $U$ is not necessarily 
equal to zero, and we therefore use $P$ (Eq.~\ref{eq_polarization})
instead of $P_{\rm s}$.} of the heterogeneous model planet shown in 
Fig.~\ref{fig17} with cloud coverages equal to 0.0 (cloud--free), 
0.5, and 1.0 (completely cloudy). The cloud properties are given 
in the figure caption.
Not surprisingly, $F$ of the fully cloudy planet is higher than 
that of the cloud--free planet. The lowest $F$ is that
of the partly cloudy planet, while one would expect a value in 
between the fluxes of the two other planets. 
Upon closer inspection, however, it appears that the poles of the 
partly cloudy planet are (partly) covered by clouds, suppressing
$F$, because the clouds are less bright than the snow and ice surfaces.
The polarization signal of the partly covered planet is between those 
of the other planets, because the polar regions do not contribute 
a significantly different polarization signal compared to the clouds.

\begin{table}
\begin{center}
\begin{tabular}{c||r|r|r|r}
Type & phase & $b_{\rm cloud}$ & $z_{\rm bot}$ (km) & $z_{\rm top}$ (km) \\ \hline
1 & liquid &  0.2 \hspace*{0.03cm} &  4.0  \hspace*{0.3cm} &   6.0  \hspace*{0.3cm}   \\
2 & liquid &  5.0 \hspace*{0.03cm} &  4.0  \hspace*{0.3cm} &   6.0  \hspace*{0.3cm}   \\
3 & liquid & 10.0 \hspace*{0.03cm} &  4.0  \hspace*{0.3cm} &   6.0  \hspace*{0.3cm}   \\
4 & liquid & 50.0 \hspace*{0.03cm} &  4.0  \hspace*{0.3cm} &   6.0  \hspace*{0.3cm}   \\
5 & liquid & 10.0 \hspace*{0.03cm} &  6.0  \hspace*{0.3cm} &   8.0  \hspace*{0.3cm}   \\
6 & solid  &  0.5 \hspace*{0.03cm} &  10.0  \hspace*{0.3cm} &  12.0  \hspace*{0.3cm}   \\
\end{tabular}
\caption{The properties of the six model cloud types.} 
\label{tab1}
\end{center}
\end{table}
 
As already shown in Fig.~\ref{fig9}, horizontally homogeneous planets
present a wide variation in band strengths in $P$ versus band depths in $F$.
Next, we will look at the variation that can be expected for horizontally
inhomogeneous planets.
In contrast to the curves in Fig.~\ref{fig18} which have been computed
for a horizontally inhomogeneous planet with patchy clouds and surface
albedo, for this, we will compute the variation of band strengths and depths
by taking weighted sums of horizontally homogeneous planets to limit the
computation times while maximizing the variations. \citet{2012A&A...546A..56K}
investigated the differences between 'true' horizontal inhomogeneities and
the weighted sum approach, and while for individual planets, the differences
can be significant, statistically both methods yield the same results.

For a given value of $\eta$, we thus compute the flux vector of a 
planet using $\pi {\bf F}= (1/7) \Sigma_{i}^{7} \pi {\bf F}_i$,
with ${\bf F}_i$ the flux vector of a horizontally homogeneous planet.
The horizontally homogeneous planets used for the modeling can be
either without clouds or completely covered by one of 6 cloud types.
Table~\ref{tab1} shows the properties of the 6 cloud types. 
Types 1 - 5 consist of the same liquid water cloud particles that 
we used before. Type~6 consists of water ice crystals with scattering
properties taken from \citet{2012A&A...548A..90K}.
We use the 6 surface albedo's $A_{\rm s}$ specified in the caption of
Fig.~\ref{fig17}), and as before assume Lambertian surface reflection,
because we are mainly interested in the effects of the clouds, and 
because the influence of a polarizing surfaces such as Fresnel reflection,
is relatively small \citep[][]{2008A&A...482..989S}.

\begin{figure*}
\figurenum{19}
\plottwo{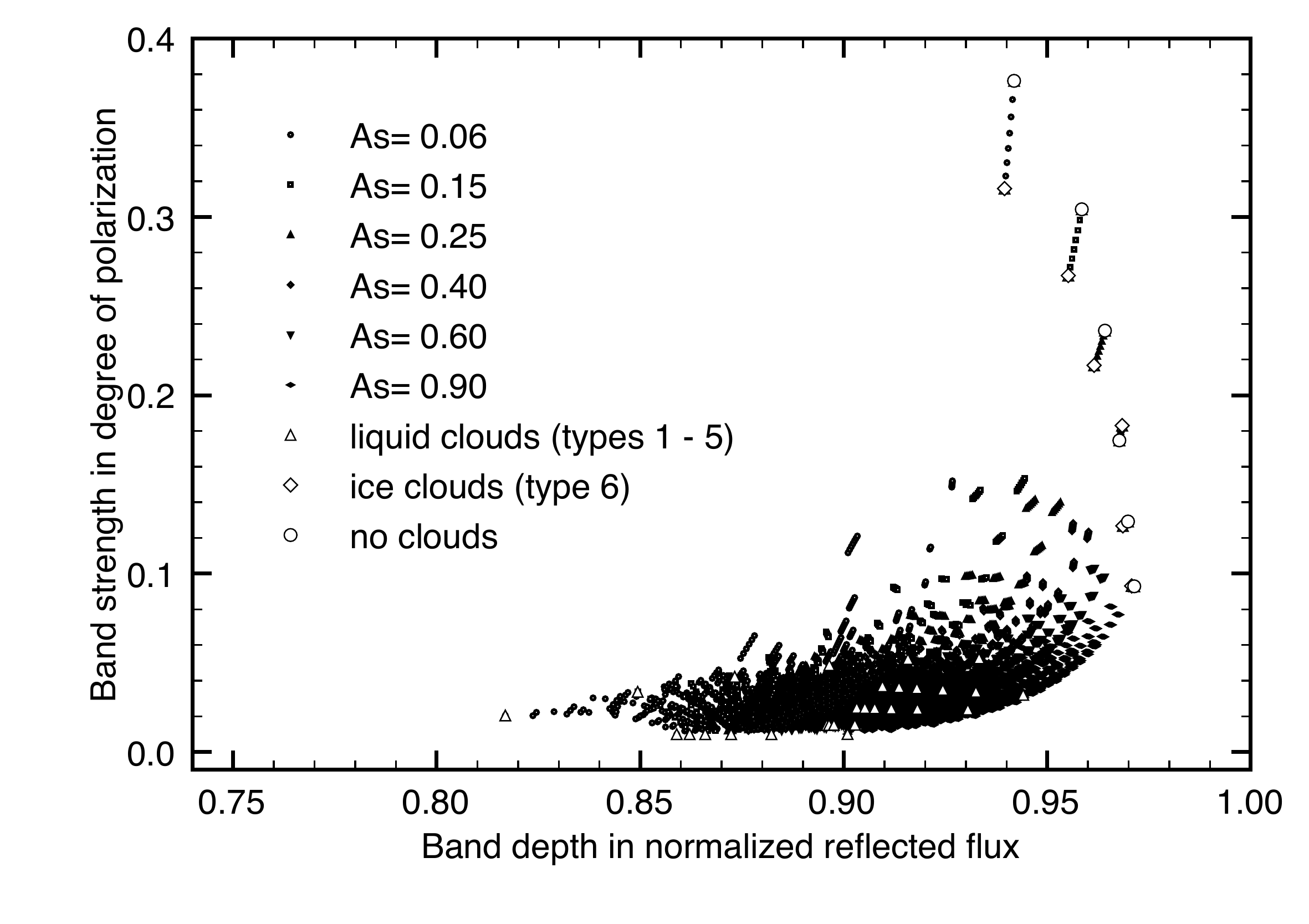}{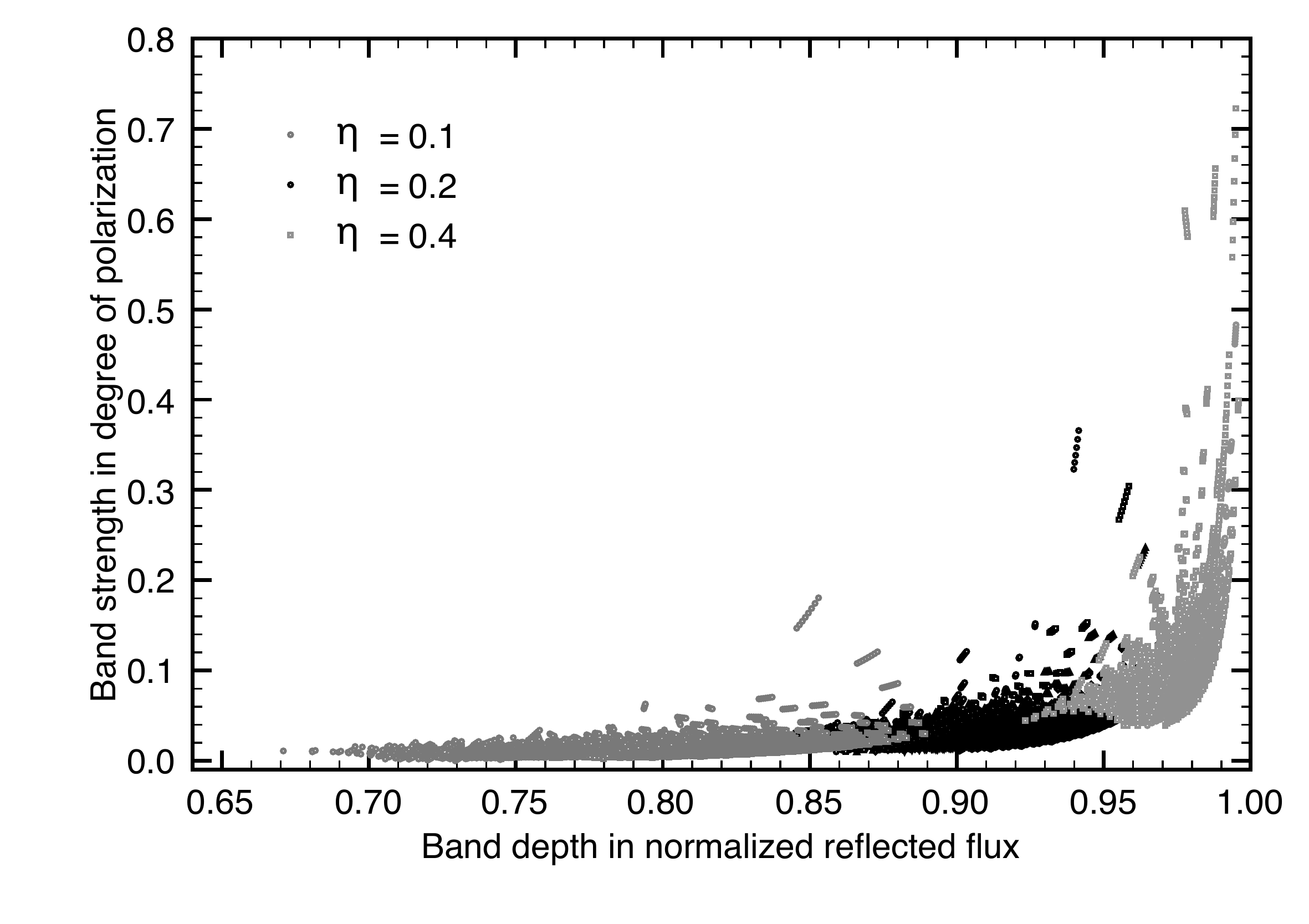}
\caption{Similar to Fig.~\ref{fig9} for $\eta=0.2$ (left) and 
         $\eta=0.1$, $0.2$, and 0.4 (right), $\alpha=90^\circ$. 
         Each model planet has 6~possible values of $A_{\rm s}$,  
         cloud--free and/or cloudy pixels (of the 6~types in Tab.~\ref{tab1}).
         Left: the white symbols pertain to horizontally homogeneous 
         planets. The small, dark symbols pertain to horizontally 
         inhomogeneous planets.
         Right: the data points for $\eta=0.2$ partially 
         over--plotted by those for $\eta=0.1$ (on the left side) 
         and 0.4 (on the right side).}
\label{fig19}
\end{figure*}

To avoid ending up with a huge number of data points, we decided to 
use a single value for $A_{\rm s}$ for each model planet, which still
gives us 10348 flux vectors of planets.
Figure~\ref{fig19} shows data points for $\eta=0.2$, 0.1, and 0.4.
Several data points can be seen to line up, in particular the data points 
for cloud--free planets and planets with ice clouds. These lines 
of data points pertain to the 'transition' from a homogeneous planet 
with a given surface albedo without clouds to a homogeneous planet 
with the same surface albedo but with clouds, with partially 
cloudy planets in between.
Planets with optically thin clouds would fill the region between 
the densest part of the cluster of data points and the high $P$ 
points for the planets without clouds and with ice clouds
(see also Fig.~\ref{fig9}).

The data points in Fig.~\ref{fig19} clearly illustrate the 
variation of the band depth in $F$ for a single value of $\eta$ due 
to the differences in $A_{\rm s}$ and the cloud types. Although increasing
$\eta$ generally increases the band depth in $F$, there is a huge overlap 
of data points for different values of $\eta$. 
When relatively small band depths in $F$ are observed, however, one would 
know that $\eta$ is small, even without knowledge on the presence of clouds 
and their properties or $A_{\rm s}$.

The vertical extend of the data points in Fig.~\ref{fig19}
is a measure of the added value of measuring the band strength in $P$.
The band strength could provide extra information about $b_{\rm cloud}$,
the cloud coverage, and $A_{\rm s}$, in particular for optically thinner 
clouds. As can be seen, for higher values of $\eta$, the range of band 
strengths and thus the added value of $P$ increases strongly at this phase 
angle of 90$^\circ$. Conversely, large band strengths in $P$ combined with
large band depths in $F$ would indicate high values of $\eta$. 

The model planets that provided the data points for Fig.~\ref{fig19} are 
quasi--horizontally inhomogeneous: their $F$ and $P$ values are 
representative for horizontally inhomogeneous planets, but they do not
account for, for example, the effects of localized and patchy clouds 
(the spectra in Fig.~\ref{fig18} do account for these effects).
In Fig.~\ref{fig20}, we illustrate the variation in the band depth in $F$
and the band strength in $P$ for planets with homogeneous surfaces 
($A_{\rm s}= 0.15$ or 0.6) and patchy clouds of the types 1, 5, 
and 6 (the ice clouds). Each planet has one type of clouds. 
Given a planet with a certain cloud coverage,
we computed $F$ and $P$ for 200~randomly selected different 
shapes and locations of the cloud patches (this method is described 
in detail in \citet{Rossi2017}). 
The error bars indicate the 1--$\sigma$ variability in the 
computed signals for each planet (the variability equals zero for
the fully cloudy planets).

Figure~\ref{fig20} shows that for a given cloud coverage CC and type, 
the actual distribution of the clouds across the planetary disk
usually also influence the bands in $F$ and $P$. 
The variability decreases with increasing CC, because a large amount
of clouds allows less spatial variation.
Above a dark surface ($A_{\rm s}=0.15$), high altitude, liquid water 
clouds with $b_{\rm cloud}=10$ (type 5, bottom left), show a smaller variance 
in both $P$ and $F$ than thin, low altitude clouds (type 1, top left). 
Increasing the surface albedo of the latter planet (top right) 
decreases the variability both in $P$ and in $F$ for all cloud coverages,
while not strongly changing the average values. The spatial distribution 
of the thin, high altitude ice clouds (type 6, bottom right) has a 
negligible influence on the bands, for every $\eta$.

The large number of free parameters on a planet introduces significant
degeneracies in the relation between band depths in $F$ and band strengths
in $P$. A detailed investigation of the data points and the underlying 
planet models as shown in Figs.~\ref{fig19} and~\ref{fig20} not only 
at $\alpha=90^\circ$ but also at other phase angles,
and across a wider wavelength region (in particular, including shorter
wavelengths that are more sensitive to the gas above the clouds)
could lead to the development of a retrieval algorithm for future 
observations, but is outside of the scope of this paper.


\section{Discussion}
\label{sect:discussion}

The detection of absorption bands and the subsequent determination of 
column densities or mixing ratios of atmospheric biosignatures 
such as O$_2$, H$_2$O and CH$_4$ are crucial tools in the search for
habitable environments and life on exoplanets. The strength of a gaseous
absorption band in a visible planetary spectrum depends on the mixing 
ratio of the gas under investigation, on other atmospheric constituents 
and vertical structure, and on the surface properties. 

On Earth, oxygen arises from large--scale photosynthesis and is well--mixed
up to high altitudes. Because we have detailed knowledge about the Earth's
O$_2$ mixing ratio and vertical distribution, a routine Earth--observation
method for determining cloud top altitudes is the measurement of the depth 
of the O$_2$ A--band in the flux of sunlight that is reflected by 
completely cloudy regions \citep[see][etc.]{saiedy1965,1998GeoRL..25.3159V,koelemeijer2001,preusker2007,lelli2012,desmons2013}.
Knowledge of cloud top altitudes is important for climate studies and also
to determine the mixing ratios of gases like ozone, H$_2$O and CH$_4$, 
that are strongly altitude dependent, and whose spectral signatures are 
also affected by the presence and properties of clouds.
For exoplanets, cloud parameters and mixing ratios are unknown,
and the aim is to determine both. 

Our computations show that there will be significant degeneracies
if only fluxes of reflected starlight are being used. 
Assuming that a planet is horizontally 
homogeneous, and observed at a phase angle of 90$^\circ$,
a measured absorption band depth in flux $F$ could be fitted
with different O$_2$ mixing ratios $\eta$ depending on the assumed 
cloud optical thickness (Fig.~\ref{fig12}) and cloud top altitude 
(Fig.~\ref{fig16}). The surface albedo appears to be less of an
influence (Fig.~\ref{fig5}). Here, we assumed that only the relative
fluxes are available, because absolute exoplanet fluxes can only be
obtained when the planet radius and distances to its star and observer
are known (the flux of a small bright planet can equal that of a 
large dark planet).

Degeneracies are even more of a problem if the planet is assumed to 
be horizontally inhomogeneous. Our computations for planets with different
mixing ratios, and patchy clouds with various coverage percentages and 
spatial distributions, optical thicknesses and altitudes, and thermodynamical
phase (liquid or ice; the phase could be derived from the cloud altitude if
the atmospheric temperature profile is known),
show a wide range of band depths in the flux with significant overlap 
between the signals of various model planets. Including surface pressures
different than that on Earth, clouds made of other condensates than water, 
and atmospheric hazes and/or other aerosol, would increase the range of 
possible fits to the observations.
 
Measuring both the flux and the polarization of starlight that is
reflected by a planet could help to reduce the degeneracies, because
the band strength in polarization has a different sensitivity to 
the atmospheric parameters and the surface albedo than the band depth
in reflected flux. Indeed, assuming a horizontally homogeneous planet,
and a phase angle of 90$^\circ$,
measuring $P$ would help to retrieve the cloud optical thickness,
although for optically thin clouds, the surface albedo would influence
the signal, too. The sensitivity of the polarization does depend
on the mixing ratio: the larger $\eta$, the larger the range of 
$b_{\rm cloud}$ polarimetry is sensitive to. 
The polarization also holds information on the 
cloud top altitude, although the mixing ratio should be known for
an actual retrieval. Here, a combination with flux measurements could
help. Retrieving cloud top altitudes with an error of $\pm 2$~km would
require polarization measurements with a precision of 1 -- 2$\%$.

Assuming horizontally inhomogeneous planets, the range of planet
parameters that would fit a certain combination of flux and 
polarization measurements obviously increases. To diminish degeneracies,
the observational precision should be increased, as subtle differences 
in flux and polarization could help to distinguish between different models.
Measurements at several phase angles would also provide more information,
as in particular the degree of polarization in and outside the band
depend strongly on the atmospheric parameters. Indeed, measurements 
at the rainbow phase angle (38$^\circ$) would not only help to determine
whether cloud particles are made of liquid water, but also provide
information on the cloud optical thickness. \citet{1974SSRv...16..527H} 
show a number of plots of the single scattering polarization of particles 
with different compositions. An investigation into what would be the
distinguishing 'rainbow' phase angle for clouds made of such particles
would help to plan observations, but is outside the scope of this paper.

Another method to diminish degeneracies would be to perform observations
at a range of wavelengths, in particular, short wavelengths are more
sensitive to scattering by gas and small particles than longer 
wavelengths, and 
could thus provide more information about the amount of gas above the
clouds (i.e.\ the cloud top altitude), and possibly about the cloud 
patchiness (by reflecting more light from the cloud--free regions).
Longer wavelengths are less scattered by the gas and small particles,
and thus give a better view of the clouds and surfaces. Because of 
the spectral information in both the flux and the polarization, not
only in absorption bands but also in the continuum, it
is thus essential to use narrow band observations rather than
broadband observations. 

\begin{figure*}
\figurenum{20}
\plottwo{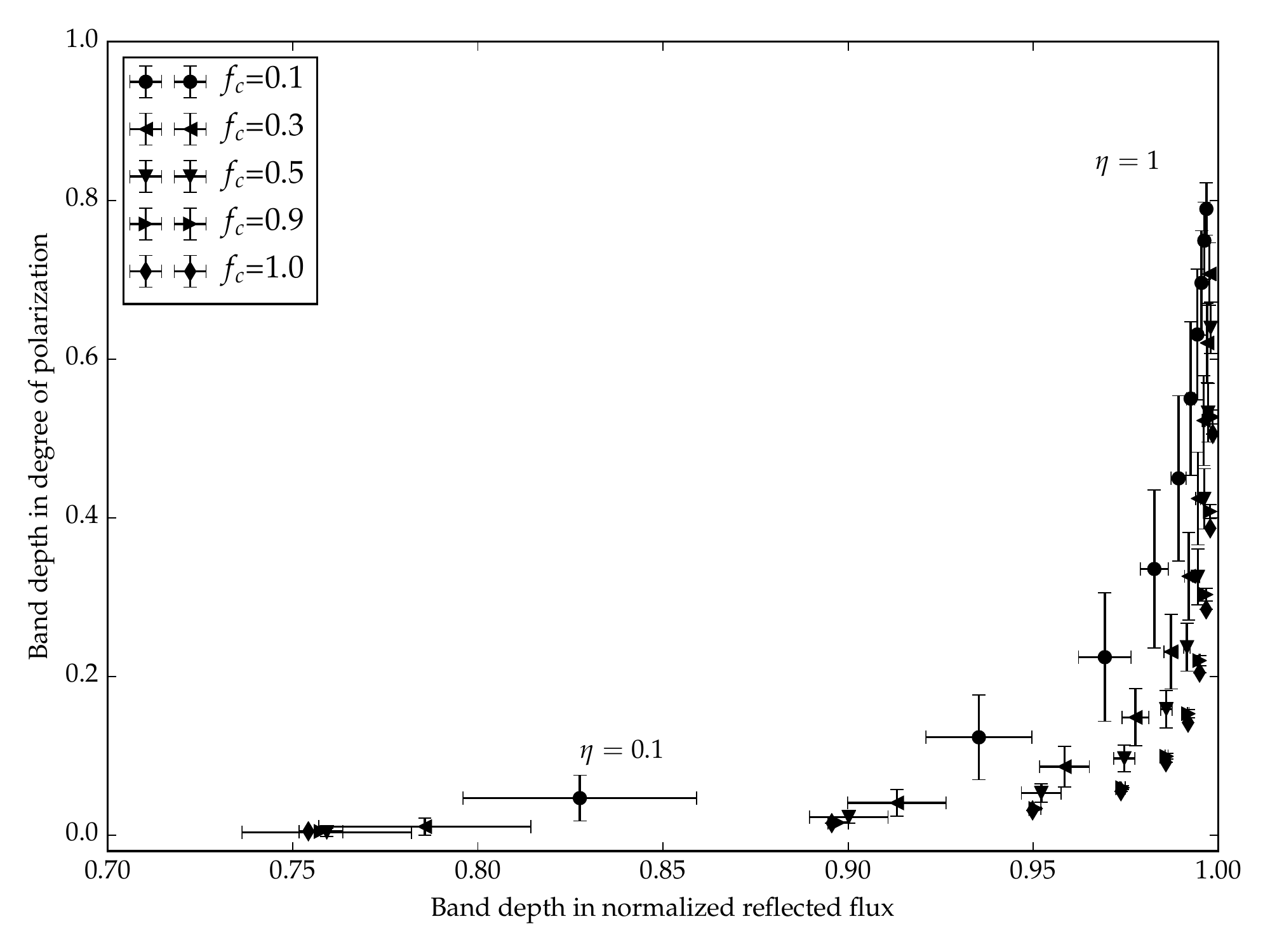}{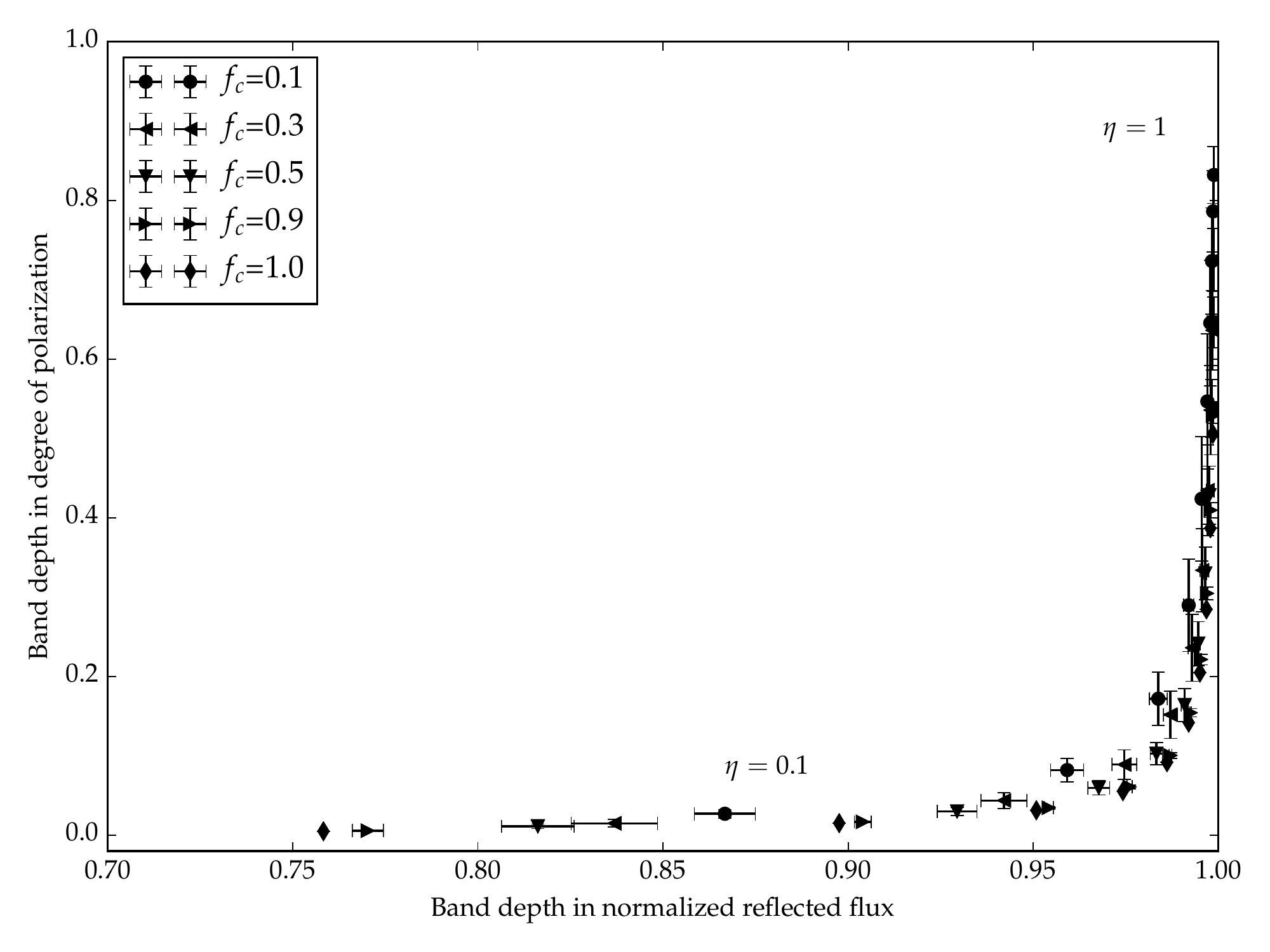}
\plottwo{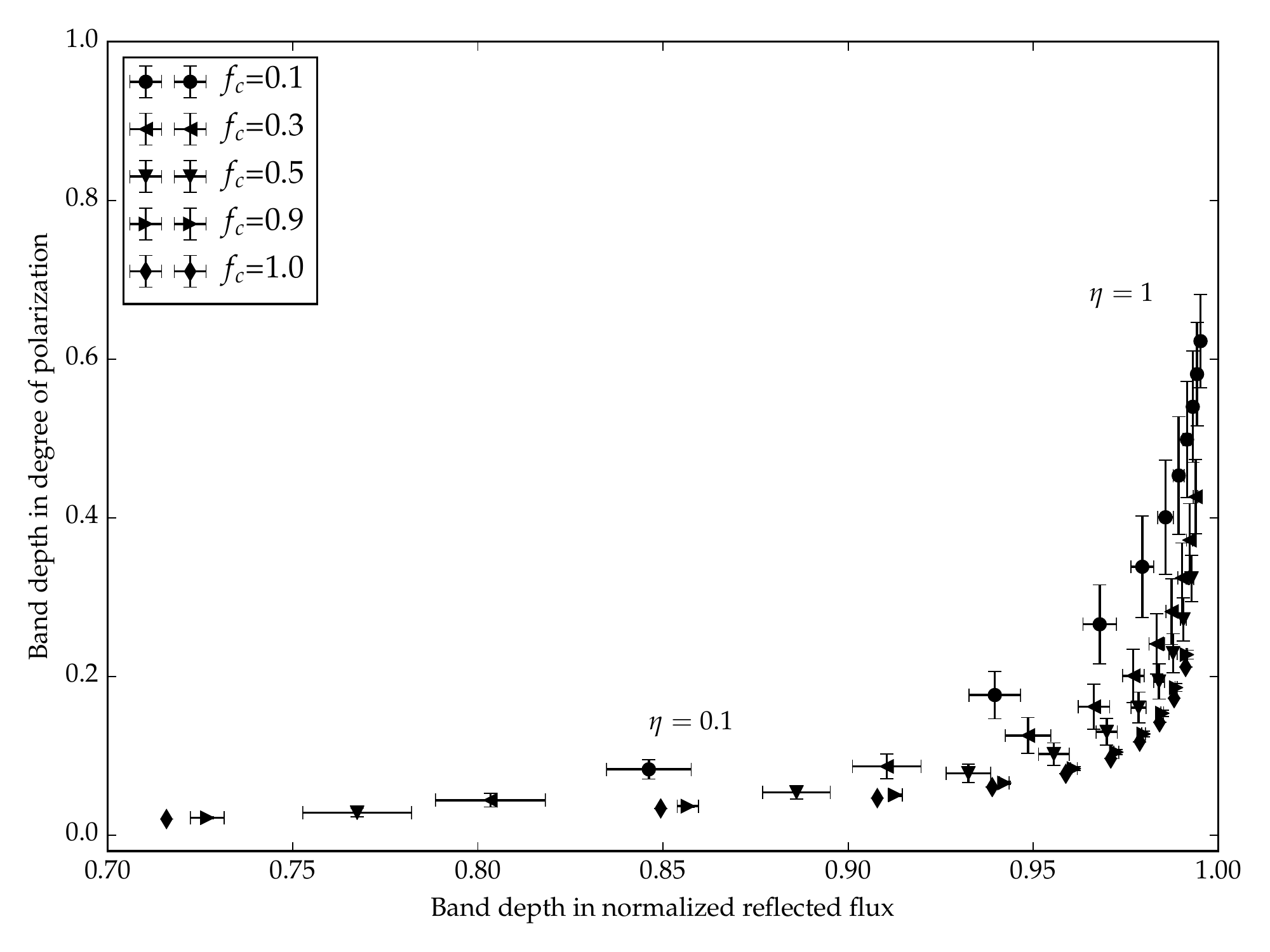}{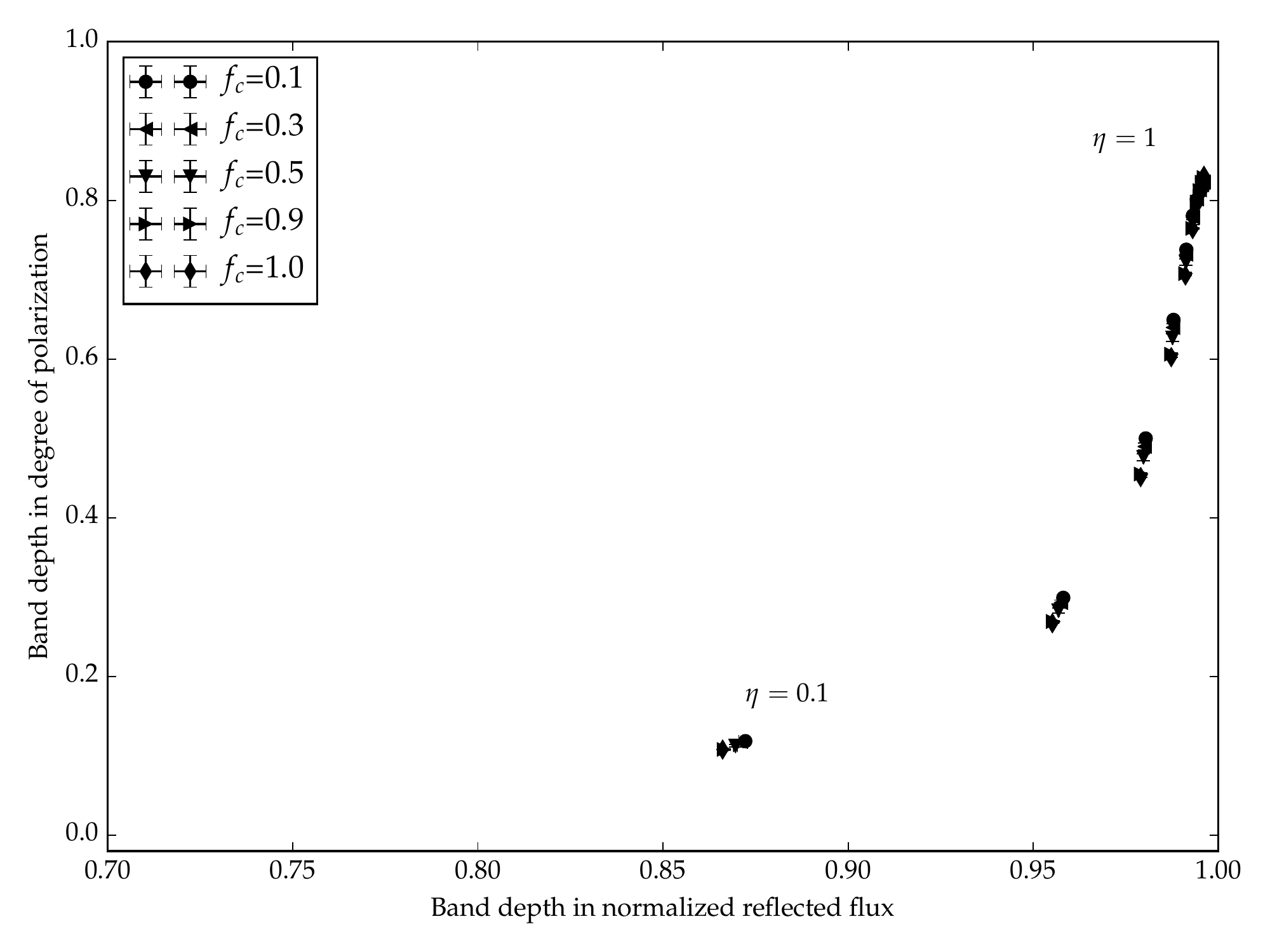}
\caption{Similar to Fig.~\ref{fig19} except for planets with patchy
         clouds, and $\alpha=90^\circ$. 
         Each planet has one value for $A_{\rm s}$ and one cloud type
         (see Tab.~\ref{tab1}).
         Top left: $A_{\rm s}= 0.15$, type 1. Top right:
         $A_{\rm s}= 0.6$, type 1. Bottom left: $A_{\rm s}=0.15$,
         type 5. Bottom right: $A_{\rm s}= 0.15$, type 6 (ice clouds).
         The error bars indicate the variance in the band depth in $F$
         and in the band strength in $P$ for the same value of CC 
         ($f_{\rm c}$ in the plot), but for different distributions of 
         the clouds.
         Mixing ratio $\eta$ increases from 0.1 to 1.0 in steps of 0.1.}
\label{fig20}
\end{figure*}

A high temporal resolution would also help to identify the contributions 
of time varying signals such as the rotation of a planet with a 
horizontally inhomogeneous surface (a regular variation, not addressed
in this paper) and/or weather patterns changing the cloud coverage
(as on Earth, the cloud coverage could also depend on the surface
properties, such as the presence of mountains). Our computations
of the variability in the O$_2$ A--band depth and strength in flux 
and polarization due to different shapes and locations of cloud patches
(Fig.~\ref{fig20}) illustrate which observational accuracy would be needed 
to resolve such variations, which depend on the cloud coverage and the
mixing ratio.

In our models, the surfaces reflect Lambertian, i.e.\ depolarizing.
Various types of natural surfaces reflect linearly polarized
light, and sometimes also circularly polarized 
\citep[see, e.g.][and references therein]{2017JQSRT.189..303P}.
Including such polarizing surfaces into our planet models would 
strongly increase the number of free parameters, and while the 
influence of the polarizing surfaces could influence the continuum
polarization for the regions on a planet that are cloud--free with
at most a few percent provided the surface albedo is not too low
\citep[see][for results for a planet completely covered
by a flat, Fresnel reflecting ocean surface]{2008A&A...482..989S}. 
For cloudy regions no influence of the surface polarization is expected.

While our computations show that measuring the polarization would
decrease the degeneracies in retrievals of mixing ratios of gases such as
the biosignature O$_2$, they can also be used to optimize the design of
instruments and/or telescopes. Indeed, many spectrometers
are sensitive to the polarization of the observed light: the measured
flux will depend on the polarization of the incoming light. Because the 
polarization will usually vary across a gaseous absorption band, any
instrumental polarization sensitivity will change the shape and depth of 
measured absorption bands. Unless carefully corrected for (which is difficult
when the incoming polarization is not known, even when the instrumental
polarization is accurately known), such changes will influence the retrieved
gas mixing ratios. Typical absorption band strengths in our polarization computations
can be combined with (estimated) instrumental polarization sensitivities to 
estimate and possibly minimize the errors that could result.

These errors will strongly depend on the spectral resolution of the observations:
the higher the spectral resolution, the higher the observable degree of
polarization. This can be seen in Fig.~\ref{fig3} and also in the various
figures that show $F$ and $P$ (or $P_{\rm s})$ as functions of $b_{\rm abs}$,
the atmospheric absorption optical thickness. Thus the higher the spectral
resolution, the more information can be obtained from observations, but the
more care should be taken to account for instrumental polarization.

Current large ground--based telescopes with broadband photometric and polarimetric 
capabilities for exoplanet observations are SPHERE/VLT and GPI/Gemini North. 
Narrow--band spectropolarimetry might become available in the near--future on 
EPICS/ELT and proposed space--telescopes such as WFIRST, LUVOIR or HabEX,
and provide us with the first spectra of the O$_2$ A--band to be used
for the characterization of exoplanets (absorption line resolving
Doppler detections of exoplanetary O$_2$ might indeed succeed earlier
and provide lower limits for the O$_2$ column density).
Then the difficulty will be to discard false--positives detections of O$_2$.
Indeed, for instance, dioxygen may not necessarily be a biosignature 
\citep[][]{2014ApJ...785L..20W,2014ApJ...792...90D,2015ApJ...812..137H,
2016ApJ...819L..13S,2016ApJ...821L..34S}. 
Lifeless, terrestrial planets in the habitable zone of any type of star 
can develop oxygen--dominated atmospheres through photolysis of H$_2$O
\citep[e.g.][]{2014ApJ...785L..20W} or CO$_2$ \citep[e.g.][]{2015ApJ...812..137H}. 
However,  O$_2$ due to CO$_2$ photolysis would not be detectable 
with current and planned space-- and ground--based instruments 
on planets around F and G type stars. Planets around K--stars and 
especially M-stars, may produce detectable abiotic O$_2$ because of 
the low UV--flux from their parent stars \citep[][]{2015ApJ...812..137H}. 
Also, strong O$_4$ features that could be visible in transmitted spectra 
at 1.06 and 1.27~$\mu$m or in UV/VIS/NIR reflected light spectra by 
a next generation direct--imaging telescopes such as LUVOIR/HDST 
or HabEx could be a sign of an abiotic dioxygen--dominated atmosphere 
that suffered massive H--escape 
\citep[see][]{2016ApJ...819L..13S,2016ApJ...821L..34S}. 
These considerations highlight the importance of wide spectral coverage for 
future exoplanet characterization missions  \citep[][]{2014ApJ...792...90D},
and the combination of flux and polarization measurements in order to
obtain as much information as possible from the limited numbers of photons.

\section{Summary}
\label{sect4}

We have computed the O$_2$ A--absorption band feature in 
flux and polarization spectra of starlight that is reflected by Earth--like 
exoplanets and investigated its dependence planetary parameters.
The O$_2$ A--band covers the wavelength region from about 755 to 775~nm, 
and it is the strongest absorption band of diatomic oxygen in the visible. 
Observations of the depth of this band in the reflected flux and the
strength of the band in polarization could be used to derive the O$_2$ 
mixing ratio in the atmosphere of an exoplanet and with that provide insight 
into the possible level of photosynthetic processes on the planet.

In our model computations, we assumed that O$_2$ is well--mixed
and that the clouds on a planet consist of water, like on Earth. We varied
the O$_2$ mixing ratio $\eta$, the planetary surface albedo $A_{\rm s}$,
the cloud optical thickness $b_{\rm cloud}$, the cloud top altitude $z_{\rm top}$, 
and the cloud fraction and spatial distribution over the planet. 
We computed both the flux $F$ and the degree of linear polarization $P$ 
(or $P_{\rm s}$) of the starlight that is reflected by the model planet. 
Because of the difficulty of measuring absolute fluxes and polarization 
of exoplanets (and without accurate knowledge of the distance to an exoplanet
and its size, a measured reflected flux cannot be directly related to the 
albedo of the planet),
we focus on the relative difference between $F$ in the deepest part of the 
O$_2$ A--band (around 760.4~nm) and in the continuum outside the band 
(at 755~nm), with the latter normalized to one. For $P$, which 
is independent of distances and planetary radii because it is a relative measure, 
we focus on the absolute difference between the polarization in the deepest part
of the band and that in the continuum. We use a spectral bin width of 0.5~nm
for the spectral computations, but values for narrower bins can be derived
from computations of $F$ and $P$ as functions of the 
atmospheric absorption optical thickness.

While, as expected, $F$ in the band is always smaller than $F$ in the 
surrounding continuum, $P$ in the band is higher than in the continuum in 
most of our model computations, with strong variations in the band strength
depending on the atmospheric and surface parameters. 
For absorption line resolving observations, $P$ will show  even more 
variation with respect to its continuum value, especially in the deepest
absorption lines and with vertical structure in the atmosphere, such 
as the presence of high altitude ice clouds.

Our computations lead us to the following main conclusions: \\

\noindent $\bullet$
For $\alpha= 90^\circ$ and a given value of $\eta$, the band depth in 
(relative) $F$ is very {\em in}sensitive to the surface albedo 
$A_{\rm s}$ if it is larger 
than about 0.1, and the cloud optical thickness $b_{\rm cloud}$ if it is larger 
than about 0.5.\\ 

\noindent $\bullet$
For $\alpha=90^\circ$, the band depth in (relative) $F$ is sensitive to 
the cloud top altitude $z_{\rm top}$, although the sensitivity decreases 
for $\eta > 0.2$ and $z_{\rm top} < 10$~km). \\

\noindent $\bullet$ 
The band strength in $P$ is very sensitive to $A_{\rm s}$, except when
$\eta > 0.7$, and for $30^\circ < \alpha < 150^\circ$. If $A_{\rm s} \approx 0$,
the band strength is smaller than 0.04 (for a spectral bin width of 0.5~nm).\\ 

\noindent $\bullet$
The band strength in $P$ is very sensitive to $b_{\rm cloud}$ as long
as $b_{\rm cloud}$ is smaller than about 5 for $\eta = 0.2$. 
The band strength increases with $\eta$. If $\eta=0.2$ and $b_{\rm cloud} > 5$, 
the band strength is smaller than 0.04 
(for 0.5~nm wide bins, and $\alpha=90^\circ$). \\

\noindent $\bullet$
The band strength in $P$ decreases with $z_{\rm top}$. For $\eta=0.2$, 
the maximum band strength (for $z_{\rm top}= 2$~km) is about 0.10 at
$\alpha=90^\circ$. \\

\noindent $\bullet$
For partly cloudy planets with or without horizontally inhomogeneous surface
albedo's below the atmosphere, the degeneracies are even larger. Variations 
in $F$ and $P$ measured over time, could help to identify whether observed
band depths and strengths are due to clouds or surface features. \\

\noindent $\bullet$
Measuring both $F$ and $P$ appears essential to reduce the existing degeneracies
due to the effects of the surface albedo, cloud optical thickness and altitude,
and O$_2$ mixing ratios. Increasing the wavelength coverage (short and long
wavelengths) and the phase angle
range of the observations will also help to reduce degeneracies. In particular,
phase angles close to the rainbow angle, i.e. $38^\circ$, would not only
help with identifying the presence of liquid water particles but 
the strength of the O$_2$ A--band in polarization at that phase angle
can be used for cloud optical thickness retrieval. \\

\noindent $\bullet$
The band depth in $F$ will increase and the band strength in $P$ will become 
stronger (in vertically homogeneous atmospheres), with increasing spectral
resolution, i.e.\ increasing spectral bin width. Absorption--line resolving
observations of $P$ would reveal a large variation of the behavior of $P$
across the deepest absorption lines, in particular in the presence of 
high--altitude clouds or hazes. \\
          
The O$_2$ A--band is the strongest spectroscopic feature of dioxygen detection in 
reflected starlight. The retrieval of $O_2$ mixing ratio will be difficult due to 
degeneracies that can occur for various combinations of cloud optical thickness, 
cloud altitude, (similarly for hazes) and surface properties. Cloud composition,
which we haven't included in our study, would also play a role.
To reduce the degeneracies, the flux and the polarization should be used together 
to provide information on the cloud optical thickness, in particular if measured 
at a range of phase angles and wavelengths.


\section{Acknowledgments}

We would like to thank the anonymous reviewer for the elaborate comments
that allowed us to improve the quality of this paper.
This work was initiated a the LOA, Lille 1 University and completed during
Thomas Fauchez NASA Postdoctoral Program (NPP). This paper is not funded or 
sponsored by NASA.


\end{document}